\documentclass[11pt]{article}
\pdfoutput=1 

\usepackage{jhep_like} 

\usepackage{tensor}
\usepackage{braket}
\usepackage{comment}

\usepackage{dsfont}
\usepackage{bbm}
\usepackage{tikz}
\usepackage{upgreek}

\usepackage[normalem]{ulem}
\usepackage{amsmath}
\usepackage{enumerate}
\usepackage{epsfig}
\usepackage{mathbbol}

\newcommand\half{{\ensuremath{\frac{1}{2}}}}

\newcommand{\be}{\begin{equation}}
\newcommand{\ee}{\end{equation}}
\newcommand{\bea}{\begin{eqnarray}}
\newcommand{\eea}{\end{eqnarray}}
\newcommand{\bega}{\begin{gather}}
\newcommand{\eega}{\end{gather}}
\newcommand{\nn}{\nonumber\\}
\newcommand{\bi}{\begin{itemize}}
\newcommand{\ei}{\end{itemize}}
\newcommand{\ben}{\begin{enumerate}}
\newcommand{\een}{\end{enumerate}}
\newcommand{\bca}{\begin{cases}}
\newcommand{\eca}{\end{cases}}
\newcommand{\bln}{\begin{align}}
\newcommand{\eln}{\end{align}}
\newcommand{\bst}{\begin{split}}
\newcommand{\est}{\end{split}}
\def\ie{\begin{equation}\begin{aligned}}
\def\fe{\end{aligned}\end{equation}}
\newcommand{\bma}{\le(\begin{matrix}}
\newcommand{\ema}{\end{matrix}\ri)}
\newcommand{\bwt}{\begin{widetext}}
\newcommand{\ewt}{\end{widetext}}

\newcommand\ha{{\half}}

\def\le{\left}
\def\ri{\right}

\newcommand\sH{{\ensuremath{{\mathcal H}}}}

\newcommand\sN{{\ensuremath{{\mathcal N}}}}
\newcommand\sO{{\ensuremath{{\mathcal O}}}}
\newcommand\sP{{\ensuremath{{\mathcal P}}}}

\newcommand\sS{{\mathcal S}}

\newcommand{\Tr}{\text{Tr}}
\newcommand{\tr}{\text{tr}}

\usepackage{amsthm}

\newtheorem{theorem}{Theorem}[section]

\begin{document}

\title{Petz recovery from subsystems in conformal field theory}
\author[a]{Shreya Vardhan,}
\author[b]{ Annie Y. Wei,} 
\author[a]{ and Yijian Zou}
\affiliation[a]{Stanford Institute for Theoretical Physics, Stanford University, Stanford, CA 94305}
\affiliation[b]{Center for Theoretical Physics, 
Massachusetts Institute of Technology, Cambridge, MA 02139}

\abstract{ 
We probe the multipartite entanglement structure of the vacuum state of a CFT in 1+1 dimensions, using recovery operations that attempt to reconstruct the density matrix in some region  from its reduced density matrices on smaller subregions. We use an explicit recovery channel known as the twirled Petz map, and  
study distance measures such as the fidelity, relative entropy, and trace distance between  the original state and the recovered state. 
One setup we study in detail involves three contiguous intervals $A$, $B$ and $C$ on a spatial slice, where we can view these quantities as measuring correlations between  $A$ and $C$ that are not mediated by the region $B$ that lies between them. 
We show that each of the distance measures is both UV finite and independent of the operator content of the CFT, and hence depends only on the central charge and the cross-ratio  of the intervals. We evaluate  these universal quantities numerically using lattice simulations in critical spin chain models, and derive their analytic forms in the limit where $A$ and $C$ are close using the OPE expansion. 
In the case where $A$ and $C$ are far apart, we find a surprising non-commutativity of the replica trick with the OPE limit. For all values of the cross-ratio, the fidelity is strictly better than 
a general information-theoretic lower bound in terms of the conditional mutual information. We also  compare the mutual information between various subsystems in the original and recovered states, which leads to a more qualitative understanding of the differences between them. 
Further, we introduce  generalizations of the recovery operation to more than three adjacent intervals, for which the fidelity is again  universal with respect to the operator content. 
} 

\maketitle


\section{Introduction} 
\label{sec:intro}

In the last two decades, entanglement has provided many valuable insights into quantum many-body systems, quantum field theories, and quantum gravity. A quantity called entanglement entropy has played a central role in these studies. Given a state $\rho$ and a subsystem $A$ with complement $\bar A$, the entanglement entropy is defined as 
\be \label{ent}
S(\rho_A) = - \Tr[\rho_A \log \rho_A], \quad \rho_A = \Tr_{\bar A}\rho \, .
\ee
From an information-theoretic perspective, if the state is pure, $\rho = \ket{\psi}\bra{\psi}$, then the entanglement entropy is the answer to an intuitive operational question: what is the largest number of bell pairs between $A$ and $\bar A$ that can be extracted by acting only with local operations and classical communication (LOCC) between $A$ and $\bar A$?~\footnote{This interpretation holds in the limit of an asymptotically large number of copies of the system.}

The first studies of entanglement entropy in many-body systems considered its behaviour in conformal field theories \cite{larsen, rico, calabrese}. For a single interval in a conformal field theory in (1+1) spacetime dimensions, the entanglement entropy of the vacuum state was found to take a simple and universal form, independent of the operator content of the theory. For an interval of length $R$ in a CFT with central charge $c$ and UV cutoff $\epsilon$, the vacuum entanglement entropy is given by  
\be \label{sr_uni} 
S(R) = \frac{c}{3}\log\le(\frac{R}{\epsilon}\ri) \, . 
\ee
This universal expression has turned out to be vastly useful, with applications ranging from identifying a $c$-function that behaves monotonically under RG flow in quantum field theory \cite{casini} to motivating a key element of the AdS/CFT correspondence known as the Ryu-Takayanagi formula \cite{rt}.


While the entanglement entropy quantifies bipartite entanglement in pure states, much remains to be understood about the multipartite entanglement structure in quantum many-body systems, even in familiar settings like the vacuum state of a (1+1)-D CFT. A variety of quantities have been introduced in order to probe this structure, including the mutual information~\cite{tonni1, tonni2}, negativity~\cite{neg1,neg2}, and reflected entropy~\cite{faulkner_ref}. However, unlike the entanglement entropy, these quantities do not have a clear operational interpretation. 

In this paper, we study a set of new information-theoretic measures associated with multiple regions in the vacuum state of a (1+1)-D CFT. These measures  address operational questions about how well we can reconstruct the  density matrix of a combined system from the reduced density matrices of smaller subsystems. We will find that these measures are independent of the matter content of the CFT, and depend only on its central charge. The  universality of \eqref{sr_uni} can be understood from the fact  that the entanglement entropy of a single interval is determined by a two-point function of primary twist operators. The measures  we study in this paper are $n$-point functions of twist operators for $n \geq 4$, but nevertheless turn out to be universal. 

\begin{figure}
    \centering
    \includegraphics[width=\textwidth]{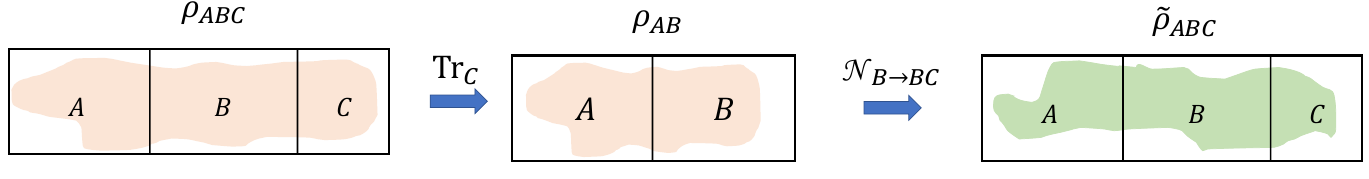}
    \caption{We trace out the subsystem $C$ from a state $\rho_{ABC}$, and then attempt to recover the original state with a channel $\sN_{B \rightarrow BC}$  that acts non-trivially only on $B$.}
    \label{fig:recov}
\end{figure}

For most of this paper, we focus on the operational question shown in Fig. \ref{fig:recov}. Suppose we have a state $\rho_{ABC}$ on the union of three subsystems $A$, $B$ and $C$. We trace out one of the subsystems $C$, and are left with the reduced state $\rho_{AB}$. We would then like to recover an approximation to the state $\rho_{ABC}$,  with the restriction that we can act with some non-trivial quantum channel on $B$, but not on $A$. How close can the recovered state 
$\sN_{B \rightarrow BC}(\rho_{AB})$ be to the state $\rho_{ABC}$? 

Intuitively, the restricted set of operations we allow can generate direct correlations between $B$ and $C$, but any correlations that they generate between $A$ and $C$ must be mediated by $B$.   This operational question is thus a concrete way of probing correlations between the subsystems $A$ and $C$ that are not mediated by $B$. 
Ideally, we would want to minimize the distance between the two states for all possible choices of the channel $\sN_{B\rightarrow BC}$. It is not clear how to carry out such a minimization procedure in practice, so we will instead make use of an explicit channel called the twirled Petz map, which can be defined for any state:
\be \label{twirled}
{\sP^{(\lambda)}}_{B \rightarrow BC} (\cdot)= \rho_{BC}^{\ha -\frac{i\lambda}{2}}\rho_{B}^{-\ha+\frac{i\lambda}{2}} (\cdot) \rho_{B}^{-\ha- \frac{i\lambda}{2}}\rho_{BC}^{\ha + \frac{i\lambda}{2}}
\ee
The parameter $\lambda$ can be any real number. The $\lambda=0$ case is called the Petz map.

\begin{figure}[!h]
    \centering
    \includegraphics[width=\textwidth]{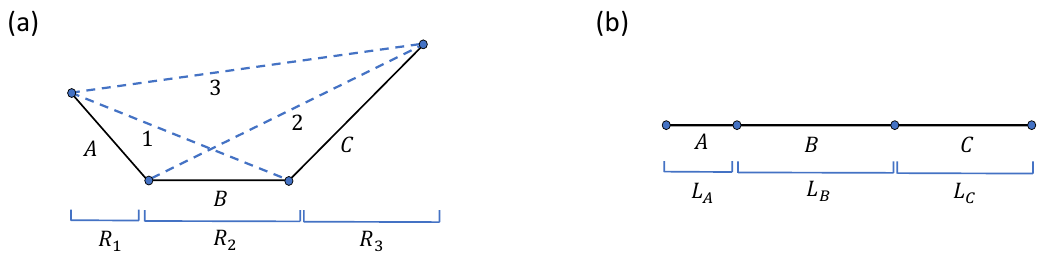}
    \caption{(a) shows a setup in conformal field theory in 1+1 dimensions where the recovery operation of Fig. \ref{fig:recov} can be carried out perfectly. $B$ is a spatial slice, $A$ and $C$ are null, and 1, 2 and 3 are spacelike. (b) shows the  setup we study in this paper, where $A$, $B$ and $C$ are spacelike intervals and the recovery is imperfect.}  \label{fig:regions}
\end{figure}

Previous works   \cite{cmarkov_1, cmarkov_2, ch_lectures} identified  certain setups in quantum field theory where the above recovery operation works perfectly. As shown in \cite{hayden, petz1, petz2}, 
perfect recovery for the setup of Fig.~\ref{fig:recov} in any quantum state $\rho_{ABC}$ is equivalent to the 
vanishing of a quantity called the conditional mutual information (CMI), defined as  
\be 
I(A:C|B) = I(A:BC)- I(A:B)= I(C:AB) - I(C:B)
\ee
where $I(P:Q)$ is the mutual information 
\be 
I(P:Q) = S(\rho_{P})+S(\rho_Q) - S(\rho_{PQ}) \, . 
\ee
 The CMI can be written more explicitly in terms of entropies as 
\be 
I(A:C|B)= S(\rho_{AB})+S(\rho_{BC}) - S(\rho_B) - S(\rho_{ABC}) \, .   \label{cmidef}
\ee
One setup where the CMI vanishes is when $A$ and $C$ are null regions on either side of a spacelike interval $B$ in a (1+1)-D CFT, shown in Fig.~\ref{fig:regions}~(a). To see this, note that since $AB$ is related to the spacelike slice 1 by unitary evolution, 
\be 
S(\rho_{AB}) = S(\rho_1) = \frac{c}{3}\log\le(\frac{\sqrt{R_2 (R_2+2R_1)}}{\epsilon}\ri)
\ee
where in the last inequality, we have used \eqref{sr_uni} 
together with the fact that the vacuum entanglement entropy in some region in a relativistic theory is a function only of the Lorentz-invariant length of the region. Similarly using $S(\rho_{BC})=S(\rho_2)$ and $S(\rho_{ABC})=S(\rho_3)$ and substituting into \eqref{cmidef}, we find that 
the CMI vanishes.  

As emphasized in \cite{ch_lectures}, the fact that $A$ and $C$ are null regions is important for the perfect recoverability in this setup. The statement of perfect recoverability  and the vanishing of the CMI are also equivalent to a third statement that $\rho_{ABC}$ has a ``Markov state'' structure \cite{hayden}, which in particular implies that the reduced density matrix on $AC$ is separable between these two systems.  Due to the Reeh-Schlieder theorem~\cite{reeh,witten_notes}, the reduced density matrix for any two regions of non-zero volume in the vacuum state of a QFT cannot be  separable~\cite{distillable}. Indeed, if we consider the case shown in Fig.~\ref{fig:regions}~(b), where $A$, $B$, and $C$ are three adjacent regions on a spatial slice, we can use \eqref{sr_uni} to see that the CMI is non-zero and is given by 
\be \label{cmi}
I(A:C|B)= - \frac{c}{3}\log(1- \eta), \quad \eta = \frac{L_A L_C}{(L_A+L_B)(L_B+L_C)} \, 
\ee
indicating that perfect recovery is not possible in general.
When $L_B$ is much greater than at least one out of $L_A$ and $L_C$, so that the cross-ratio $\eta$ is much smaller than 1, the CMI is small and we can use certain information-theoretic inequalities to give a non-trivial lower bound on how well the recovery works~\cite{sutter, junge, wilde}. We review these inequalities in Section \ref{sec:def}.

To better understand the entanglement structure of the vacuum state, it is useful to directly quantify the imperfect recoverability in the setup of Fig.~\ref{fig:regions}~(b) for arbitrary sizes of $A$, $B$ and $C$. 
This is our goal in the present paper, where we 
study the difference  between $\rho_{ABC}$ and the recovered state 
\be
\tilde \rho_{ABC}^{(\lambda)} = {\sP^{(\lambda)}}_{B \rightarrow BC} (\rho_{AB}) \label{tildedef}
\ee
by a few different distance measures: the fidelity, trace distance, relative entropy, and a one-parameter generalization of the Renyi entropy.~\footnote{To simplify the notation below, we will sometimes drop the superscript $(\lambda)$ and simply write $\tilde\rho_{ABC}$.} In principle, we should be careful in applying the map \eqref{twirled} in a quantum field theory, since the reduced density matrices appearing in it are not well-defined without putting a UV cutoff on the theory using some lattice regularization. Despite this, we will find 
that the various distance measures between $\rho_{ABC}$ and $\tilde \rho_{ABC}$ are independent of the UV cutoff and are finite in the continuum limit. This shows that although $\tilde \rho_{ABC}$ is constructed from reduced density matrices, it is able to capture the large amount of short-distance entanglement between adjacent subsystems that is present in $\rho_{ABC}$.    Due to conformal invariance, we further find that these measures depend on the interval lengths only through the cross-ratio $\eta$  defined in \eqref{cmi}.  

More remarkably still, we find that each of these distance measures between $\rho_{ABC}$ and $\tilde \rho_{ABC}$  depends on the specific CFT being studied only through its central charge. This universality can be seen by using replica tricks which allow us to express these quantities as analytic continuations of four-point functions of twist operators for an integer number of copies of the CFT.
Standard techniques for studying such correlation functions involve mapping them to a partition function on a  covering space~\cite{lunin}. The genus of the covering space is determined by the structure of the twist operators, and turns out to be zero for the quantities studied here, leading to a universal result. 

To find these universal functions numerically, we use lattice simulations in critical spin chain models including the Ising model and the free fermion and free boson CFTs. We find that the $\lambda=0$ case of \eqref{twirled} corresponds to the best recovery.
As $L_B$ becomes much smaller than both $L_A$ and $L_C$, so that $\eta \to 1$, both the logarithm of  the fidelity $F(\rho_{ABC}, \tilde \rho_{ABC})$ and the relative entropy $D(\rho_{ABC}, \tilde \rho_{ABC})$ diverge,~\footnote{We define $F(\rho,  \sigma)$ and $D(\rho||\sigma)$ in Sections \ref{sec:def} and \ref{sec:rel_ent} respectively.}
\begin{align} 
&-\log F(\rho_{ABC}, \tilde \rho_{ABC}) = -\frac{c}{9}\log(1-\eta) + \sO(1)  \label{18}\\
&D(\rho_{ABC}|| \tilde \rho_{ABC}) = -\frac{c}{3}\log(1-\eta) + \sO(1)  , \quad \quad \quad \eta \to 1\, . \label{19}
\end{align}
The coefficient of the logarithmically divergent contribution to the relative entropy is the same as that of the CMI in \eqref{cmi}, while the coefficient in $-\log F$ is smaller, consistent with the bounds of \cite{junge, sutter, wilde}. 
We explain these limiting behaviours using the OPE expansions in the twist operator formalism. This formalism also allows us to relate the $\sO(1)$ contributions in both \eqref{18} and \eqref{19} to other entanglement quantities, some of which are not obviously related to the fidelity or relative entropy from a general information-theoretic perspective. 

In the limit of small $\eta$, where at least one out of $L_A$ and $L_C$ is much smaller than $L_B$, we numerically find at $\lambda =0$ that both $-\log F$ and $D$ approach 0 with a quadratic dependence on $\eta$, 
\begin{align}
&-\log F(\rho_{ABC}, \tilde \rho_{ABC}) = f_2 \, c \,  \eta^2 + \sO(\eta^4)  \label{fs}\\
&D(\rho_{ABC}|| \tilde \rho_{ABC}) = d_1 \, c\, \eta^2  + \sO(\eta^4), \quad\quad\quad \eta \to 0 \, . \label{ds}
\end{align}
for some numerically determined universal constants $f_2$ and $d_1$. In contrast, the upper bound on \eqref{fs} from \cite{junge, sutter, wilde} is linear in $\eta$. 
We also study $-\log F$ at other values of $\lambda$, and find  that it approaches zero with a $\lambda$-dependent power as 
$\eta \to 0$. Using the OPE expansion for this limit in the replica formalism, we are able to correctly see that both quantities are zero in the strict $\eta=0$ limit. However, this formalism can also be used to argue that all coefficients in the expansion of $-\log F$ or $D$ away from the strict $\eta = 0$ limit should vanish, which contradicts the  numerically observed power laws in \eqref{fs} and \eqref{ds}. We expect that the incorrect prediction from the replica formalism in this case likely comes from a non-commutation between the replica limit and this OPE limit; see Sec. \ref{sec:smallz} for details.


In addition to the above quantitative measures of the distance between $\rho_{ABC}$ and $\tilde \rho_{ABC}$, we can also  ask more qualitative questions about which subregions or which types of correlations account for the distance between the two states. This is useful both for understanding the structure of the original vacuum state, and for understanding the structure of the state formed by \eqref{twirled} from an information-theoretic perspective. Immediately from the definition of \eqref{twirled}, we can see that $\tilde \rho_{ABC}$ has identical reduced density matrices to $\rho_{ABC}$ on $A$ and on $BC$. However, the correlations between $A$ and $B$, and between $B$ and $C$, are different in $\tilde \rho_{ABC}$ from those in $\rho_{ABC}$. 
We consider the following  mutual information differences between the original and the recovered state: 
\be \label{delta1}
\delta I(A:B) = I(A:B)_{\rho} - I(A:B)_{\tilde \rho} \, ,  \quad  \delta I(A:C) = I(A:C)_{\rho} - I(A:C)_{\tilde \rho} \, . 
\ee
$\delta I(A:B)$
and $\delta I(A:C)$ are both UV-finite functions of the cross-ratio. $\delta I(A:B)$ depends only on the central charge of the CFT, while $\delta I(A:C)$ is also sensitive to the operator content. In the $\eta \to 1$ limit, where the intervals are close, $\delta I(A:B)$ approaches a constant non-zero value, while $\delta I(A:C)$ shows a divergence, similar to  the CMI, $-\log F$, and the relative entropy. In fact, the divergence has precisely the same form as that of the CMI. So in this limit, we see that $\tilde \rho_{ABC}$
fails to capture a diverging amount of correlations between $A$ and $C$, as well as some  finite amount of correlations between $A$ and $B$.

In the limit $\eta \to 0$ where the intervals $A$ and $C$ are far, $\delta I(A:B)$ is larger than $\delta I(A:C)$, showing that the loss of  correlations between $A$ and $B$ accounts for a larger part of the difference between $\rho_{ABC}$ and $\tilde \rho_{ABC}$ in this limit. The  behaviour of $\delta I(A:C)$ in this limit is non-universal, but still  shows some interesting general features. It is well-known that $I(A:C)$ in the vacuum state decays as \cite{tonni2} 
\be 
I(A:C)_{\rho} = \kappa \, \eta^{2 (h+\bar h)}, \quad \eta \to 0  \, . \label{2hp_i}
\ee
for some constant $\kappa$, where $h~(\bar{h})$ is the (anti-)holomorphic dimension of the lowest-dimension  primary operator after the identity. 
The mutual information between $A$ and $C$ in $\tilde \rho_{ABC}$ turns out to have precisely the same leading behaviour,  
so that $\delta I(A:C)$ has a smaller  leading power than \eqref{2hp_i}. In this sense, the leading correlations between $A$ and $C$ in the far-interval limit are mediated by $B$. 

We also introduce some natural generalizations of the recovery process of Fig.~\ref{fig:recov} to multiple intervals and multiple steps. We show general information-theoretic lower bounds on the recovery fidelity of multiple-step protocols for four subsystems in terms of the CMI, by extending the methods of \cite{junge, sutter, wilde}.  For the CFT vacuum state, we show explicitly that these quantities are universal for the case of four adjacent intervals. Based on this, and a general argument discussed in the conclusions, it is natural to conjecture that the generalization to an arbitrary number of adjacent intervals is also universal. As expected, we find quantitatively that the fidelity of the single-step recovery process is higher than that of  the multiple-step ones, which in turn are larger than the general information-theoretic lower-bounds we show for them. 

The plan of the paper is as follows. In Section \ref{sec:eig}, we study the fidelity between $\rho_{ABC}$ and $\tilde \rho_{ABC}$, and develop the general formulation in terms of twist operators which is also used with some small modifications for all other quantities studied in later sections. In Section \ref{sec:rel}, we study the relative entropy, a one-parameter generalization called the Petz-Renyi relative entropy, and the trace distance between $\rho_{ABC}$ and $\tilde \rho_{ABC}$, and compare both sides of various general information-theoretic quantities relating these distance measures. In Section \ref{sec:corr}, we explore the differences in correlations between different subsystems in the two states. In Section \ref{sec:multiple}, we 
discuss generalizations to multiple intervals and multiple steps.
We discuss a number of future directions motivated by our  results in Section \ref{sec:disc}.

Appendix \ref{app:num} provides details of the tensor network algorithms and numerical methods used to evaluate  various quantities. In Appendix \ref{app:lou}, we review the  covering space method of \cite{lunin}, derive the values of certain OPE coefficients from it, and discuss some challenges with analytic continuation using this method. In Appendix \ref{app:proof}, we show general information-theoretic bounds on multiple-step recovery tasks for four intervals. 




\section{Fidelity between $\rho_{ABC}$ and $\tilde \rho_{ABC}$}
\label{sec:eig}

In this section, we compare the state $\rho_{ABC}$ with the state $\tilde \rho_{ABC}$ obtained using the Petz map by evaluating the fidelity between the two states. We define the fidelity and discuss a general information-theoretic bound on it in Section~\ref{sec:def}, and present numerical results in Section~\ref{sec:num}. In Section~\ref{sec:rep}, we 
develop a replica formalism for this quantity and discuss  its consequences, including universality.   Sections~\ref{sec:z1} and~\ref{sec:smallz} discuss the OPE limits.

\subsection{Review of fidelity and information-theoretic bounds}
\label{sec:def}
%


In any quantum-mechanical system, the fidelity between two states $\rho$ and $\sigma$ is defined as 
\be 
F(\rho, \sigma) = || \sqrt{\rho} \sqrt{\sigma} ||_1, \quad ||X||_1 = \text{Tr}[(X^{\dagger} X)^{1/2}] \, . \label{fid}
\ee
This quantity satisfies the following upper and lower bounds: 
\be \label{bounds}
0\leq F(\rho, \sigma) \leq \Tr[\rho] \Tr[\sigma] = 1 \, . 
\ee
In particular, the upper bound holds for normalized states even in infinite-dimensional systems. 
$F(\rho, \sigma)=1$ if and only if $\rho=\sigma$,  and $F(\rho, \sigma)=0$ if 
the support of $\rho$ is orthogonal to that of $\sigma$. 
The quantity $-2\log F(\rho, \sigma)$ can  therefore be seen as a measure of distance between $\rho$ and $\sigma$, which is sometimes referred to as the min-relative entropy. For the later discussion, it is useful to note that we can also write 
\be \label{fdef2}
F(\rho, \sigma) = \Tr[(\rho \sigma)^{\ha}] 
\ee
The right-hand side is defined as the sum of the square root of the eigenvalues of $\rho\sigma$. Although $\rho\sigma$ is not a Hermitian matrix in general, its eigenvalues are equal to those of $\sqrt{\rho}\sigma\sqrt{\rho}$, and so in particular are real and nonnegative. This is due to the fact that for any two square matrices $A$ and $B$, the matrices $AB$ and $BA$ have the same eigenvalues. To see \eqref{fdef2}, take $A =\sqrt{\rho}\sigma$ and $B = \sqrt{\rho}$.

To understand physically why the fidelity is a good measure of the distance between two states, suppose we measure both states using some positive operator-valued measurement $\{ E_m \}$, generating the probability distributions $p_m = \Tr[E_m \rho_m]$ and $q_m = \Tr[E_m \sigma_m]$. The quantum fidelity between $\rho$ and $\sigma$ turns out to be the minimum of the classical fidelity $\sum_m \sqrt{p_m}\sqrt{q_m}$~ between these probability distributions over all possible choices of $\{E_m\}$. In this sense, the fidelity tells us  how well two states can be distinguished by an optimal measurement~\cite{nc}.

In any quantum system and for any subsystems $A, B, C$, for the pair of states $\rho_{ABC}$ and  $\tilde \rho_{ABC}^{(\lambda)}$, we have a lower bound on the fidelity  in terms of the CMI defined in \eqref{cmidef}. 
Recall that the strong subadditivity of entropy is the statement that \eqref{cmidef} is non-negative. As mentioned in the introduction, when the strong subadditivity inequality is saturated, i.e. $I(A:C|B)=0$, \cite{petz1, petz2, hayden} showed that 
$\tilde \rho_{ABC}^{(\lambda)} = 
\rho_{ABC}$
for any real value of $\lambda$. More generally, we have the following inequalities between the CMI  and $F(\rho_{ABC}, \tilde \rho_{ABC}^{(\lambda)})$ \cite{sutter, junge, wilde}:~\footnote{These  inequalities were shown in \cite{sutter, junge, wilde} for general states in quantum mechanical systems, which are type I von Neumann algebras. They follow from a more general inequality which strengthens the data-processing inequality for any quantum channel. 
A proof of these inequalities for general quantum channels was given for Type III von Neumann algebras, which include quantum field theories, in \cite{faulkner_alg}.} 
\be 
{\rm min}_{\lambda} - \log F\le(\rho_{ABC}, \tilde \rho_{ABC}^{(\lambda)}\ri) \leq  \ha I(A:C|B) \label{in} \, . 
\ee
and 
\begin{align} 
 &- \int_{-\infty}^{\infty} d\lambda \, \beta(\lambda) \,  \log F\le(\rho_{ABC}, \tilde \rho_{ABC}^{(\lambda)}\ri) \leq \ha I(A:C|B) \, , \label{110}
 \\ 
 & - \log F\le( \rho_{ABC}, \int_{-\infty}^{\infty} d\lambda \, \beta(\lambda) \,  \tilde \rho_{ABC}^{(\lambda)}  \ri)  \leq \ha I(A:C|B) \, , \label{111}
\end{align}
where 
\be 
\beta(\lambda) = \frac{\pi}{2} \frac{1}{1+\cosh(\pi \lambda)} \label{fi}
\ee
Note that \eqref{110} implies \eqref{111} due to concavity of the function $\log F$. 
When the CMI is small, these inequalities tell us that both the optimal case of  $\sP^{(\lambda)}$ and its average with the probability distribution \eqref{fi} work well. 

Let us now turn to our setup in Fig.~\ref{fig:regions}(b) of three adjacent intervals on a spatial slice in the vacuum state of a (1+1)-D CFT, and recall the formula \eqref{cmi} for the CMI.  
Note that the dependence of the UV cutoff in \eqref{sr_uni} gets cancelled out in the CMI, so that the latter is a well-defined quantity in the continuum. We can immediately use the above discussion to conclude certain properties of the fidelity: 
\begin{enumerate}
\item 
Although the expression for $F(\rho_{ABC}, \tilde \rho_{ABC})$ involves both positive and negative powers of various reduced density matrices, it neither diverges nor approaches zero in the continuum limit. We can see the fact that it does not diverge from the upper bound in \eqref{bounds}, since we have $\Tr[\rho]= \Tr[\sigma]=1$ even in the continuum limit.  
From \eqref{cmi} and \eqref{in}, we have a lower-bound on $\max_\lambda F(\rho_{ABC}, \tilde \rho_{ABC}^{(\lambda)})$ which does not depend on $\epsilon$. Assuming that the qualitative features are similar for different $\lambda$, we conclude that the fidelity does not approach 0 for $\epsilon \rightarrow 0$. We will see more explicitly in the following sections that this quantity is independent of the UV cutoff. 

\item In the limit $\eta \rightarrow 0$, we must have $F(\rho_{ABC}, \tilde \rho_{ABC}^{(\lambda)}) \rightarrow 1$ for all $\lambda$, since \eqref{110} implies that the average value approaches 1, which is also the largest possible value.   
\end{enumerate} 
Note further that the expression \eqref{cmi} for the CMI depends only on the central charge of the CFT and not on its detailed operator content. One natural way for the above bounds to be satisfied is if the fidelity is also universal with respect to the operator content. We will find that this is indeed the case.

\subsection{Numerical results}
\label{sec:num}

Before introducing a replica trick formalism to study $F(\rho_{ABC}, \tilde \rho_{ABC})$ analytically, let us summarize some numerical results for this quantity from lattice calculations. 
We consider lattice discretizations of three CFTs: the Ising CFT with $c=0.5$, the tricritical Ising CFT with $c=0.7$, and the compactified free boson CFT with $c=1$. We make use of critical quantum spin chains which realize these CFTs, with length $L$ and periodic boundary conditions. 

 
 The Ising CFT and the tricritical Ising CFT are realized using the O'Brien-Fendley model \cite{OBF}
\begin{equation}
    H = -\sum_{i=1}^L (\sigma^{x}_i \sigma^{x}_{i+1}+\sigma^z_i) + \lambda\sum_{i=1}^L (\sigma^{x}_i \sigma^{x}_{i+1} \sigma^z_{i+2}+\sigma^z_{i}\sigma^x_{i+1}\sigma^x_{i+2})
\end{equation}
This model has  a critical line at $0\leq \lambda \leq \lambda^{*} \approx 0.428$. At $\lambda = 0$, the model is the transverse field Ising model. At $\lambda=\lambda^{*}$, the model flows to the tricritical Ising CFT. For $0<\lambda<\lambda^{*}$, we have an RG flow from the tricritical Ising CFT to the Ising CFT. The free boson CFT is realized by the XXZ spin chain
\begin{equation}
    H = -\sum_{i=1}^L (\sigma^{x}_i \sigma^{x}_{i+1} +\sigma^{y}_i \sigma^{y}_{i+1} + \Delta \sigma^{z}_i \sigma^{z}_{i+1}),
\end{equation}
where $-1\leq \Delta<1$, and the compactification radius $R$ is related to $\Delta$ as $\Delta = \cos (2\pi/R^2)$ \cite{Alcaraz_1988}. The operator content depends on $R$ but the central charge is constant and equal to 1 along the critical line. 

 We obtain the ground state in each case using the periodic uniform matrix product state method \cite{puMPS} and compute the fidelity $F(\rho, \tilde \rho)$ numerically.  In computing the fidelity, we make use of the Uhlmann theorem \cite{Uhlmann} to reduce the numerical cost, which enables us to go to up to $L=128$ spins for the Ising CFT and $L=60$ for the free boson CFT. See Appendix \ref{app:num} for details on the numerical implementation. 

\begin{figure}[!h]
    \centering
    \includegraphics[width = 0.7\linewidth]{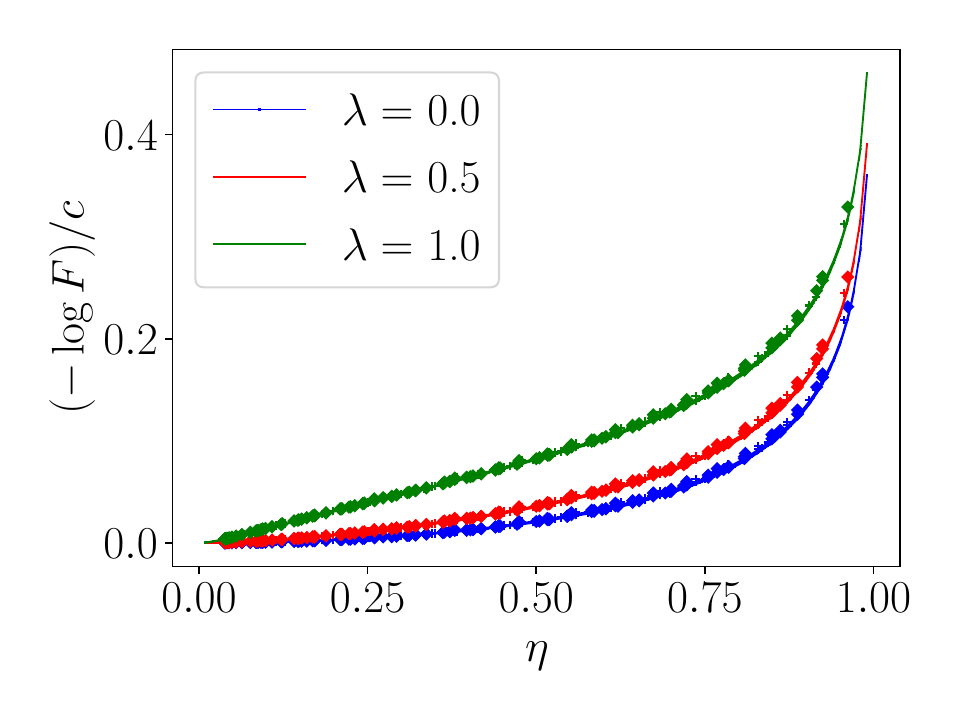}
    \caption{We show the universal behaviour of $-\log F^{(\lambda)}(\eta)/c$ in different conformal field theories.  Different colors label different values of $\lambda$. The solid line (markers not shown), diamond, and the "+" markers represent numerical data for the Ising, TCI, and XXZ ($\Delta=-0.6$) models, respectively.}
    \label{fig:Flambda}
\end{figure}

\begin{figure}[!h]
    \centering
    \includegraphics[width = 0.7\linewidth]{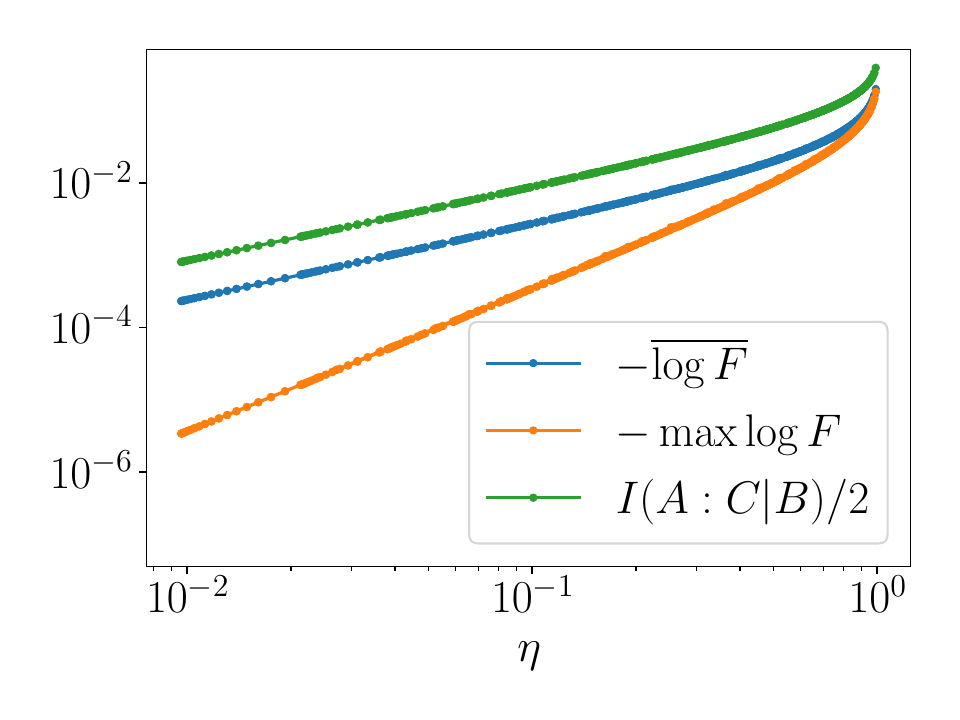}
    \caption{Behaviour of $-\log F^{(\lambda)}$ and the CMI $I(A:C|B)$ for the Ising model with $c=1/2$. The orange curve shows the $\lambda=0$ case, which has a slope $2$ as $\eta\rightarrow 0$, indicating $-\log F \propto \eta^2$. The blue curve shows the averged fidelity $-\overline{\log F}$, which has a linear dependence in $\eta$ as $\eta\rightarrow 0$ (parallel to the green CMI line). }
    \label{fig:Flambda_avg}
\end{figure}

\begin{figure}[!h]
    \centering
    \includegraphics[width = 0.7\linewidth]{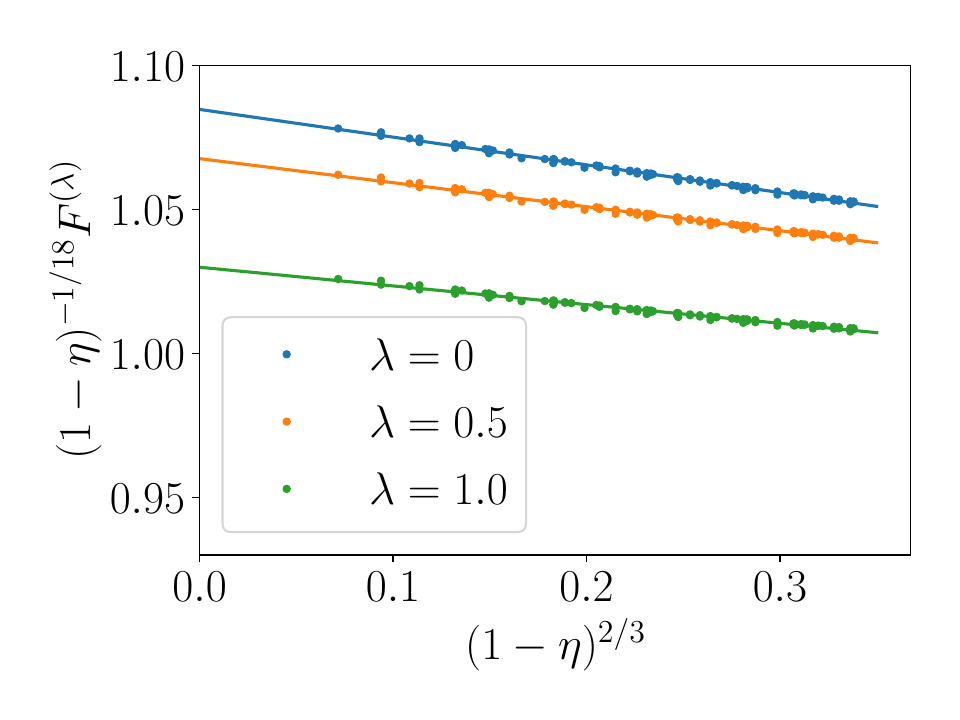}
    \caption{Behaviour of $F^{(\lambda)}(\eta)$ at large $\eta$ for the Ising model. The x-axis is $(1-\eta)^{2/3}$ and the y-axis is $(1-\eta)^{-c/9} F^{(\lambda)}$. The linear fit shows the dependence of the form Eq.~\eqref{216}, where the slope and intercept correspond to $a_1 c e^{a_0 c}$ and $e^{a_0 c}$, respectively. }
    \label{fig:Flambda2}
\end{figure}

We observe the following features of the fidelity $F(\rho_{ABC}, \tilde \rho_{ABC})$: 
\begin{enumerate}

\item In Fig. \ref{fig:Flambda}, for each model and choice of $\lambda$, we consider the quantity $F(\rho_{ABC}, \tilde \rho_{ABC}^{(\lambda)})/c$ for various different choices of $L_A, L_B, L_C$, and plot it as a function of the cross-ratio $\eta$. The cross-ratio is  
defined for a  compact system with periodic boundary conditions as 
\begin{equation}
    \eta = \frac{\sin(\pi L_A/L)\sin(\pi L_C/L)}{\sin(\pi (L_A+L_B)/L)\sin(\pi (L_B+L_C)/L)} \, . 
\end{equation}
We find that there is no dependence on the individual $L_A, L_B, L_C$ except through the cross-ratio, which in particular shows that   
$F(\rho_{ABC}, \tilde \rho_{ABC})$  is independent of the UV cutoff.  

\item $- \log F^{(\lambda)}(\eta)/c$ is a universal function which does not depend on any details of the CFT. We show in Fig.~\ref{fig:Flambda} that for each value of $\lambda$, the data for each of the three different CFTs (Ising, TCI, and XXZ ($\Delta=-0.6$)) collapses on to a single curve. We have also checked that the curve does not depend on $\Delta$ for the XXZ model, and so do not explicitly show the other choices of $\Delta$ on the plot.

\item $- \log F^{(\lambda)}(\eta)/c$ has a non-trivial dependence on $\lambda$, with the largest fidelity achieved at $\lambda =0$ for all $\eta$, and decreasing monotonically with increase in $\lambda$.

\item As shown in Fig. \ref{fig:Flambda_avg}, the inequalities with the conditional mutual information \eqref{in} and \eqref{110} are not saturated for any value of $\eta$. Both quantities 
\be 
-\log \max F = -\log F^{(\lambda=0)}  
\ee
and 
\be \label{av_f}
- \overline{\log F} = - \int_{-\infty}^{\infty} d\lambda \, \beta(\lambda) \,  \log F\le(\rho_{ABC}, \tilde \rho_{ABC}^{(\lambda)}\ri)
\ee
are strictly smaller than $I(A:C|B)/2$. ~\footnote{The average in \eqref{av_f} is computed numerically by using the range of $\log F^{(\lambda)}$ with $-3\leq \lambda \leq 3$, with the Romberg integration method at a step size $\delta\lambda = 0.1$.} 

\item 
At small $\eta$, where at least one out of $A$ and $C$ is much smaller than $B$,   $-\log F^{(\lambda=0)}$ has a perturbative expansion that starts at quadratic order, 
\begin{equation}
\label{eq:F_smalleta}
     -\log F^{(\lambda=0)}= f_2 \, c \, \eta^2 + \sO(\eta^4)  , \quad  \eta\ll 1 \, ,
\end{equation}
where $f_2 \approx 0.070$ is a numerically determined constant.  For a general value of $\lambda$, we observe that $-\log F^{(\lambda)} \propto \eta^p$ at small $\eta$, where the power $p$ decreases as $\lambda$ increases. We show this small $\eta$ behaviour in Fig. \ref{fig:Flambda_small}.

\begin{figure}[!h]
    \centering
    \includegraphics[width = 0.7\linewidth]{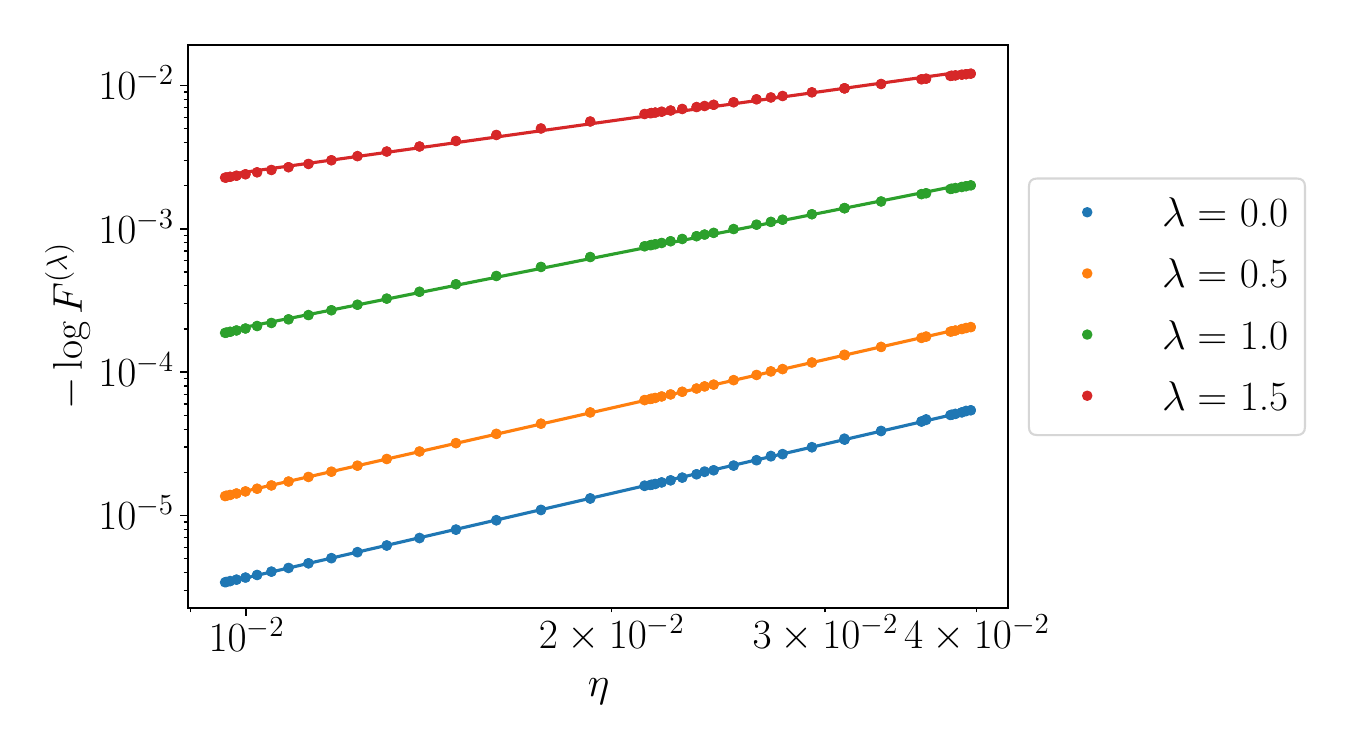}
    \caption{Behaviour of $-\log F^{(\lambda)}$ at small $\eta$, for the Ising model with $c=1/2$. $\eta$ is shown on a log scale on the $x$-axis. 
    Different colors label different values of $\lambda$. Lines represent fitting with a power law $\eta^p$, where the four lines have slope $p\approx 2.0,1.9,1.7,1.2$ for $\lambda = 0,0.5,1,1.5$, respectively.}
    \label{fig:Flambda_small}
\end{figure}

The average over $\lambda$  in \eqref{av_f} has the form
\begin{equation}
    -\overline{\log F} = \overline{f}\,  c \, \eta + \sO(\eta^2)  , \quad  \eta\ll 1, 
\end{equation}
where $\overline{f} \approx 0.055$.

Note that the expansion of the conditional mutual information \eqref{cmi} at small $\eta$ starts at linear order: 
\begin{align} 
\ha I(A:C|B) & = \frac{c}{6}\eta + \sO(\eta^2)     
\end{align} 
The inequality Eq.~\eqref{110} requires that $\overline{f}\leq 1/6$, which is satisfied but not saturated. 

\item For the regime $1 -\eta \ll 1$, where $A$ and $C$ are both much larger than $B$, we have an expansion for any $\lambda$ of the form 
\begin{equation} \label{216}
    F = e^{a_0 c} (1-\eta)^{c/9} + a_1 \, c \,  e^{a_0 c} (1-\eta)^{c/9+2/3} + \sO\le((1-\eta)^{c/9+1}\ri), \quad \ 1- \eta \ll 1
\end{equation}
See Fig.~\ref{fig:Flambda2}. The coefficients $a_0$ and $a_1$ are 
$\lambda$-dependent constants. 
So in particular the inequality \eqref{in} is again not saturated: 
\begin{align} 
\ha I(A:C|B) & = -\frac{c}{6}\log(1-\eta)  \\
 -\log F(\rho_{ABC}, \tilde \rho_{ABC}^{(\lambda)}) &= -\frac{c}{9}\log(1-\eta)  \,+ \sO(1) \, .  
\end{align} 
\end{enumerate} 


\subsection{Replica formalism and properties of the fidelity}
\label{sec:rep}

In this section, we first introduce a replica trick in order to derive the fidelity through analytic continuation. This maps the problem to studying the   partition function  on $M$ copies of the theories for some integer $M$, with different boundary conditions connecting the copies in $A$, $B$, and $C$. We show how to write this partition function  as a four-point function of twist operators associated with different permutations in $\sS_M$, the symmetric group of $M$ elements. By examining the properties of this four-point function, we explain various features of the fidelity observed in the previous section, including independence of the UV cutoff and universality with respect to the operator content. We then consider the OPE limits of this four-point function.  

\subsubsection{Replica trick and analytic continuation}

Recall that we want to evaluate the quantity 
\be \label{81}
F^{(\lambda)} = \Tr\le[ \le((\rho_{BC}^{\ha- \frac{i\lambda}{2}}\rho_B^{-\ha +\frac{i\lambda}{2}}\rho_{AB} \rho_B^{-\ha+ \frac{i\lambda}{2}} \rho_{BC}^{\ha-\frac{i \lambda}{2}}) \rho_{ABC}\ri)^{\ha}\ri] \, . 
\ee
As we discuss below, standard techniques involving partition functions on an integer number of copies of the theory will allow us to calculate a quantity of the form 
\be \label{fint}
F_{k, n_1, n_2, m_1, m_2} = \text{Tr}[(\rho_{BC}^{m_1} \rho_B^{n_1} \rho_{AB} \rho_B^{n_2} \rho_{BC}^{m_2} \rho_{ABC})^{k} ]
\ee
for integer values of each of the parameters $m_i, n_i,  k$.
We can infer  \eqref{81} from \eqref{fint} using analytic continuation, similar to the replica trick used in \cite{pssy} for  Petz recovery in a different context. Since we must analytically continue five different parameters in \eqref{fint}, let us try to spell out the assumptions carefully. 

Let us first promote $m_i, n_i$ to arbitrary complex numbers while requiring $k$ to be a positive integer. The quantity \eqref{fint} is  still well-defined for such values of the parameters: if $v_i$ are the (in general complex) eigenvalues of the matrix inside the parentheses in \eqref{fint}, then \eqref{fint} is given by $\sum_i v_i^k$. We conjecture that \eqref{fint} is an analytic function of $m_i, n_i$ in the region defined by  
\be  \label{bounded}
\text{Re}[m_1]>0, \quad \text{Re}[m_2]>0, \quad \text{Re}[n_1]>- \text{Re}[m_1] - a, \quad \quad \text{Re}[n_2]>- \text{Re}[m_2] - a, 
\ee
for some $a > 0$.~\footnote{We checked this statement numerically in some special cases. For example, for $k=1$, $m_1 = m_2 = \ha$ and $n_1=n_2=n$, we find that \eqref{fint} is bounded for $n\gtrsim-0.8$.} 
In particular, the case 
where we set $m_i, n_i$ equal to the values in \eqref{81} falls within this range. Provided \eqref{fint} also satisfies the growth conditions of Carlson's theorem, we can try to uniquely infer the value of 
\be 
F_k^{(\lambda)} = \Tr\le[ \le((\rho_{BC}^{\ha- \frac{i\lambda}{2}}\rho_B^{-\ha +\frac{i\lambda}{2}}\rho_{AB} \rho_B^{-\ha+ \frac{i\lambda}{2}} \rho_{BC}^{\ha-\frac{i \lambda}{2}}) \rho_{ABC}\ri)^k\ri]
\label{fkl}\ee
by analytic continuation of $m_i, n_i$ from positive integer values of these parameters.

Next, we want to consider $F_k^{(\lambda)}$ for general complex values of $k$. Recalling from the discussion below \eqref{fdef2} that the 
eigenvalues of $\rho \sigma$ for two density matrices $\rho, \sigma$ are non-negative, we see that \eqref{fkl} is well-defined for complex values of $k$.  
 We further expect based on numerical evidence that  $F_k^{(\lambda)}$ is  upper-bounded by 1 for $\text{Re}(k)>0$, and analytic  in this regime.  Then provided we can find an analytic continuation of $F_k^{\lambda}$ that satisfies Carlson's theorem, we can evaluate the quantity \eqref{81}.

Below we will sometimes refer to this analytic continuation of all five variables as the {\it replica limit}: 
\be \label{rep}
m_1 \rightarrow \ha - \frac{i\lambda}{2}, \quad n_1 \rightarrow -\ha + \frac{i\lambda}{2}, \quad m_2 \rightarrow \ha + \frac{i\lambda}{2}, \quad n_2 \rightarrow -\ha - \frac{i\lambda}{2}, \quad  \quad k \rightarrow \ha \, . 
\ee

\subsubsection{Representation in terms of twist operators}
\label{sec:twist}
Let us now set up the calculation of the quantity  \eqref{fint} for positive integer values of all parameters in a QFT in 1+1 dimensions. We consider the theory on the manifold $\mathbf{R}^2$.
Results for the CFT on a  cylinder, which was considered in the numerics, can be derived by conformal mapping.   

\begin{figure}[!h]
    \centering   \includegraphics[width=\textwidth]{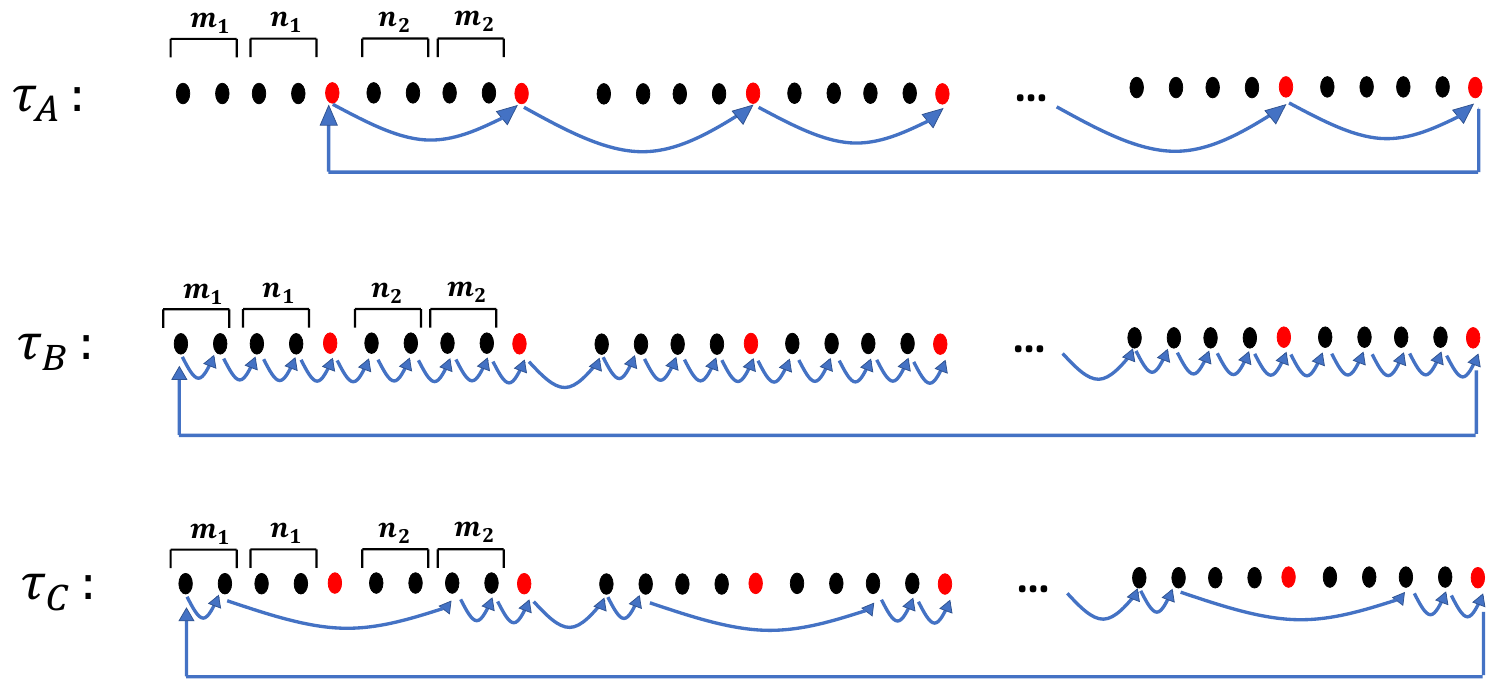}
    \caption{We show the permutations in \eqref{perm1}-\eqref{perm3}. In all such figures, the copies are labelled from 1 to $Nk$ from left to right unless explicitly stated otherwise. There are $k$ groups of $N=n_1+m_1+n_2+m_2+2$ copies. The arrows show how the copies map to each other under the permutation. Copies that are not connected by arrows  map to themselves.}
    \label{fig:perms}
\end{figure}

Recall that the matrix elements of the vacuum state $\rho=\ket{0}\bra{0}$ of a QFT between two states $\ket{\phi}$ and $\ket{\phi'}$ are given by 
\begin{align} \label{bc_1}
\braket{\phi| 0} \braket{0|\phi'} 
= \frac{1}{Z_1} \, \le( \int_{\tau = -\infty}^{\psi(\tau= 0)=\phi} D \psi \, e^{- I_E [\psi, \partial_{\mu} \psi] } \ri) \times \le( \int_{\psi(\tau = 0)=\phi'}^{\tau = \infty} D \psi \, e^{- I_E [\psi, \partial_{\mu} \psi] } \ri)
\end{align} 
where $I_E$ is the Euclidean action, and 
\be 
 Z_1 = \int D \psi e^{-I_E[\psi, \partial_{\mu}\psi]} \, . 
\ee
\eqref{fint} is then given by a Euclidean path integral on $Nk$ copies of the theory, with $N = m_1+n_1 +m_2+n_2+2$. On the $I$-th copy, we have a matrix element 
$\braket{\phi_I| 0} \braket{0|\phi'_I}$.
The partial traces and matrix multiplications in \eqref{fint}  indicate that in each of the subsystems $P=$  $A$, $B$, $C$, and the complement $\overline{ABC}$,   we should identify each $\phi_I'$ with some $\phi_I$. More explicitly, we have the conditions $\phi'_I(\vec x) = \phi_{\tau_P(I)}(\vec x)$ for certain permutations $\tau_P \in \sS_{Nk}$. Since $\overline{ABC}$ is traced out in each copy of $\rho$,  $\tau_{\overline{ABC}}$ is the identity permutation. The permutations in $A$, $B$, and $C$ are given by (see Fig. \ref{fig:perms} for diagrams)
\begin{align} \label{perm1}
\tau_A &= (m_1+ n_1 +1 , ~ N ,  ~ N + m_1 + n_1 + 1, ~ 2N, ~ ... ~, (k-1)N + m_1 + n_1 +1, ~ k N ) \\
\tau_B &= (1, 2, ... \, , Nk -1, Nk ) \label{bdef}\\
\tau_C &= (1, 2, ... \, , m_1, ~~ m_1 + n_1+n_2+2, m_1 + n_1+n_2+3, \, ... \, , \,   N, \nn
&~~~N+1, N+2, ... \, , N+m_1, ~~ N+m_1 + n_1+n_2 +2, N+m_1 + n_1+n_2 +3, \, ... \, , 2N, \nn
& ~~~~~~~~~~~... Nk)  \label{perm3}
\end{align} 
Here we use the cycle notation for each permutation, so for instance $\tau = (1, 2, 3)$ in $\sS_5$ refers to the permutation which sends 1 to 2, 2 to 3, 3 to 1, and the remaining elements 4 and 5 to themselves. In enumerating the cycles of various permutations below, we will often  explicitly discuss only cycles of length greater than 1. Each of the permutations $\tau_{A, B, C}$ consists of a single  cycle, which has length $2k$, $Nk$, and $(m_1+m_2+1)k$ respectively.  

\begin{figure}[!h]
\centering
\includegraphics[width=\textwidth]{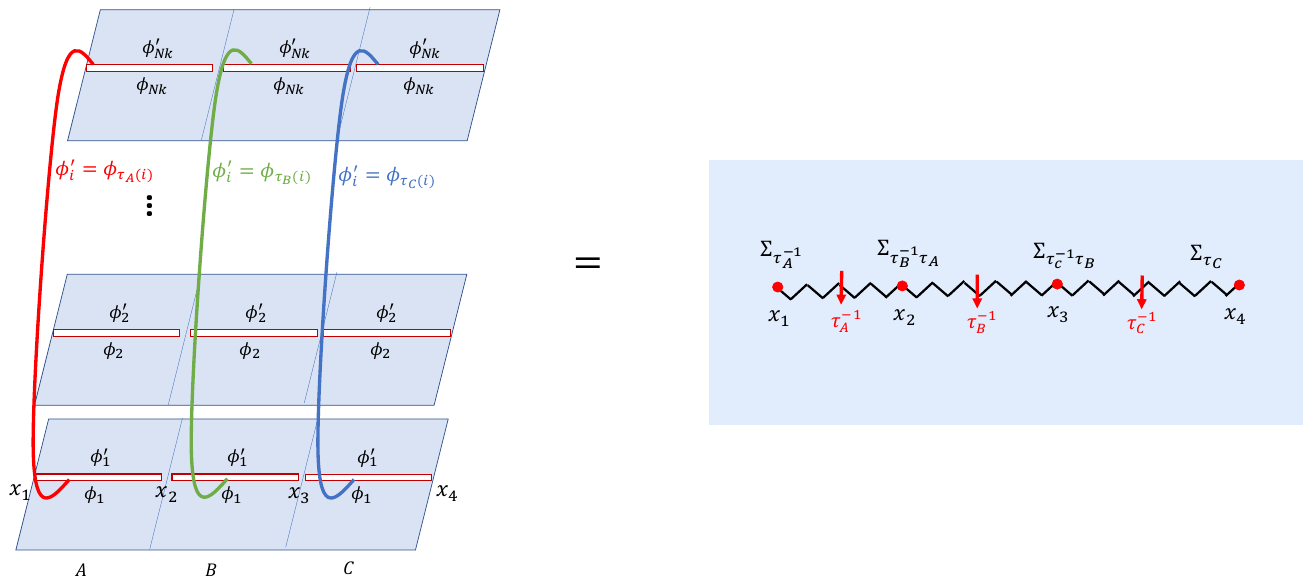}
\caption{The Euclidean path integral for \eqref{fint} on $Nk$ copies, with boundary conditions relating the different copies specified by $\tau_{A, B, C}$, is shown on the left. This can be reinterpreted as a four-point function of twist operators in a theory with $Nk$ copies of the fields living on ${\bf R}^2$, as shown on the right.}
\label{fig:twist_ops}
\end{figure}

\begin{figure}[!h]
    \centering   \includegraphics[width=\textwidth]{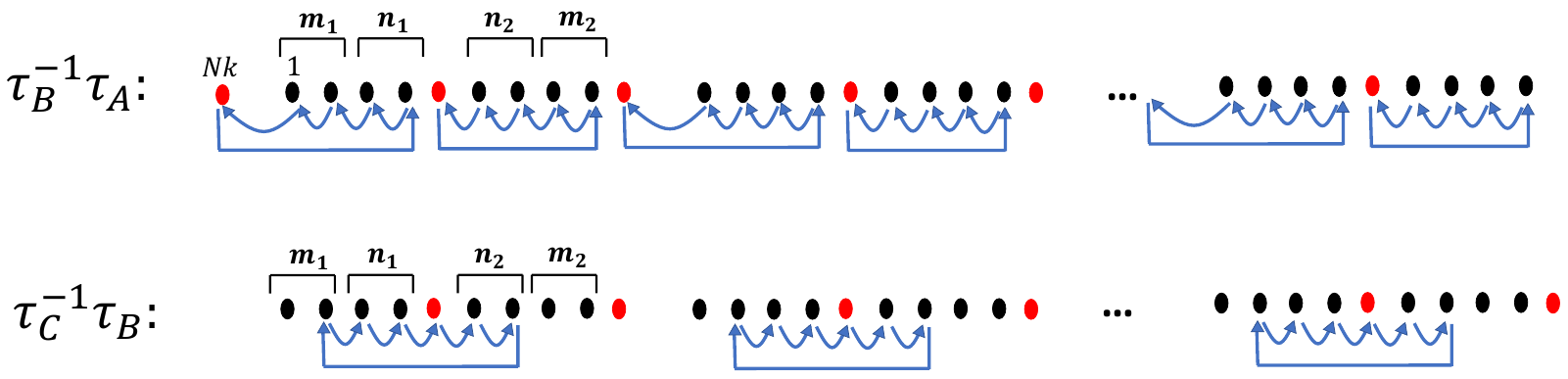}
    \caption{We show the permutations  \eqref{tba} and \eqref{tcb}, using similar notation to Fig. \ref{fig:perms}. Note that in $\tau_B^{-1}\tau_A$, we put the $Nk$-th copy to the left of the first.}
    \label{fig:comp_tau}
\end{figure}

\begin{figure}[!h]
\centering
\includegraphics[width=0.7\textwidth]{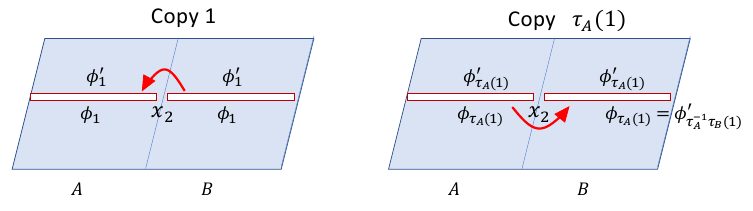}
\caption{We show an example of how to identify the twist operator inserted at $x_i$, by considering a closed path around $x_i$ in the presence of the boundary conditions in the left figure of Fig. \ref{fig:twist_ops}.}
\label{fig:perm}
\end{figure}

The path integral for \eqref{fint} thus involves  $Nk$ copies of the fields $\psi_I(x^{\mu})$, whose combined action is given by 
\be 
I_E^{\rm tot}[\psi_1, \partial_{\mu}\psi_1, ...,  \psi_{Nk}, \partial_{\mu}\psi_{Nk}] = \sum_{I=1}^{Nk} I_E [\psi_I, \partial_{\mu}\psi_I] \, . \label{itot}
\ee
with the boundary conditions shown in the left figure of Fig. \ref{fig:twist_ops}.
As shown on the right of Fig. \ref{fig:twist_ops}, the boundary conditions can be seen as the effect of certain twist  operators inserted in the path integral at the points $x_i$. For a permutation $\tau \in \sS_n$, we define the twist operator $\Sigma_{\tau}(p)$ as an operator that implements the following boundary condition on a small circle around the point $p$: 
\be 
\Sigma_{\tau}(p): ~~\psi_I (x^{\mu}) \mapsto \psi_{\tau(I)}(x^{\mu}) ~~~\text{on going anticlockwise around }p\, .\label{sigmadef} \, 
\ee 
Then we have the following representation for the path integral:  
\begin{align}  
&F_{k, n_1, n_2, m_1, m_2} = \braket{\Sigma_{\tau_A^{-1}}(x_1) \Sigma_{\tau_B^{-1}\tau_A}(x_2)\Sigma_{\tau_C^{-1}\tau_B}(x_3)\Sigma_{\tau_C}(x_4) }  \label{4pt} \\
& \quad \quad= \frac{1}{(Z_1)^{N k}} \times \int \, \prod_I D \psi_I \, e^{- I_E^{\rm tot}} \,  \Sigma_{\tau_A^{-1}}(x_1) \Sigma_{\tau_B^{-1}\tau_A}(x_2)\Sigma_{\tau_C^{-1}\tau_B}(x_3)\Sigma_{\tau_C}(x_4) \, . \label{4p2} 
\end{align} 
The twist operators inserted at $x_2$ and $x_3$ are associated with the permutations   
\begin{align}  \label{tba}
\tau_B^{-1} \tau_A = &(m_1 + n_1 ,~ m_1 + n_1 -1 ,~ ...,~1, Nk ) ~( N-1,~N-2,~ ... , m_1+n_1+1)  \nn
& (N + m_1+n_1, N+ m_1+n_1, ..., N) ~ (2N-1,~ 2N-2, ..., N + m_1+n_1+1) \nn
& ... ((k-1)N+ m_1 + n_1 ,~ ... ,~
 (k-1)N)~(kN-1,~  ... ,~ (k-1)N+m_1+n_1+1) 
 \end{align} 
 and 
 \begin{align} \label{tcb}
\tau_C^{-1} \tau_B  =& (m_1 , m_1+1, ...,  m_1 +n_1+ n_2 +1) \nn
 & (N + m_1 , N + m_1+1, ..., N+   m_1 +n_1 +n_2 +1) \nn & ... ((k-1)N + m_1 , (k-1)N + m_1 +1, ..., (k-1)N + m_1 +n_1 + n_2 +1 )
\end{align} 
These permutations are shown in Fig. \ref{fig:comp_tau}. Note in particular that: 
\begin{itemize}
    \item 
$\tau_B^{-1}\tau_A$ has $k$ cycles of length $m_1+n_1+1$, and $k$ cycles of length $m_2+n_2+1$.
\item $\tau_C^{-1} \tau_B$ has $k$ cycles of length $n_1+n_2+2$. 
\end{itemize}
In order to see that the relevant twist operators should be the ones appearing in \eqref{4pt}, we can start with the field  $\phi_I'$ and go in a small anticlockwise circle around $x_i$ in the left figure of Fig. \ref{fig:twist_ops}, and see which $\phi_{J}'$ it gets mapped to. We show an example in Fig. \ref{fig:perm}. 

In later sections, we will study other quantities such as the 
relative entropy between $\rho_{ABC}$ and $\tilde \rho_{ABC}$, or the entanglement entropy of various subsystems in $\tilde \rho_{ABC}$. By similar reasoning, we will express these quantities as four-point functions of the form \eqref{4pt}, with the difference that  $\tau_{A, B, C}$ refer to some other permutations depending on the quantity. 

\subsubsection{Consequences of the twist operator representation}
\label{sec:comments}
Let us now make some comments both on the formal properties and the physical consequences of the expression \eqref{4pt} or \eqref{4p2}: 
\begin{enumerate}
    \item In a 1+1 D CFT, each $\Sigma_{\tau}$ defined by \eqref{sigmadef} is a primary scalar operator. If $\tau$ has $t$ cycles with $a_1, a_2, ..., a_t$ elements respectively, then the holomorphic dimension of $\Sigma_{\tau}$ is 
\be 
\Delta_{\tau} = \sum_{i=1}^{t} \Delta_{(a_i)}, \quad \Delta_{(n)} \equiv \frac{c}{24} n \le(1- \frac{1}{n^2} \ri) \, .\label{dim}
\ee
Hence, the holomorphic dimensions of the operators inserted at the points $x_i$ are respectively 
\begin{align} 
& \Delta_{1} = \Delta_{\tau_A^{-1}} = \frac{c}{24}(2k) \le(1- \frac{1}{(2k)^2}\ri), \label{d1}\\ 
&\Delta_2 = \Delta_{\tau_B^{-1}\tau_A} = k\frac{c}{24}(m_1 + n_1 + 1) \le(1- \frac{1}{(m_1 + n_1 + 1)^2}\ri) \nonumber \\
&\quad \quad \quad \quad \quad \quad + k\frac{c}{24}(m_2 + n_2 + 1) \le(1- \frac{1}{(m_2 + n_2 + 1)^2}\ri), \\
& \Delta_3 =\Delta_{\tau_C^{-1}\tau_B} = k\frac{c}{24}( n_1 + n_2+  2) \le(1- \frac{1}{(n_1 + n_2 + 2)^2}\ri), \\ \label{d4}
 &\Delta_4 = \Delta_{\tau_C} = \frac{c}{24}(m_1 + m_2 + 1)k\le(1- \frac{1}{(m_1 + m_2 + 1)^2k^2}\ri) \, . 
 \end{align} 
These operators are scalars, so their antiholomorphic dimensions, $\bar \Delta_1$, ..., $\bar \Delta_4$, are also given by \eqref{d1}-\eqref{d4}. 
 
 \item 
In the replica limit \eqref{rep}, note that the lengths of all cycles of each of the permutations \eqref{perm1}-\eqref{perm3} and \eqref{tba}-\eqref{tcb} become 1. As a result, the  dimensions in \eqref{d1}- \eqref{d4} are  zero in this limit. Naively, \eqref{4pt} in the replica limit  seems to reduce to a four-point function of identity operators. Despite this, we find that the fidelity has a non-trivial behaviour as a function of $\eta$, as we discuss more explicitly in Section \ref{sec:z1}. 

 \item The twist operators in \eqref{4pt} are not local operators: as discussed in the literature on $CFT^{\otimes M}/\sS_M$ orbifolds, we get genuine local operators only on summing over  operators corresponding to all permutations in a given conjugacy class. The correlation function is therefore well-defined only on specifying the branch cuts of the twist operators, as we have done in Fig. \ref{fig:twist_ops}. Due to this non-locality, the correlation function is not invariant under arbitrary reorderings of the twist operators, but it is invariant under cyclic reorderings, as discussed in \cite{dei_eb, rastelli}. 
 The OPE of any two such   operators will also depend on their ordering.  
\item As discussed in \cite{lunin}, we can evaluate any correlation function of twist operators such as \eqref{4pt} by 
mapping the original manifold ${\bf R}^2$, on which the fields are multi-valued due to the presence of twist operators, to a  ``covering space'' on which the fields become single-valued. The covering space has a metric $ds^2$ with non-trivial curvature, and can also have non-trivial topology. By the Weyl anomaly, 
\begin{align} 
 &\int \, \prod_I D \psi_I \, e^{- I_E^{\rm tot}} \,  \Sigma_{\tau_A^{-1}}(x_1) \Sigma_{\tau_B^{-1}\tau_A}(x_2)\Sigma_{\tau_C^{-1}\tau_B}(x_3)\Sigma_{\tau_C}(x_4) \nn 
 & ~ = \int_{ds^2} D\psi e^{- I_{E, g}[\psi]} 
 = e^{S_L} \int_{d\tilde s^2} D\psi e^{- I_{E}[\psi]} \, . \label{sl} 
\end{align} 
where $d\tilde s^2$ is some simple metric on the covering space. $S_L$ is the Liouville action that relates the partition functions with metrics $ds^2$ and $d\tilde s^2$. Its only dependence on the specific CFT involved is through an overall factor of the central charge $c$; the remaining factor is determined entirely by the structure of the permutations. We review these methods in some more detail in Appendix~\ref{app:lou}.  
\item The genus of the covering space is determined by the cycle structure of the permutations in the correlation function. 
Suppose the total number of cycles involved in all the twist operators of a given correlation function is $l$, the total number of copies of the theory that appear in one or more of the $l$ cycles is $s$, and  the length of the $i$'th cycle is $n_i$. Then the genus of the covering space is given by the {\it Riemann-Hurwitz formula}: 
\be \label{rhf_1}
g = \frac{1}{2}\sum_{i=1}^l(n_i-1) - s +1 \, . 
\ee
Putting the data for the permutations in \eqref{4pt} into this formula, we find $g=0$ for any  $n_i, m_i, k$.
Hence we can take $\tilde g$ in \eqref{sl} to be the flat metric on ${\bf R}^2$. Then putting \eqref{sl} into \eqref{4p2}, we find   
\be 
F_{k, n_1, n_2, m_1, m_2} = e^{S_L} Z_1^{1-Nk} \, . 
\ee
In the replica limit, $Nk = 1$, so 
$-\log F(\rho_{ABC}, \tilde \rho_{ABC}^{(\lambda)})$ is equal to the limit \eqref{rep} of $- S_L$. As discussed in the previous point, the only dependence on the specific CFT in this quantity is the overall factor of the central charge.  This explains the universality of $-\log F/c$ that we found numerically in Fig. \ref{fig:Flambda}. 
    \item For positive integer values of $m_i, n_i, k$, the quantity \eqref{4pt} has an explicit dependence on the UV cutoff, similar to the Renyi entropies. To see this, note that the boundary condition \eqref{sigmadef} should be imposed in a small circle of radius $\epsilon$ around the point $p$. The Liouville action depends on this UV cutoff $\epsilon$. In order to cancel out the dependence on $\epsilon$, we can divide \eqref{4pt} by the quantity 
    \be 
    \sN = \braket{\Sigma_{\tau_A^{-1}}(0) \Sigma_{\tau_A}(a)}^{\ha} \braket{\Sigma_{\tau_B^{-1} \tau_A}(0) \Sigma_{\tau_A^{-1}\tau_B}(a)}^{\ha} \braket{\Sigma_{\tau_B^{-1}\tau_C}(0) \Sigma_{\tau_C^{-1}\tau_B}(a)}^{\ha} \braket{\Sigma_{\tau_C}(0) \Sigma_{\tau_C^{-1}}(a)}^{\ha}
    \ee
for some arbitrary length $a$. See for instance Appendix D of \cite{avery} for an explanation of this cancellation. Note that $\sN$ can be expressed in terms of the density matrix $\rho_{Q}$ on an interval $Q$ of length $a$ as 
\be 
\sN = \Tr[\rho_Q^{2k}]^{\ha} \Tr[\rho_Q^{m_1+n_1+1}]^{\frac{k}{2}} \Tr[\rho_Q^{m_2+n_2+1}]^{\frac{k}{2}} \Tr[\rho_Q^{n_1+n_2+2}]^{\frac{k}{2}} \Tr[\rho_Q^{(m_1+m_2+1)k}]^{\ha}
\ee
In the replica limit \eqref{rep}, we simply have $\sN= \Tr[\rho_Q]=1$, which indicates that this limit of $S_L$ does not depend on the UV cutoff. This is consistent with the vanishing of the dimensions \eqref{d1}-\eqref{d4} in the replica limit.   This explains the cutoff-independence of the fidelity, which we anticipated from upper and lower bounds in Section \ref{sec:def} and observed numerically in Section \ref{sec:num}.

    \item Due to conformal invariance, since \eqref{4pt} is a four-point function of  primary scalar operators, it takes the following form \cite{yellowbook}:
\be 
F_{k, n_1, n_2, m_1, m_2}  = g_{1, 2, 3, 4}(\eta, \bar \eta)\prod_{i<j} |x_{ij}|^{2\Delta/3 - 2\Delta_i - 2\Delta_j} \label{4p_g}
\ee
where $x_{ij} = x_i - x_j$, $\Delta = \sum_{i=1}^4\Delta_i$, and 
\be \label{uvdef}
\eta = \frac{x_{12} \, x_{34}}{x_{13} \, x_{24}} = \frac{L_A L_C}{(L_A + L_B)(L_C+L_B)} . 
\ee
In the replica limit, the prefactor involving $|x_{ij}|$ becomes 1 as all $\Delta_i$ are zero, so the fidelity only depends on the cross ratio, as we observed numerically. Note in particular that the fidelity is unaffected on interchanging the values of $L_A$ and $L_C$, despite the asymmetric roles of $A$ and $C$ in the definition \eqref{twirled} of the twirled Petz map.  This invariance is an interesting feature of this quantity specific to conformally invariant theories.  

\end{enumerate}

In principle, the covering space method introduced in \cite{lunin} and used in points 4 and 5 above can be used to find the expression for $-\log F(\eta)/c$ if we can evaluate and analytically continue $S_L$. An  important ingredient of this calculation is finding the covering map that takes us from the base space to the covering space. In Appendix \ref{app:lou}, we write down a parameterization of the covering map for integer values of $m_i, n_i, k$. However, in order to fix certain constants in the map which appear in $S_L$, we need to solve polynomial equations of arbitrarily high degree for general $m_i, n_i, k$. No general analytic solutions can be found for such equations, which in turn prevents us from obtaining an analytic expression for $S_L$ in terms of $m_i, n_i, k$ using this method. 

In the next two subsections, we study the two OPE limits $\eta \rightarrow 1$ and $\eta \rightarrow 0$ of the four-point function \eqref{4pt}, which are  more tractable than finding the expression for arbitrary $\eta$. We will find that  we are able to explain all numerical observations of the  $\eta\rightarrow 1$ limit, and find and confirm some interesting interpretations of the coefficients of the expansion in that limit, but the $\eta\rightarrow 0$ limit has some subtle features.

\subsection{$\eta\rightarrow 1$ limit}
\label{sec:z1}

The cross-ratio $\eta$ approaches 1 in the limit where $L_A$ and $L_C$ are both much larger than $L_B$. When $\eta$ becomes exactly equal to 1, the region $B$ disappears, and the states $\rho_{ABC}$,  $\tilde \rho_{ABC}$ reduce to $\rho_{AC}$,  $\rho_A \otimes \rho_C$. The fidelity $F(\rho_{AC}, \rho_A \otimes \rho_C)$ is equal to zero in the continuum limit $\epsilon \to 0$, so we should find that $F(\rho_{ABC},\tilde\rho_{ABC})$ approaches zero as $\eta \rightarrow 1$. In this section, we will find the way in which this quantity approaches zero by using its OPE expansion.

Recall that the correlation function of interest is
\be 
\braket{\Sigma_{\tau_A^{-1}}(x_1) \Sigma_{\tau_B^{-1}\tau_A}(x_2)\Sigma_{\tau_C^{-1}\tau_B}(x_3)\Sigma_{\tau_C}(x_4) } \, \label{corrf}  
\ee
Since various other quantities we study will also be expressible as  four-point functions of this kind, let us write down a series expansion allowing the permutations $\tau_{A, B, C}$ to be general, and later specialize to the specific permutations \eqref{perm1}-\eqref{perm3} for the fidelity. 

As $x_2 \rightarrow x_3$, we can use the OPE of $\Sigma_{\tau_B^{-1}\tau_A}$ and $\Sigma_{\tau_C^{-1}\tau_B}$,  
\be 
\Sigma_{\tau_B^{-1}\tau_A}(x_2) \Sigma_{\tau_C^{-1}\tau_B}(x_3) = \sum_{O_p} x_{23}^{\Delta_p-\Delta_2-\Delta_3}{\bar x_{23}}^{\bar\Delta_p-\Delta_2-\Delta_3}  f_{23{\bar O_p}}\le(1 + \sO(x_{23}\partial_{x})\ri)\, O_p(x_2) \, . \label{exp} 
\ee
Here 2 and 3 are shorthand for the operators $\Sigma_{\tau_B^{-1}\tau_A}$ and $\Sigma_{\tau_C^{-1}\tau_B}$, and $f_{A B C}$ is the  coefficient appearing in the three-point function $\braket{A(x_1) B(x_2) C(x_3)}$.
Since $A$, $B$, $C$ are twist operators, the ordering in the three-point function is important. We consider the $\braket{A(x_1) B(x_2) C(x_3)}$ with $x_1< x_2 < x_3$, and put the branch cuts between $x_1,~x_2$, and between $x_2,~x_3$. $\bar O_p$ is the primary  operator with which $O_p$ has a non-zero two-point function. (For non-gauge-invariant twist operators, $O_p \neq \bar O_p$.)

Putting the expansion \eqref{exp} into the four-point function \eqref{4pt}, we find 
\begin{align} \label{series}
&\braket{\Sigma_{\tau_A^{-1}}(x_1) \Sigma_{\tau_B^{-1}\tau_A}(x_2)\Sigma_{\tau_C^{-1}\tau_B}(x_3)\Sigma_{\tau_C}(x_4) } \nn
&\quad \quad \quad = f_1(x_{ij}, \Delta_i) \sum_p \frac{f_{23 \bar O_p} f_{1 O_p 4}}{c_{O_p }} (1-\eta)^{\Delta_p}(1-\bar \eta)^{\bar \Delta_p} g_{\Delta_4, \Delta_1, \Delta_2, \Delta_3; \Delta_p, \bar \Delta_p}(1-\eta, 1- \bar \eta)
\end{align} 
where $c_{O_p}$ is the overall coefficient in the two-point function of $O_p$ and $\bar O_p$, the overall factor involving $x_{ij}$ is 
\be\label{f1def}
f_1(x_{ij}, \Delta_i)= 
\frac{1}{|x_{23}|^{2(\Delta_2+\Delta_3)}|x_{14}|^{2(\Delta_1+\Delta_4)}} \le(\frac{|x_{13}|}{|x_{34}|}\ri)^{2(\Delta_4-\Delta_1)} \le(\frac{|x_{34}|}{|x_{24}|}\ri)^{2(\Delta_2-\Delta_3)}  
\ee
and $g$ is the  conformal block for any (1+1)-D CFT which takes into account contributions from descendants, and has an expansion of the form 
\be \label{block_a_1}
g_{\Delta_1 , \Delta_2, \Delta_3, \Delta_4 \, ; \, \Delta_{p}, \bar \Delta_p}(\eta, \bar \eta) =
(1 +\sum_{k=1}^{\infty} B^{(k)}_{\Delta_1 , \Delta_2, \Delta_3, \Delta_4 \, ; \, \Delta_{p}} \eta^{k})(1 +\sum_{k=1}^{\infty} B^{(k)}_{\Delta_1 , \Delta_2, \Delta_3, \Delta_4 \, ; \, \bar\Delta_{p}} {\bar\eta}^{k})
\ee
where $B^{(k)}_{\Delta_1 , \Delta_2, \Delta_3, \Delta_4 \, ; \, \Delta_{p}}$ are universal coefficients  that are fixed by conformal symmetry, and that have been evaluated for instance in \cite{cft_long}.

\begin{figure}[!h]
    \centering
    \includegraphics[width=0.8\textwidth]{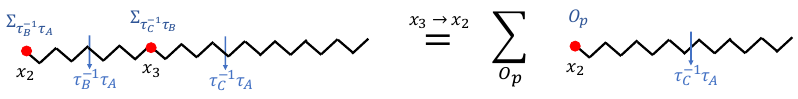}
    \caption{Each operator appearing in the OPE of $\Sigma_{\tau_B^{-1}\tau_A}$ and $\Sigma_{\tau_C^{-1}\tau_B}$ should  give rise to the same  branch cut at long distances as the combined effect of $\Sigma_{\tau_B^{-1}\tau_A}$ and $\Sigma_{\tau_C^{-1}\tau_B}$.}
    \label{fig:branch_cuts}
\end{figure}

Let us now identify the operators $O_p$ appearing in the OPE \eqref{exp}, whose dimensions will determine the behaviour of \eqref{corrf} in the $\eta \rightarrow 1$ limit. Since the twists associated with $\Sigma_{\tau_B^{-1}\tau_A}$ and $\Sigma_{\tau_C^{-1}\tau_B}$ do not cancel with each other, all primary operators appearing in the OPE must also implement the same twisted boundary conditions at infinity as the combined effect of these operators. 
See Fig. \ref{fig:branch_cuts}. Below we will label the primary operators appearing in the OPE with $p = a, b, c, ...$. The lowest-dimension operator of this kind is $O_a = \Sigma_{\tau_C^{-1}\tau_A}$. 

\begin{figure}
    \centering
\includegraphics[width=\textwidth]{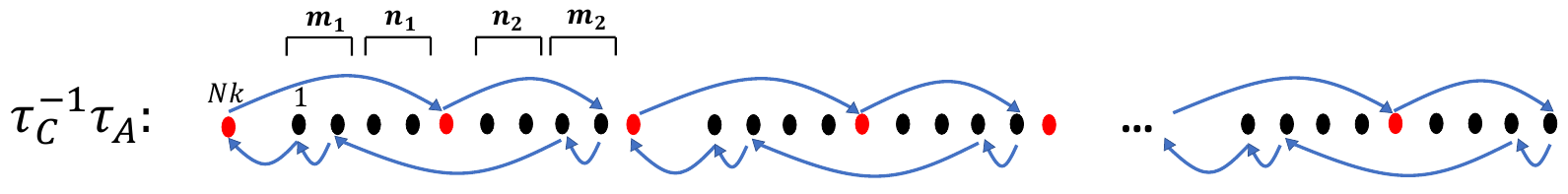}
    \caption{The permutation $\tau_C^{-1}\tau_A$ from \eqref{ac}.}
    \label{fig:tauca}
\end{figure}

The discussion so far applied to any correlation function of the form \eqref{corrf}. Let us now consider the specific case of the fidelity, by taking the permutations $\tau_{A,B,C}$  to be the ones in \eqref{perm1}-\eqref{perm3}. Then $\tau_C^{-1}\tau_A$ is the following permutation, 
shown in Fig. \ref{fig:tauca}:  
\begin{align} 
\tau_C^{-1}\tau_A =& (N-1, N-2, ..., N - m_2+1, m_1, m_1-1, ..., 1, Nk, m_1+n_1+1) \nn
& (2N-1, 2N-2, ..., 2N - m_2+1, N+m_1, N+m_1-1, ..., N+1, N, N+m_1+n_1+1) \nn
&... (Nk -1, ..., Nk-m_2+1, N(k-1)+m_1, \nn
&\quad  N(k-1)+m_1-1, ..., N(k-1)+1, N(k-1), N(k-1)+m_1+n_1+1) 
 \label{ac}
\end{align} 
$\tau_{C}^{-1}\tau_A$ has $k$ non-trivial cycles, each of length $m_1+m_2+2$, and its dimension is therefore 
\be \label{del0}
\Delta_{a} = \bar \Delta_a =  k \frac{c}{24}\le(m_1+m_2+2- \frac{1}{m_1+m_2+2}\ri) \, . 
\ee
The remaining  primary operators that create the boundary conditions in Fig. \ref{fig:branch_cuts} are excitations of $\Sigma_{\tau_C^{-1}\tau_A}$ with fractional modes. Recall that in a 1+1-D CFT without twisted boundary conditions, we can give the following mode decomposition of any quasiprimary field with dimension $h$:
\be 
\phi(z) = \sum_{m \in {\bf Z}} z^{-m-h} \phi_{m}  \, . 
\ee
If we have $q$ copies of the quasiprimary field $\phi_I$, which are multiple-valued due to the presence of a twist operator $\Sigma_{\tau_q}(0)$ associated with 
$\tau_q=(1, ..., q)$, then the mode decomposition includes fractional powers of $z$: 
\be 
\phi_I(z) = \sum_{m \in {\bf Z}} z^{-\frac{m}{q}-h} e^{-2\pi i \frac{m}{q}(k-1)} \phi_{m/q} \, .  
\ee
For the holomorphic and antiholomorphic parts of the stress tensor $T(z)$, $\bar T(\bar z)$ the associated modes are labelled $L_{m/q}$, $\bar L_{m/q}$. The commutation relations of the operators $L_{m/q}$ and $\phi_{m/q}$ can be worked out by mapping these operators to the covering space on which the fields become single-valued; see \cite{roump, peet} for details. Using this algebra, we find that each $\phi_{-m/q}\Sigma_{\tau_q}$ for $-h \leq m \leq q-1$ and each $L_{-m/q}\Sigma_{\tau_q}$ for $-2 \leq m \leq q-1$ is a primary operator.  For a twist operator with multiple cycles such as $\Sigma_{\tau_C^{-1}\tau_A}$, we can independently dress any subset of cycles with fractional modes. For instance, the lowest-dimension fractional mode operators are ones where a single cycle $i$ is dressed. We list a few such operators and their holomorphic and anti-holomorphic dimensions.   
\begin{align} 
&O_b = \phi^{(i)}_{-\frac{h}{m_1+m_2+2}, -\frac{\bar h}{m_1+m_2+2}} \Sigma_{\tau_C^{-1}\tau_A}, \quad \Delta_b = \Delta_{a}+\frac{h}{m_1+m_2+2}, \quad \bar \Delta_{b} = \Delta_{a}+\frac{\bar h}{m_1+m_2+2} \, ,  \label{opl}\\
&O_c = L^{(i)}_{-\frac{2}{m_1+m_2+2}} \Sigma_{\tau_C^{-1}\tau_A}, \quad \Delta_c = \Delta_a+\frac{2}{m_1+m_2+2}  , \quad \bar \Delta_{c} = \Delta_a \, ,  \label{opl1}\\
&O_d = \bar L^{(i)}_{-\frac{2}{m_1+m_2+2}} \Sigma_{\tau_C^{-1}\tau_A}, \quad \Delta_d = \Delta_a, \quad \quad \quad \quad \quad \quad \quad \quad \bar \Delta_{d} = \Delta_{a} +\frac{2}{m_1+m_2+2}  \, . 
\label{opbl}
\end{align}
Now if the operator $O_b$ appeared in the OPE, it would give a contribution to \eqref{series} which depends on the dimension $h$, and hence on the operator content of the theory. Since we argued in the previous subsection that the quantity $F_{m_1, n_1, m_2, n_2, k}$ should not not have any such dependence, 
the product $f_{23 \bar O_p} f_{1 O_p 4}$
for all such non-universal operators must be zero.~\footnote{As an aside,  the OPE coefficients $f_{23O_p}$ and $f_{1O_p4}$ do not individually have to be zero for all $O_p$. Consider the correlation functions  $\braket{\Sigma_{\tau_B^{-1}\tau_A}(x_1) \Sigma_{\tau_C^{-1}\tau_B}(x_2)  \Sigma_{\tau_B^{-1}\tau_C}(x_3) \Sigma_{\tau_A^{-1}\tau_B}(x_4)}$ and $\braket{\Sigma_{\tau_C^{-1}}(x_1) \Sigma_{\tau_A}(x_2) \Sigma_{\tau_A^{-1}}(x_3) \Sigma_{\tau_C}(x_4)}$, which have series expansions in terms of $f_{23\bar O_p}$ and $f_{1O_p 4}$ respectively. These quantities do not have genus zero and can therefore have non-universal contributions in their OPE expansion. So it seems that we can have $f_{23O_p}$ be non-zero for some non-universal $O_p$ as long as $f_{1O_p4}$ is zero, and vice versa.} The lowest-dimension dressed operators contributing to \eqref{series} are then \eqref{opl1} and \eqref{opbl}. 

Putting together the contributions from $O_a$, $O_c$, $O_d$, the first two powers appearing in the $\eta \rightarrow 1$ expansion of $F_{m_1, n_1, m_2, n_2, k}$ are 
\be 
\frac{F_{m_1, n_1, m_2, n_2, k}}{f_1(x_{ij}, \Delta_i)} \approx \frac{f_{23\bar O_a} f_{1O_a4}}{c_{O_a}} (1-\eta)^{2\Delta_{a}} + \le(\frac{f_{23\bar O_c} f_{1O_c4}}{c_{O_c
}} + \frac{f_{23\bar O_d} f_{1O_d4}}{c_{O_d}}  \ri) (1-\eta)^{2\Delta_{a} + \frac{2}{m_1+m_2+2}} \label{large_eta}  
\ee
with $\Delta_a$ defined in \eqref{del0}. In the replica limit \eqref{rep}, $f_1(x_{ij}, \Delta_i)$ becomes 1, and the powers of $(1-\eta)$ in the leading and subleading terms become $c/9$ and $c/9+2/3$ respectively. These agree with the powers observed numerically in Fig. \ref{fig:Flambda2}. The descendants in the conformal block of $\Sigma_{\tau_C^{-1}\tau_A}$ give further subleading corrections starting at $\sO(1-\eta)$. 

Let us now turn to understanding the coefficients in \eqref{large_eta}. In the first coefficient, $f_{1O_a4}$ and $f_{23 \bar O_a}$ are both proportional to three-point functions of twist operators, and $c_{O_a}$ is proportional to a two-point function of twist operators. As mentioned  in  section \ref{sec:comments} and reviewed in more detail in Appendix \ref{app:lou_rev}, each of these correlators can be written in the form $e^{S_L}$ where $S_L$ is the Liouville action associated with some covering map, which is proportional to $c$. The genus of the covering surface for each of these cases is zero. So the coefficient in front of $(1-\eta)^{2\Delta_{a}}$ can be written as $e^{a_0 c}$ for some universal constant $a_0$, as noted in \eqref{216}. 

Let us now try to derive this constant $a_0$. First note that the coefficient  $c_{O_a}$ can be related to the Renyi entropy of an interval:
\be
\braket{\Sigma_{\tau_C^{-1}\tau_A}(a_1)\Sigma_{\tau_A^{-1}\tau_C}(a_2)} = \le(\Tr[\rho_A^{m_1+m_2+2}]\ri)^k = \frac{c_{O_a}}{|a_{12}|^{4\Delta_{a}}} \label{261}
\ee
where $a_1$ and $a_2$ are the endpoints of some interval $A$, and $a_{12}=a_1-a_2$.  
Taking the replica limit, we see that $c_{O_a}$ is related to the third Renyi entropy of a single interval, 
\be 
c_{O_a} = \Tr[\rho_A^3]^{\ha} \, |a_{12}|^{2c/9} \, . \label{codef}
\ee
Next, consider the factor $f_{1O_a4}$, which is defined by the following
three-point function (with $y_1<y_2<y_3$ on some spatial slice according to our conventions):  
\begin{align}
\braket{\Sigma_{\tau_{A}^{-1}}(y_1) \Sigma_{\tau_C^{-1}\tau_A}(y_2) \Sigma_{\tau_C}(y_3)} &= \frac{f_{1O_a4}}{|y_{12}|^{2(\Delta_1 + \Delta_a - \Delta_4)} |y_{23}|^{2(\Delta_a + \Delta_4 - \Delta_1)} |y_{13}|^{2(\Delta_1 + \Delta_4 - \Delta_a)} } \label{f14def}
\end{align}
Say $R$ is the interval between $y_1$ and $y_2$, and $S$ the interval between $y_2$ and $y_3$. By inverting the reasoning that we used in Section \ref{sec:twist} to express a quantity involving reduced density matrices as a correlation function of twist operators, we can
rewrite \eqref{f14def} as 
\be 
\braket{\Sigma_{\tau_{A}^{-1}}(y_1) \Sigma_{\tau_C^{-1}\tau_A}(y_2) \Sigma_{\tau_C}(y_3)} = \Tr[(\rho_R\rho_S^{m_1+m_2}\rho_{RS})^k] \, . \label{fm1}
\ee
We are interested in the replica limit of this expression, where we have 
\be
f_{1O_a4} = \Tr[\le((\rho_R\otimes \rho_S) \, \rho_{RS}\ri)^{\ha}] \le(\frac{|y_{23}||y_{12}|}{|y_{13}|}\ri)^{c/9} \, . \label{f14dens}
\ee
From \eqref{fdef2}, the quantity on the RHS can be identified to be the fidelity between $\rho_R\otimes \rho_S$ and $\rho_{RS}$ for two adjacent intervals $R$ and $S$. Recall from the discussion at the beginning of this section that in the strict $\eta = 1$ limit, the region $B$ vanishes and $\rho_{ABC}$, $\tilde \rho_{ABC}$ reduce to $\rho_{AC}$,  $\rho_A\otimes \rho_C$ for adjacent intervals $A, C$. It is therefore reasonable that the universal coefficient associated with the fidelity between these two limiting states contributes to $e^{a_0 c}$.  
In Appendix \ref{app:lou_fid}, we use the Liouville action to find  
\be 
\frac{f_{1O_a4}}{\sqrt{c_{O_a}}} =   e^{0.12062 \, c} \, . \label{const}
\ee
which agrees with the numerically evaluated value from \eqref{f14dens}, $\frac{f_{1O_a4}}{\sqrt{c_{O_a}}}= e^{0.11465 \, c}$, up to finite size corrections.~\footnote{We evaluate the numerical value of the ratio  $\frac{f_{1O_a4}}{\sqrt{c_a}}$ as this is cutoff-independent, unlike the individual factors $f_{1O_a4}$ and ${\sqrt{c_a}}$ which depend on the UV regularization. Note that the $y_{ij}$ refer to the lengths of various intervals and not to the lengths divided by the UV cutoff.} 

Let us next try to understand the constant $f_{23 \bar O_a}$. By similar reasoning to \eqref{fm1}, we have 
\be
\langle\Sigma_{\tau_B^{-1}\tau_A}(y_1) \Sigma_{\tau_C^{-1}\tau_B}(y_2) \Sigma_{\tau_A^{-1}\tau_C}(y_3)\rangle = \le(\Tr_S\le[\Tr_R[\rho_{RS}^{m_1+1}\rho_R^{n_1}] \Tr_R[\rho_{RS}^{m_2+1}\rho_{R}^{n_2}] \ri]\ri)^k \label{m1n1}
\ee
Taking the replica limit, 
\be 
f_{23\bar O_a} = \le(\Tr_S\le[\Tr_R\le(\rho_{RS}^{\frac{3}{2}-\frac{i\lambda}{2}}\rho_R^{-\ha+\frac{i\lambda}{2}}\ri)\Tr_R\le(\rho_{RS}^{\frac{3}{2}+\frac{i\lambda}{2}} \rho_{R}^{-\ha-\frac{i\lambda}{2}}\ri)\ri]\ri)^{\ha} \le(\frac{|y_{23}||y_{13}|}{|y_{12}|}\ri)^{c/9} \label{f23dens}
\ee
Unlike \eqref{f14dens}, \eqref{f23dens} is not an obviously identifiable information-theoretic quantity. One simple observation we can make is that in the case where $\rho_{RS}$ is a product state $\rho_R \otimes \rho_S$, we have 
\be 
\frac{\le(\Tr_S\le[\Tr_R\le(\rho_{RS}^{\frac{3}{2}-\frac{i\lambda}{2}}\rho_R^{-\ha+\frac{i\lambda}{2}}\ri)\Tr_R\le(\rho_{RS}^{\frac{3}{2}+\frac{i\lambda}{2}} \rho_{R}^{-\ha-\frac{i\lambda}{2}}\ri)\ri]\ri)^{\ha}}{(\Tr[\rho_S]^3)^{\ha}} =1 \,  \quad \text{ if }\quad \rho_{RS} = \rho_R \otimes \rho_S \, , \label{qty2}
\ee
so it may be possible to interpret the log of the LHS of \eqref{qty2} as a measure of entanglement between $R$ and $S$ in $\rho_{RS}$. 
It would be interesting to better understand this quantity and the reason why it appears in this limit of  $F(\rho_{ABC},\tilde\rho_{ABC})$. 

We can further try to derive the numerical value of the cutoff-independent quantity $\frac{f_{23\bar O_a}}{\sqrt{c_a}}$ using the Liouville action, similar to \eqref{qty2}.  
In Appendix \ref{app:lou_23}, we outline the calculation of \eqref{m1n1} using the Liouville action for integer values of $m_i, n_i, k$. We find that one of the parameters which appears in the Liouville action can only be derived by solving a polynomial equation whose degree increases with $m_i, n_i$. As a result, no analytic solution can be found for general $m_i, n_i$, and we cannot obtain $\frac{f_{23\bar O_a}}{\sqrt{c_a}}$ by analytic continuation using this method. Indeed, as we discuss in the appendix, the Liouville action and the equations for the covering map seem to depend only on $\lambda$-independent combinations of the parameters $m_i, n_i$. So if there had been a solution that we could analytically continue, this would have led to inconsistency with the explicit $\lambda$-dependence in this quantity that we show below.  This case presents a simpler version of the issues encountered in trying to obtain an analytic continuation of the four-point function \eqref{corrf} using the Liouville action. 

\begin{figure}[!h]
    \centering
    \includegraphics[width = 0.5\linewidth]{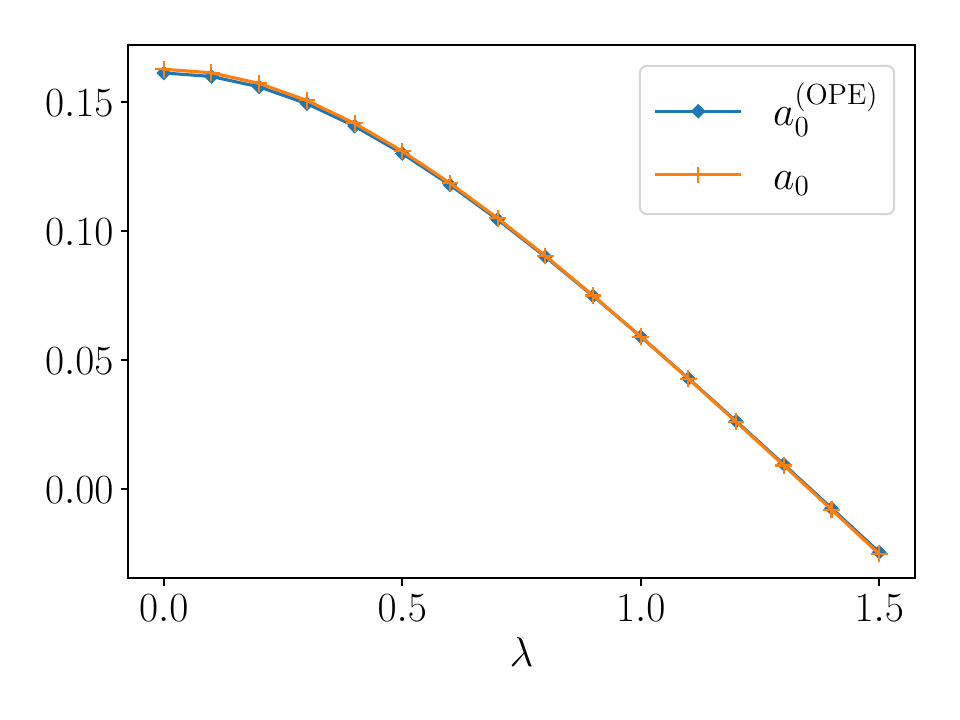}
    \caption{We compare the coefficient $a_0$ in \eqref{216} to the prediction from OPE expansion $a_0^{\rm (OPE)}$ in \eqref{a0_ope}. The quantities on the RHS of \eqref{codef}, \eqref{f14dens}, \eqref{f23dens} are also evaluated numerically.}
    \label{fig:ope_comparison}
\end{figure}

We can, however, check that the constant 
\be \label{a0_ope}
a_0^{\rm (OPE)} = \frac{1}{c}\log \le(\frac{f_{23\bar O_a} f_{1O_a4}}{c_{O_a}} \ri)\, , 
\ee
obtained using the quantities on the RHS of \eqref{codef}, \eqref{f14dens}, and \eqref{f23dens}, agrees  with the constant $a_0$ obtained from the numerical fitting of $F(\rho_{ABC}, \tilde \rho_{ABC})$ in \eqref{216}. We check this in Fig. \ref{fig:ope_comparison}, finding excellent agreement for all $\lambda$. 
So we find that although we cannot obtain the numerical value of $a_0$ using analytic continuation, the analytic continuation works well at the level of relating it to other entanglement quantities.  

It was discussed in \cite{rastelli, dei_eb} that correlation functions of twist operators are invariant under cyclic reorderings. In particular, for a three-point function of twist operators, this implies that the constant $f_{ABC}$ should be equal to $f_{CAB}$. However, the two correlation functions $\braket{A(y_1) B(y_2) C(y_3)}$ and 
$\braket{C(y_1) A(y_2) B(y_3)}$ can have very different expressions in terms of reduced density matrices. For instance, the coefficient $f_{1 O_a4}$ that we related to the fidelity of $\rho_{RS}$ and $\rho_{R}\otimes \rho_S$ in \eqref{f14dens} can also be written as  
\be 
f_{1O_a 4} = f_{41 O_a} = \le(\frac{|y_{23}||y_{13}|}{|y_{12}|}\ri)^{2\Delta_{a}} \,  
\Tr_{R}[\le(\Tr_{S}[\rho_{RS}^{2} \rho_{S} ]\ri)^{\ha}]  \, . \label{cyc1}
\ee
Similarly, 
\be \label{cyc2}
f_{2 3 \bar O_a} = f_{\bar O_a 23} = \le( \frac{|y_{12}||y_{13}|}{|y_{23}|}\ri)^{2\Delta_a} \le(\Tr[ \rho_{R}^{\ha+ \frac{i\lambda}{2}} \rho_{RS} \rho_{R}^{\ha-\frac{i\lambda}{2}} \rho_{S}^{-\ha+ \frac{i\lambda}{2}} \rho_{RS} \rho_{S}^{-\ha- \frac{i\lambda}{2}}]\ri)^{\ha} 
\ee
We also checked the relations \eqref{cyc1} and \eqref{cyc2} numerically. Understanding the physical reason for such relations between different combinations of the reduced density matrices, which are likely unique to conformal field theories, would be an interesting question for future work.

For the coefficient of the second term in \eqref{large_eta}, there does not seem to be an obvious way to relate it to some function of the density matrices, but we can show using the general expressions for such OPE coefficients in terms of the covering map (see for instance Section 4.2 of \cite{peet}) that it is equal to $a_1 \, c\, e^{a_0 c}$ for some universal constant $a_1$.


\subsection{$\eta \rightarrow 0$ limit}
\label{sec:smallz}

Let us now consider the $\eta \ll 1$ OPE limit, which corresponds to either  $L_{A}\ll L_B$, or $L_{C} \ll L_B$, or both. Note that the $L_A \ll L_B \approx L_C$ and $L_C \ll L_B \approx L_A$ cases have different interpretations due to the different roles played by $A$ and $C$ in $\tilde \rho_{ABC}$, but the answer is the same for both.  Recall that from the discussion of the lower bound in Section \ref{sec:def}, we expect the fidelity to approach 1 in this limit.

We can now use the OPEs between $O_1, O_2$ and $O_3, O_4$ to get the following series: 
\begin{align} \label{series}
&\braket{\Sigma_{\tau_A^{-1}}(x_1) \Sigma_{\tau_B^{-1}\tau_A}(x_2)\Sigma_{\tau_C^{-1}\tau_B}(x_3)\Sigma_{\tau_C}(x_4) } \nn
&\quad \quad \quad = f_2(x_{ij}, \Delta_i) \sum_p \frac{f_{12 \bar O_p} f_{ O_p 34}}{c_{O_p }} \eta^{\Delta_p}{\bar \eta}^{\bar \Delta_p} g_{ \Delta_1, \Delta_2, \Delta_3, \Delta_4; \Delta_p, \bar \Delta_p}(\eta, \bar \eta)
\end{align} 
where 
\be 
f_2(x_{ij}, \Delta_i) = \frac{1}{|x_{12}|^{2(\Delta_1+\Delta_2)} |x_{34}|^{2(\Delta_3+\Delta_4)}} \le(\frac{|x_{24}|}{|x_{14}|}\ri)^{2(\Delta_1-\Delta_2)} \le(\frac{|x_{14}|}{|x_{13}|}\ri)^{2(\Delta_3-\Delta_4)}
\ee
and the coefficients $f_{ABC}, c_{O_p}$
and functions $g$ are defined as in the previous subsection, but note that the order of dimensions appearing in $g$ has changed. The set of operators $O_p$ contributing to the expansion are also different from the case in the previous subsection. The lowest-dimension $O_p$ appearing in the OPE of $\Sigma_{\tau_A^{-1}}$ and $\Sigma_{\tau_B^{-1}\tau_A}$ is $ \Sigma_{\tau_B^{-1}}$.

For the fidelity, $\tau_B$ is defined in \eqref{bdef}, so $\Sigma_{\tau_B^{-1}}$ has dimension 
\be 
\Delta_{\tau_B} = \frac{c}{24} \le(N k - \frac{1}{Nk} \ri) \, . 
\ee
We can see that $\Delta_{\tau_B}$ is zero in the replica limit \eqref{rep}, indicating that the fidelity approaches a constant in this limit. In order to see that this constant is 1, we can interpret the quantities $f_{12\tau_B}$, $f_{\tau_B^{-1} 34}$, and $c_{\tau_B}$ in terms of the density matrix. With $y_1<y_2<y_3$ on some spatial slice, and $R$ and $S$ defined as the intervals $y_1y_2$ and $y_2y_3$, we have   
\begin{align} 
\langle\Sigma_{\tau_A^{-1}}(y_1) \Sigma_{\tau_B^{-1}\tau_A}(y_2)& \Sigma_{\tau_B}(y_3)\rangle = \text{Tr}[(\rho_{S}^{m_1+n_1} \rho_{RS} \rho_{S}^{m_2+n_2}\rho_{RS})^k] \nn 
&= \frac{f_{12\tau_B}}{|y_{12}|^{2(\Delta_1 + \Delta_2 - \Delta_{\tau_B})}|y_{23}|^{2(\Delta_2 + \Delta_{\tau_B} - \Delta_{1})}|y_{13}|^{2(\Delta_1 + \Delta_{\tau_B} - \Delta_{\tau_2})}} \label{268}
\end{align}
and  
\begin{align} \label{269}
\langle\Sigma_{\tau_B^{-1}}(y_1) \Sigma_{\tau_C^{-1}\tau_B}(y_2) & \Sigma_{\tau_C}(y_3)\rangle =\text{Tr}[(\rho_{RS}^{m_1+m_2+1}\rho_{R}^{n_1+n_2+1})^k] \nn
& = \frac{f_{\tau_B^{-1}34}}{|y_{12}|^{2(\Delta_{\tau_B} + \Delta_3- \Delta_{4})}|y_{23}|^{2(\Delta_3 + \Delta_4 - \Delta_{\tau_B})}|y_{13}|^{2(\Delta_{\tau_B} + \Delta_4 - \Delta_{3})}}
\end{align}
Also, 
\be 
\braket{\Sigma_{\tau_B^{-1}}(y_1)\Sigma_{\tau_B}(y_2)} = \Tr[\rho_R^{Nk}] = \frac{c_{\tau_B}}{|y_{12}|^{4\Delta_{\tau_B}}} \, . \label{ctau}
\ee
Taking the replica limit of \eqref{268}, \eqref{269}, and \eqref{ctau}, we find 
\be 
f_{12\tau_B} = f_{\tau_B^{-1}34} =\Tr[\rho_{RS}] =1, \, \quad \quad c_{\tau_B} = \Tr[\rho_R] =1 \, \quad  \quad  \quad \text{(replica limit)}\,  .  \label{all_1}
\ee
Putting this into \eqref{series}, we find that the fidelity approaches 1 as $\eta \rightarrow 0$.

For integer values of the parameters $m_i, n_i, k$, the corrections to the leading behaviour of \eqref{series} come from the descendants of $\Sigma_{\tau_B^{-1}}$, and from other universal primary operators $O_p$ in the twisted sector of $\Sigma_{\tau_B^{-1}}$.   In the replica limit, where all dimensions $\Delta_{1}, ..., \Delta_4$ as well as $\Delta_{\tau_B}$ go to zero, all coefficients $B^{(k)}_{\Delta_1, \Delta_2, \Delta_3, \Delta_4; \Delta_{\tau_B}}$ in the conformal block appear to go to zero.~\footnote{We have checked that this is true for the  expressions in \cite{cft_long} and \cite{perl} up to $k=6$.}

In principle, we could have fractional modes of $\Sigma_{\tau_B^{-1}}$ appearing in the OPE: the lowest-dimension such operator consistent with universality  would be $L_{-\frac{2}{Nk}}\Sigma_{\tau_B^{-1}}$, with dimension $\Delta_{\tau_B}+ \frac{2}{Nk}$. However, we can provide the following simple argument that  all conformal block coefficients and OPE coefficients should go to zero on taking the replica limit of the $m_i$, $n_i$ parameters, even when $k$ is some positive integer.  To see this, consider the following four-point functions and their OPE expansions: 
\begin{align} 
&\braket{\Sigma_{\tau_A^{-1}}(x_1) \Sigma_{\tau_B^{-1}\tau_A}(x_2) \Sigma_{\tau_A^{-1}\tau_B}(x_3) \Sigma_{\tau_A}(x_4)} = \Tr[(\rho_B^{m_1+n_1}\rho_{ABC} \rho_B^{m_2+n_2} \rho_{ABC})^k] \label{272}\\
&  \quad \quad \quad = f_2(x_{ij}, \Delta_1, \Delta_2, \Delta_2, \Delta_4) \sum_p \frac{(f_{12 \bar O_p})^2}{c_{O_p }} \eta^{\Delta_p}{\bar \eta}^{\bar \Delta_p} g_{ \Delta_1, \Delta_2, \Delta_2, \Delta_1; \Delta_p, \bar \Delta_p}(\eta, \bar \eta) \label{273}
\end{align}
and 
\begin{align} 
&\braket{\Sigma_{\tau_C^{-1}}(x_1) \Sigma_{\tau_B^{-1}\tau_C}(x_2) \Sigma_{\tau_C^{-1}\tau_B}(x_3) \Sigma_{\tau_C}(x_4)} = \Tr[(\rho_{ABC}^{m_1+m_2+1} \rho_B^{n_1+n_2+1})^k] \label{274}\\
&  \quad \quad \quad = f_2(x_{ij}, \Delta_4, \Delta_3, \Delta_3, \Delta_4) \sum_p \frac{(f_{O_p 34})^2}{c_{O_p }} \eta^{\Delta_p}{\bar \eta}^{\bar \Delta_p} g_{ \Delta_4, \Delta_3, \Delta_3, \Delta_4; \Delta_p, \bar \Delta_p}(\eta, \bar \eta) \label{275}
\end{align}
To get the RHS of \eqref{272} and \eqref{274}, we have again used the logic of Section \ref{sec:twist}. 
Now taking the replica limit in these expressions, we find that both four-point functions above simply reduce to $\Tr[\rho_{ABC}^{2k}]$,  so  they  should become proportional to $\frac{1}{|x_{14}|^{2\Delta_{(2k)}}}$. We can check that in \eqref{273} and \eqref{275}, this  dependence on $x_{14}$  simply comes from combining the overall factor of $f_2$ with the contribution from $O_p = \Sigma_{\tau_B^{-1}}$. Hence, the coefficients of all other contributions must become zero in the replica limit. Indeed, we can observe by comparing \eqref{272}, \eqref{274} with \eqref{268}, \eqref{269} that 
the argument  for these coefficients of subleading contributions reducing to zero is closely related to the argument for the leading coefficient to be 1 in the replica limit. 

The natural conclusion from the above discussion would be that the fidelity approaches 1 faster than any power of $\eta$ in the $\eta \to 0$ limit, similar to discussions of the negativity in the far-interval limit~\cite{neg1, neg2}. But recall from the numerical results of Sec. \ref{sec:num} that we found $F(\rho_{ABC}, \tilde \rho_{ABC})\propto \eta^2$ in this limit, up to very small values of the fidelity close to $10^{-6}$. It seems likely, then, that the expression we get from first taking the OPE limit and then analytically continuing in $m_i, n_i$ (as we do here) does not agree with the answer we get from first analytically continuing and then expanding for small $\eta$. The latter procedure should be the correct one, but cannot be implemented with the Liouville action methods due to the difficulties discussed in Appendix \ref{app:lou_4pt}.

\section{Relative entropy and trace distance}
\label{sec:rel}

In this section, we will consider two other measures of the distance between $\rho_{ABC}$ and $\tilde \rho_{ABC}$: the relative entropy and the trace distance.  We study the relative entropy in Sec. \ref{sec:rel_ent}, where the calculation and results turn out to be qualitatively quite similar to the discussion of the fidelity.  We discuss the behaviour of the trace distance in  \ref{sec:trace}. We compare both sides of various general  inequalities relating the fidelity, trace distance, and relative entropy in Sec. \ref{sec:compare}. 

\subsection{Relative entropy} \label{sec:rel_ent}

The relative entropy is the following measure of distance between two states $\rho$ and $\sigma$:   
\be \label{rel}
D(\rho||\sigma)=\Tr[\rho \log \rho]- \Tr[\rho \log \sigma] \, . 
\ee
A one-parameter generalization called the Petz-Renyi relative entropy has also been widely discussed in the literature~\cite{uhl, quasi}:
\be 
D_{\alpha}(\rho||\sigma)= \frac{1}{\alpha-1} \log \text{Tr}[\rho^{\alpha}{\sigma}^{1-\alpha}] \,, \quad \alpha \in [0, 1] \label{dalpha}
\ee
\eqref{rel} is the limit of \eqref{dalpha} as $\alpha \rightarrow 1$. The $D_{\alpha}$ are monotonically increasing in $\alpha$. 
We can see that both \eqref{rel} and \eqref{dalpha} are zero if and only if $\rho = \sigma$, and diverge if the support of $\rho$ is orthogonal to that of  $\sigma$. These behaviours are similar to those noted for $-\log F(\rho, \sigma)$ in Section \ref{sec:def}. 
However, unlike the fidelity, the relative entropies for $\alpha \neq 1/2$ are not symmetric between $\rho$ and $\sigma$. One consequence of this asymmetry is that these quantities diverge whenever the support of $\rho$ is not contained in the support of $\sigma$, but not necessarily the other way round. In our discussion below, we take $\rho=\rho_{ABC}$ and $\sigma = \tilde \rho_{ABC}$: this is the natural choice given that in the limit $L_B \rightarrow 0$, $\rho_{ABC}$ approaches $\rho_{AC}$ while $\tilde\rho_{ABC}$ approaches $\rho_A \otimes \rho_C$. The case $\alpha=1/2$ is symmetric and corresponds to Holevo's just-as-good fidelity, which is equal to the fidelity between the canonical purifications of $\rho$ and $\sigma$.


The relative entropy can also be given an interpretation in terms of distinguishing $\rho$ and $\sigma$ by measurements~\cite{stein_1, stein_2}, but the setup is  different from the one discussed for the fidelity in Sec. \ref{sec:def}.   Suppose we are given $n$ copies of a certain state, and know that the state is one out of $\rho$ and $\sigma$. We want to carry out some two-outcome measurement $\{M, I-M\}$ on $n$ copies, and guess  that the state is $\rho$ if the outcome is $M$, and $\sigma$ otherwise. In general, we cannot choose $M$ such that both $p_1$, the probability  of a ``type 1'' error where we are given $\rho$ but guess $\sigma$, and $p_2$, the probability of a ``type 2'' error where are given $\sigma$ but guess $\rho$, are identically zero. If  we choose the optimal $M$ to minimize $p_2$, while also requiring that $p_1$ vanishes as $n \rightarrow \infty$, then we get an exponential decay $p_2 = e^{-n D(\rho||\sigma)}$ as $n\rightarrow \infty$. In \cite{mosonyi}, the Petz-Renyi relative entropies for other values of $\alpha \in [0,1]$ are also given interpretations in terms of similar protocols, which require more stringent conditions on the decay of $p_1$ with $n$.   

\begin{figure}[!h]
    \centering
    \includegraphics[width=\textwidth]{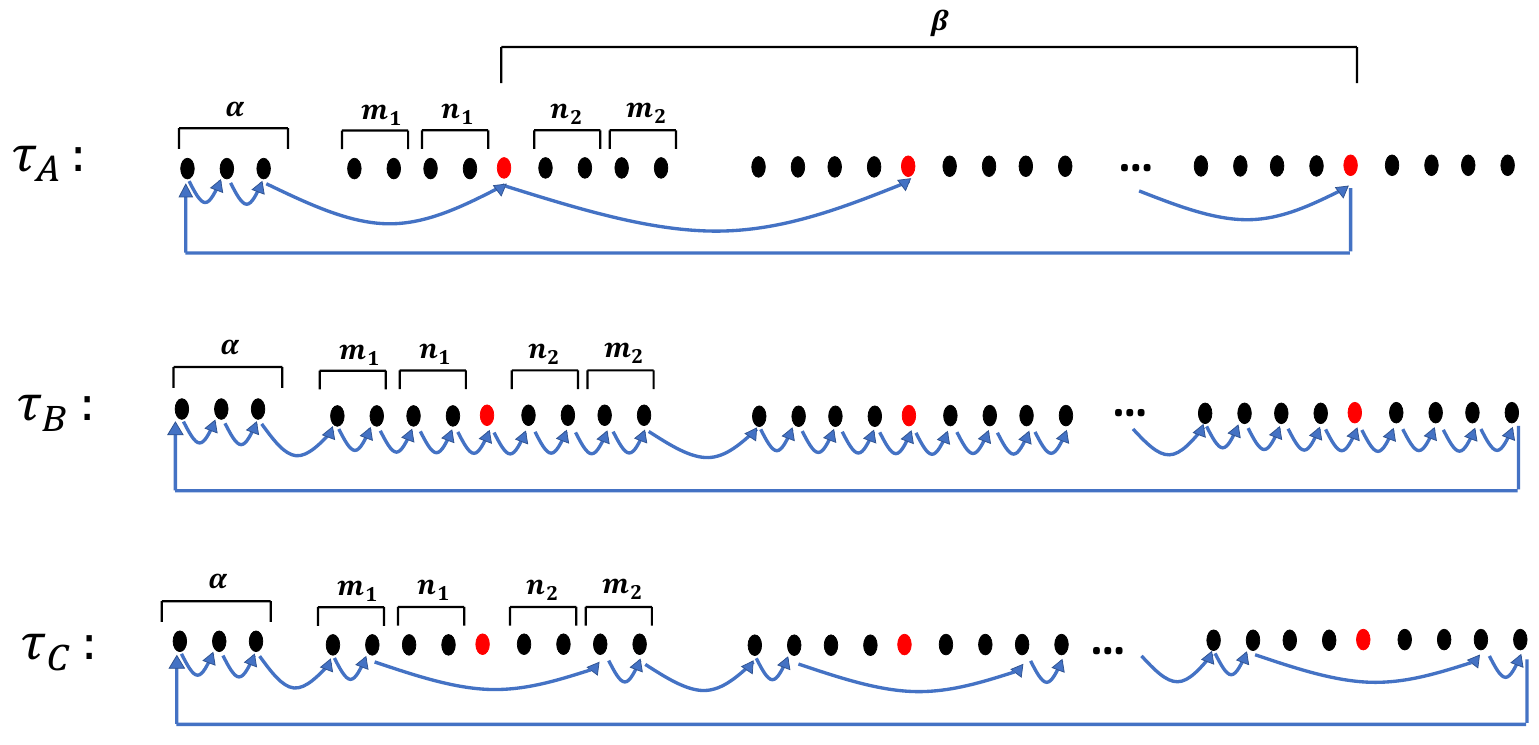}
    \caption{The quantity $\Tr[\rho^{\alpha}_{ABC}(\tilde \rho_{ABC})^{1-\alpha}]$ can be expressed as a four-point function of twist operators, $\braket{\Sigma_{\tau_A^{-1}}(x_1)\Sigma_{\tau_B^{-1}\tau_A}(x_2)\Sigma_{\tau_C^{-1}\tau_B}(x_3) \Sigma_{\tau_C}(x_4)}$, where $\tau_{A, B, C}$ are as shown in the figure.}
    \label{fig:rel_figs1}
\end{figure}

\begin{figure}[!h]
    \centering
    \includegraphics[width=\textwidth]{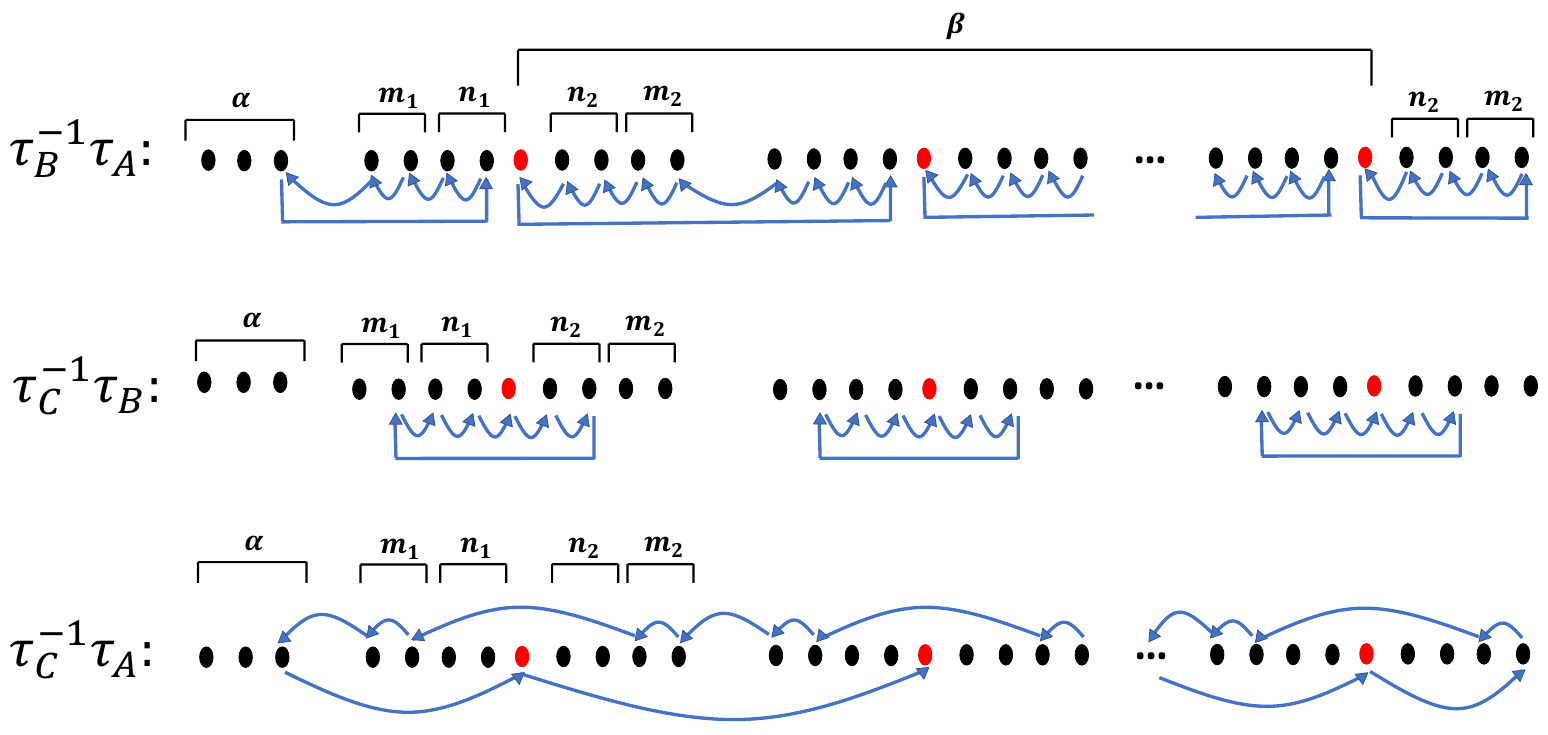}
    \caption{We show the cycle structures of the relevant permutations obtained by composing those in Fig. \ref{fig:rel_figs1}.}
    \label{fig:rel_figs2}
\end{figure}

Despite these differences between the fidelity and $D_{\alpha}$, we will find that the $D_{\alpha}$ shares several features of the fidelity for the states $\rho_{ABC}$, $\tilde \rho_{ABC}$ in our setup. To evaluate \eqref{dalpha}, we can use the following replica trick~\cite{jonah}: first evaluate 
\be 
D_{\alpha, \beta} = \frac{1}{\alpha-1} \log \Tr[\rho_{ABC}^{\alpha}{\tilde \rho_{ABC}}^{\beta}]
\label{dab}
\ee
for integer $\alpha$ and $\beta$, and then take the replica limit $\beta \rightarrow 1-\alpha$. We also replace the powers appearing in $\tilde \rho_{ABC}$ with integers $m_i,~n_i$ as before.
By similar manipulations to those in Section \ref{sec:twist}, we can express  \eqref{dab} as a four-point function of the form \eqref{corrf}, now associated with twist operators on $M=\alpha + \beta(m_1+m_2+n_1+n_2+1)$ copies.  We use the same notation $\tau_{A, B, C}$ for the permutations as in Sec. \ref{sec:eig}, but they now refer to the permutations shown in Fig. \ref{fig:rel_figs1} and \ref{fig:rel_figs2}. In particular, 
\begin{itemize}
\item $\tau_A$ has 1 cycle with $\alpha + \beta$ elements. 
\item $\tau_B^{-1}\tau_A$ has one cycle with $m_1+n_1+1$ elements, one cycle with $m_2+n_2+1$ elements, and $\beta-1$ cycles with $m_1+n_1+m_2+n_2+1$ elements. 
\item $\tau_C^{-1}\tau_B$ has $\beta$ cycles with $n_1+n_2+2$ elements. 
\item $\tau_C$ has one  cycle with $\alpha + \beta(m_1+m_2)$ elements. 
\end{itemize}

\begin{figure}[!h]
    \centering
    \includegraphics[width = 0.49\linewidth]{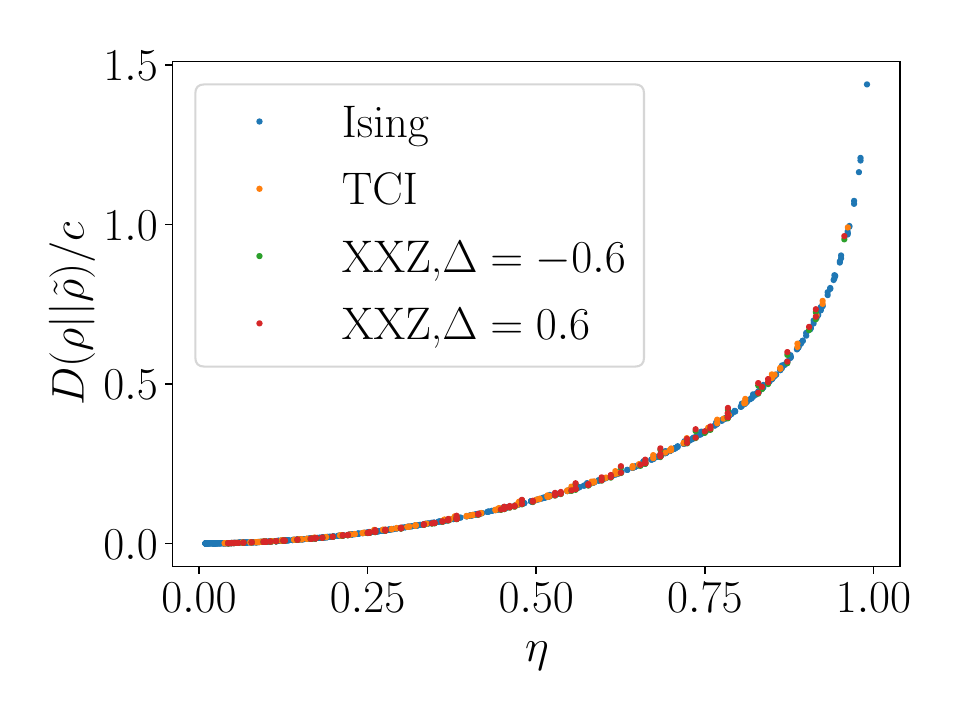}
    \includegraphics[width = 0.49\linewidth]{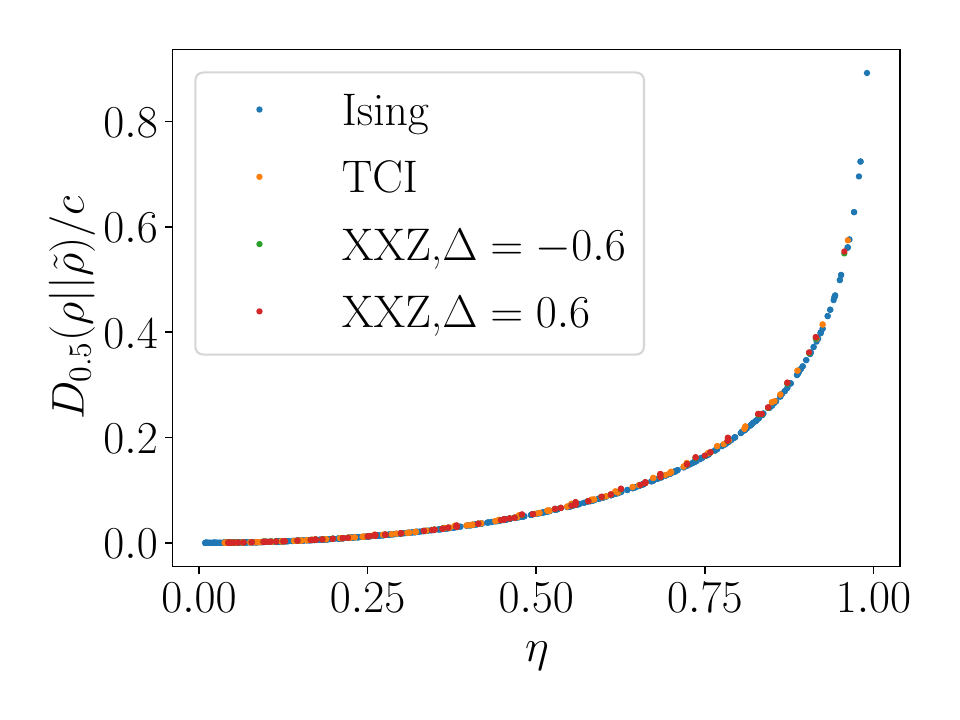}
    \caption{Petz-Renyi relative entropy between $\rho_{ABC}$ and $\tilde{\rho}_{ABC}$. Left: $\alpha=1$, corrsponding to the usual relative entropy $D(\rho||\tilde{\rho}) = \tr(\rho\log\rho - \rho\log\tilde{\rho})$. Right: $\alpha = 0.5$, corresponding to Holevo's just-as-good fidelity, $D_{0.5}(\rho||\tilde{\rho}) = \tr(\sqrt{\rho}\sqrt{\tilde{\rho}})$.}
    \label{fig:RRE_collapse}
\end{figure}

By plugging these cycle structures into \eqref{dim}, we find that the dimensions $\Delta_1, ..., \Delta_4$ all go to zero on taking the replica values of $m_i, n_i$ and $\beta = 1-\alpha$ for any $\alpha$. By similar reasoning to point 6 in Sec. \ref{sec:comments}, we can see that $D_{\alpha}(\rho||\sigma)$ is independent of the UV cutoff. Plugging the cycle structures into the Riemann-Hurwitz formula \eqref{rhf_1}, we find that the genus of the covering space is zero. The total number of copies also goes to 1 in the replica limit for any $\alpha$, so $D_{\alpha}(\rho||\sigma)/c$ is the same function of $\eta$ in any CFT. We confirm this universal behaviour using numerical computations in the lattice models introduced in Sec. \ref{sec:num}.  See  Fig. \ref{fig:RRE_collapse} for an illustration. We restrict to the case $\lambda=0$ for numerical results in this section.

\subsubsection{$\eta \to 1$ limit}
\begin{figure}
    \centering  \includegraphics[width=\textwidth]{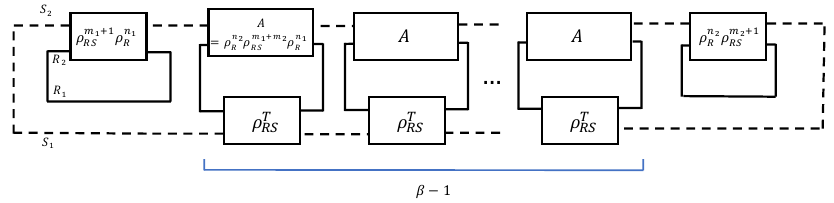}
    \caption{Tensor network representation of the quantity corresponding to the OPE coefficient $f_{23\alpha}$ for integer values of $\beta$. Each $A$ stands for the matrix $\rho_R^{n_2} \rho_{RS}^{m_1+m_2} \rho_R^{n_1}$, which becomes $\rho_R^{-\ha-i\frac{\lambda}{2}} \rho_{RS} \rho_R^{-\ha+i\frac{\lambda}{2}}$. The dashed lines represent index contractions in $S$, and the solid lines represent index contractions in $R$.}
    \label{fig:f23}
\end{figure}

Let us now consider behaviour of $D_{\alpha}$ in the OPE limit $\eta \to 1$. The discussion of Sec. \ref{sec:z1} up to Fig. \ref{fig:branch_cuts} still applies here. 
But now the lowest-dimension twist operator $\Sigma_{\tau_C^{-1}\tau_A}$ appearing in  \eqref{series} is 
as shown in Fig. \ref{fig:rel_figs2}. This permutation has  one cycle of length $(m_1+m_2+1)\beta+1$.  In the replica limit for $D_{\alpha}$, the   dimension of this operator becomes  
\be \label{drep}
\Delta_{\alpha} = \frac{c}{24}\le(3 - 2\alpha - \frac{1}{3 - 2 \alpha}\ri) \, . 
\ee
This gives a leading behaviour $\Tr[\rho^{\alpha} \tilde \rho^{1-\alpha}] = e^{b_0 c} (1-\eta)^{2\Delta_{ \alpha}}$ for some universal constant $b_0$ which can in principle be obtained from the Liouville actions for certain three-point functions.  
This coefficient can again be related to quantities associated with density matrices on two adjacent intervals. Like in \eqref{a0_ope}, we can again write  
\be 
e^{b_0 c} = \frac{f_{1\alpha 4} f_{23 \alpha}}{c_{\alpha}}
\ee
We can obtain the expressions for $f_{1{\alpha}4}$ and $c_{\alpha}$ in terms of the density matrix in a similar way to \eqref{codef} and \eqref{f14dens}, in terms of the density matrices on two adjacent intervals $R= y_1 y_2$ and $S = y_2 y_3$: 
\begin{align} 
&f_{1\alpha4} = \Tr[\rho_{RS}^{\alpha} (\rho_{R}\otimes \rho_S)^{1-\alpha}] \le( \frac{|y_{12}||y_{23}|}{|y_{13}|}\ri)^{2\Delta_{\alpha}} \label{f1alpha}\\
& c_{\alpha} = \Tr[\rho_R^{3-2\alpha}] \,  |y_{12}|^{4\Delta_{\alpha}}
\end{align}
From \eqref{f1alpha},  we see that like in the case of the fidelity, one of the factors contributing to $e^{b_0c}$ is related to the Petz-Renyi relative entropy between $\rho_{RS}$ and $\rho_{R}\otimes \rho_S$ in this limit.  
The OPE coefficient $f_{23\alpha}$ is somewhat more complicated; the corresponding function of the reduced density matrices on $R$ and $S$ for integer values of $\beta$ is shown using a tensor network representation in Fig. \ref{fig:f23}. This can be expressed in a relatively compact way in terms of two copies of the Hilbert space on $RS$, $\sH_{R_1 S_1} \otimes \sH_{R_2S_2}$, as follows: 
\begin{align}
f_{23 \alpha} &\le( \frac{|y_{12}|}{|y_{13}||y_{23}|}\ri)^{2\Delta_{\alpha}} 
 = \bra{{\rm MAX}_{S_1S_2}}\bigg[\Tr_{R_2}[\rho_{R_2S_2}^{\frac{3}{2}-\frac{i\lambda}{2}}\rho_{R_2}^{- \frac{1}{2}+ \frac{i\lambda}{2}}]\otimes \mathbf{1}_{S_1} \nn
&\le(\braket{{\rm MAX}_{R_1R_2}| \rho_{R_1S_1}^T \otimes \rho_{R_2}^{-\ha - \frac{i\lambda}{2}}\rho_{R_2S_2}\rho_{R_2}^{-\ha + \frac{i\lambda}{2}} |{\rm MAX}_{R_1 R_2}}\ri)^{-\alpha} \nn
&\Tr_{R_2}[\rho_{R_2}^{- \frac{1}{2}- \frac{i\lambda}{2}}\rho_{R_2S_2}^{\frac{3}{2}+\frac{i\lambda}{2}}]\otimes \mathbf{1}_{S_1} \bigg] \ket{{\rm MAX}_{S_1S_2}} \label{f23_comp}
\end{align}
where $\rho^T$ refers to the transpose of $\rho$, and $\ket{{\rm MAX}_{A_1 A_2}}$ is the unnormalized maximally entangled state between $A_1$ and $A_2$:  
\be \label{maxdef}
\ket{{\rm MAX}_{A_1A_2}} = \sum_{a} \ket{a}_{A_1} \ket{a}_{A_2}
\ee
for some orthonormal basis $\ket{a}$ of $A$. Also note that for a product state $\rho_{RS}=\rho_R\otimes \rho_S$, the RHS of \eqref{f23_comp} divided by $\Tr[\rho_S^{3-2\alpha}]$ is equal to 1. Again, it would be interesting to see if this quantity has some general interpretation in terms of entanglement.

The leading, divergent  behaviour of the entropy $D_{\alpha}$ in the $\eta \to 1$ limit is then given by 
\be 
D_{\alpha}(\rho_{ABC}||\tilde \rho_{ABC}) = \frac{2\Delta_{\alpha}}{\alpha-1} \log(1-\eta) + \frac{b_0c}{\alpha -1} 
\ee
The $\alpha \rightarrow 1$ limit of the above expression is 
\be \label{rel_leading}
D(\rho_{ABC}||\tilde \rho_{ABC}) = -\frac{c}{3}\log(1-\eta) + \lim_{\alpha \to 1}\frac{b_0c}{\alpha -1}
\ee
Note that \eqref{rel_leading} has the same leading divergence as the conditional mutual information. We can heuristically understand this from the fact that both $D(\rho_{ABC}||\tilde \rho_{ABC})$ and the CMI reduce to  $I(A:C)$ on setting $L_B$ to zero. 

\begin{figure}[!h]
    \centering
\includegraphics[width = 0.7\textwidth]{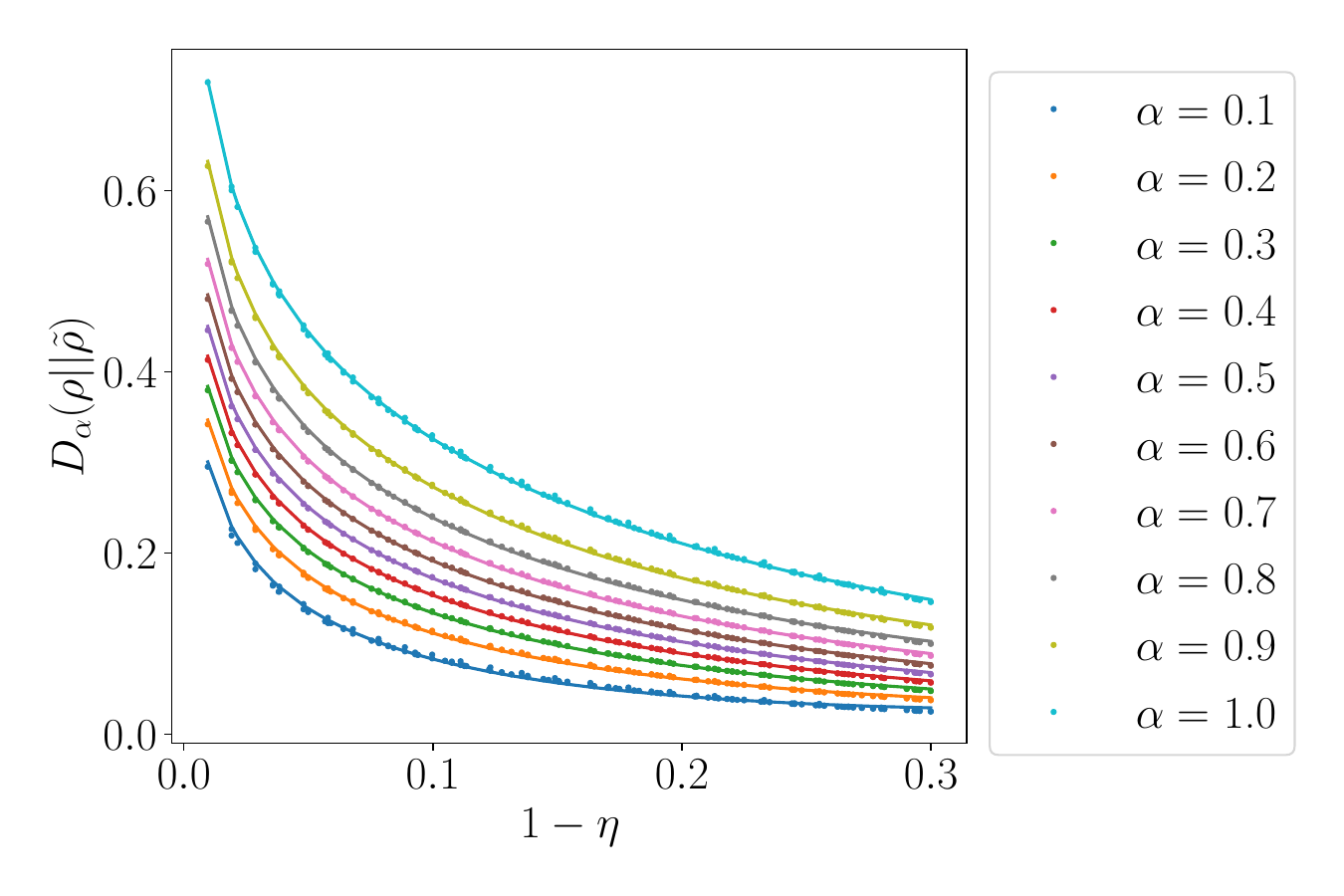}
\caption{The dots represent numerical results for $D_{\alpha}$ for the Ising model, while the solid curves show the fitting to \eqref{desc}.}
\label{fig:rel_large}
\end{figure}

There are two possible sources of leading corrections away from the $\eta \to 1$ limit: the descendants of $\tau_C^{-1}\tau_A$, and other primary operators. The next primary operator appearing in the OPE expansion is the fractional mode 
    $L_{-\frac{2}{(m_1+m_2+1)\beta+1}}\Sigma_{\tau_C^{-1}\tau_A}$, which in the replica limit has dimensions 
    \be 
    \Delta_{2,\alpha} = \Delta_{ \alpha}+\frac{2}{3-2\alpha}\,, \quad \quad  \bar \Delta_{2,\alpha} = \Delta_{ \alpha} \label{delta2}
    \ee 
   We also have its antiholomorphic version $\bar L_{-\frac{2}{(m_1+m_2+1)\beta+1}}\Sigma_{\tau_C^{-1}\tau_A}$.
For $\alpha>\ha$, the contribution is sub-dominant compared to the contribution from the linear term in the conformal block of $\Sigma_{\tau_C^{-1}\tau_A}$. 
    Including both corrections, we get   
    \be 
\Tr[\rho^{\alpha} \tilde \rho^{1-\alpha}] =  e^{b_0 c} (1-\eta)^{2\Delta_{ \alpha}}  + b_1 c e^{b_0 c} (1-\eta)^{2\Delta_{ \alpha} + \frac{2}{2(1-\alpha)+1}} + \Delta_{\alpha} e^{b_0 c} (1-\eta)^{2\Delta_{ \alpha}+1}  \label{desc}
\ee
 Here we have used the universal formula 
 \be 
 B^{(1)}_{\Delta_1 , \Delta_2, \Delta_3, \Delta_4 \, ; \, \bar\Delta_{p}} = \frac{(\Delta_{p}+\Delta_4-\Delta_1)(\Delta_{p}+\Delta_2-\Delta_3)}{2\Delta_{p}} 
 \ee
 for the first coefficient of the conformal block expansion defined in \eqref{block_a_1}.
 $b_1$ is some universal constant which would in principle be determined from a covering map. In Fig. \ref{fig:rel_large}, we compare \eqref{desc} to the numerical results for the Ising model, allowing $b_0$ and $b_1$ to be fitting parameters, and find good agreement. 

\subsubsection{$\eta \to 0$ limit}
\begin{figure}[!h]
    \centering
    \includegraphics[width = 0.55\textwidth]{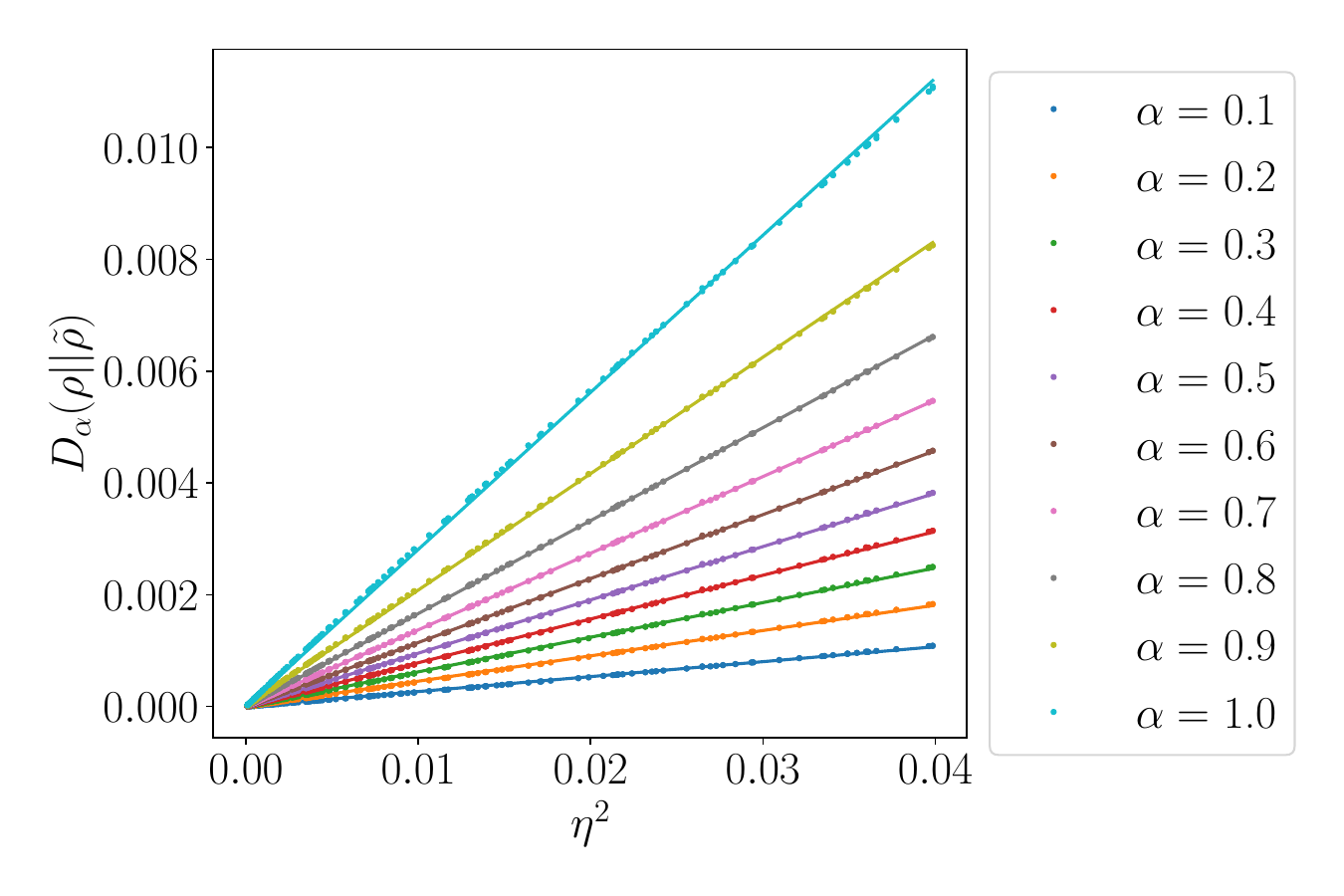}
    \includegraphics[width = 0.4\textwidth]{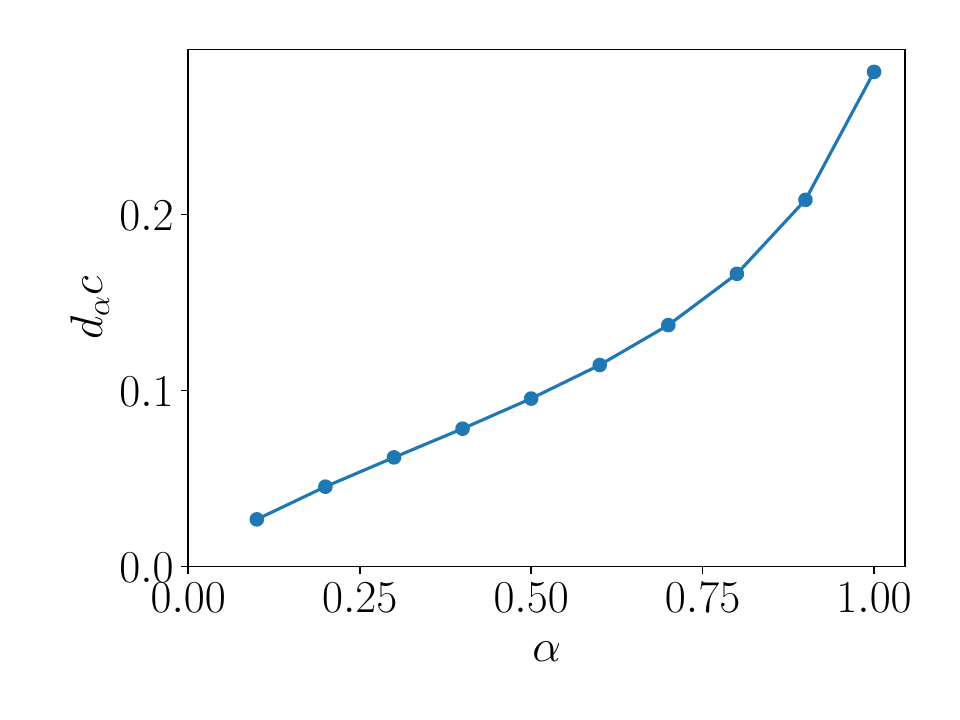}
    \caption{Numerical results for the relative entropies at small $\eta$ in the Ising model.}
\label{fig:IsingRRE}
\end{figure}

Since we expect that the difference between the states $\rho_{ABC}$ and $\tilde \rho_{ABC}$ vanishes in the $\eta \to 0$ limit, we should expect each of the $D_{\alpha}$ to approach zero in this limit. The four-point function that corresponds to $\Tr[\rho_{ABC}^{\alpha}\tilde \rho_{ABC}^{\beta}]$ should therefore approach 1 in the replica limit. The expansion of the four-point function can  be written as in \eqref{series}, and the leading operator $O_p$ in the expansion is $\Sigma_{\tau_B}$ appearing in Fig. \ref{fig:rel_figs1}, which has one cycle of length $\alpha + \beta(m_1+n_1+m_2+n_2+1)$. The dimension of this operator goes to zero on substituting $\beta=1-\alpha$ and the replica values of $m_i, n_i$, showing that $\Tr[\rho^{\alpha}\tilde \rho^{1-\alpha}]$ approaches a constant as $\eta \to 0$. We can see that the constant is 1 by very similar arguments to the discussion from \eqref{268}-\eqref{all_1}. Moreover, on considering corrections away from this limit, we run into precisely the same issue as that around \eqref{272}-\eqref{275}, arriving at the conclusion that all OPE coefficients vanish in the replica limit. In contrast, numerically we again find quadratic behaviour of each of the $D_{\alpha}$ at small $\eta$, 
\begin{equation}
\label{eq:RRE_smalleta}
    D_{\alpha}(\rho||\tilde{\rho}) = d_{\alpha} c \eta^2, \quad \eta\ll 1.
\end{equation} 
The results for the Ising model are shown in Fig. \ref{fig:IsingRRE}, where the coefficient $d_{\alpha}$ monotonically increases with $\alpha\in(0, 1]$. 

\subsection{Trace distance}
\label{sec:trace}

We now turn to the trace distance, 
\be \label{tdef}
T(\rho, \sigma) = \frac{1}{2}\Tr[\sqrt{(\rho-\sigma)^{2}}] \, . 
\ee
The trace distance takes values between 0, for the case where $\rho=\sigma$, and 1, for the case where the support of $\rho$ is orthogonal to that of $\sigma$. Like the fidelity, this is a symmetric measure of the distance between two states. Operationally, if we consider all possible projectors $P$, or all possible positive operators satisfying $P\leq 1$, the trace distance is given by 
\be 
T(\rho, \sigma) = {\rm max}_P\Tr[P(\rho -\sigma)]
\ee
It therefore tells us how well $\rho$ and $\sigma$ can be distinguished by comparing the probabilities of some measurement outcome, optimized over all possible measurements. 

\begin{figure}
    \centering
    \includegraphics[width = 0.49\linewidth]{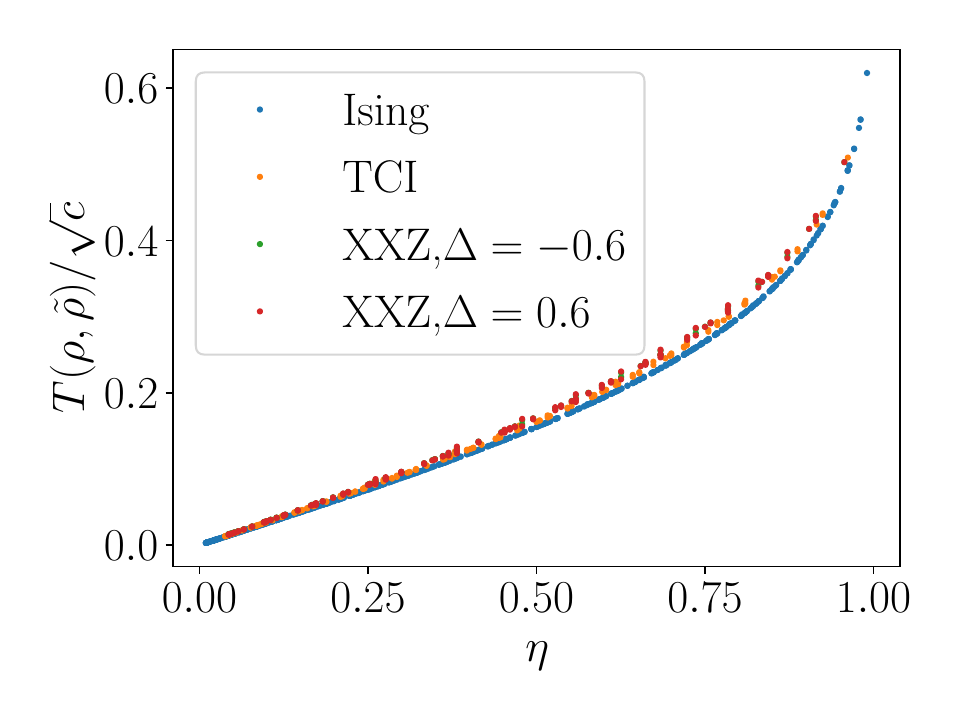}
    \includegraphics[width = 0.49\linewidth]{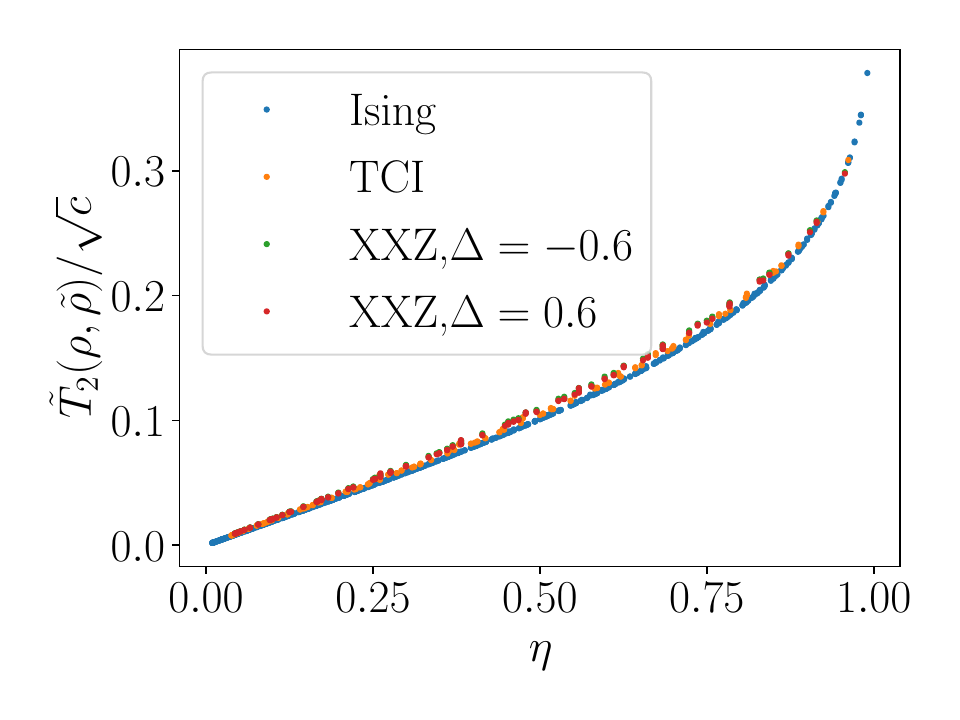}
    \caption{Distance norm $\tilde{T}_n(\rho,\tilde{\rho})$ for various models. We have shown the $n=1$ and $n=2$ cases, but other values of $n$ behave in a qualitatively similar way.}
    \label{fig:IsingTn}
\end{figure}

For the trace distance between $\rho_{ABC}$ and $\tilde \rho_{ABC}$ in the CFT vacuum state, we can 
we can use the following replica trick, introduced in \cite{trace_dist}: compute 
\be 
T_{n_e} = \frac{1}{2}\Tr[(\rho- \tilde \rho)^{n_e}] 
\label{tne}
\ee
for even $n_e$, and analytically continue to $n_e \rightarrow 1$. This replica trick allows us to see that similar to the fidelity and the relative entropy, the trace distance depends on the CFT only through its central charge and is independent of its operator content. To see this, we can note that for any $n_e$, all terms appearing in \eqref{tne} have the form 
\be \label{gen_xy}
\Tr[\rho_{ABC}^{x_1}\tilde \rho_{ABC}^{y_1}...\rho_{ABC}^{x_l}\tilde \rho_{ABC}^{y_l}]
\ee
for some integer $l$. All possible $0\leq l \leq n_e/2$ can appear, and all possible $x_i, y_i$ such that $\sum_{i=1}^l (x_i+y_i) = n_e$. The total number of copies is $s= M Y + X$, where $M = m_1+n_1+m_2+n_2+1$, $Y= \sum_{i=1}^l y_i$, and $X=\sum_{i=1}^l x_i$. We can show that \eqref{gen_xy} can be again be written as a four-point function of the form \eqref{corrf}, where   
\begin{enumerate} 
\item $\tau_A$ has one cycle with $N_{\tau_A}=X$+$Y$ elements.  
\item $\tau_C$ has one cycle with $N_{\tau_C}=X+ (m_1+m_2)Y$ elements. 
\item $\tau_B^{-1}\tau_A$ has: (i) one cycle with $m_1+n_1+1$ elements, (ii) one cycle with $m_2+n_2+1$ elements, (iii) $Y-1$ cycles with $M$ elements.
\item $\tau_C^{-1}\tau_B$ has $Y$ cycles with $n_1+n_2+2$ elements.  
\end{enumerate} 
Using the Riemann-Hurwitz formula, each such correlation function has genus zero. 
We confirm this universality in Fig. \ref{fig:IsingTn}, where the data points for the XXZ model with different compactification radius nearly coincide. 

Since \eqref{tne} involves an increasing  number of terms which each have exponential dependence on $c$, we cannot extract the $c$-dependence in the replica limit as an overall factor like in the case of $-\log F$ or the relative entropy. Let us restrict to understanding the small $\eta$ behaviour numerically. In this regime, Fig. \ref{fig:IsingTn} shows that the trace distance is proportional to $\sqrt{c}\eta$. We can also consider more general UV-regulated distance norms 
\begin{equation}
\label{eq:TD_smalleta}
    \tilde{T}_n(\rho,\tilde{\rho}) = \frac{1}{2}\frac{||\rho-\tilde{\rho}||_{n}}{||\rho||_{n}}, \quad ||\rho||_{n} = (\tr(|\rho|^{n}))^{1/n}\, 
\end{equation}
for other integers $n$, which reduces to \eqref{tdef} for $n=1$. In all cases we  find 
\begin{equation}
    \tilde{T}_n(\rho,\tilde{\rho}) = t_n \sqrt{c} \eta + O(\eta^3) ~~(\eta\ll 1).
\end{equation}
The proportional constant can be determined numerically, e.g., $t_1 \approx 0.28$.  
Based on the numerical results, $\tilde{T}_n(\rho,\tilde{\rho})/\sqrt{c}$ still weakly depends on $c$ at large $\eta$.

\subsection{Comparison between different distance measures}
\label{sec:compare}

For any two quantum states, the trace distance $T$,  fidelity $F$, and relative entropy $D$ satisfy the following inequalities
\begin{align}
    1-F \leq T \leq \sqrt{1-F^2} \label{tfbound}\\
    T \leq \sqrt{\frac{\log 2}{2} D}
\end{align}
Recall that at small $\eta$, the three distance measures obey Eqs.~\eqref{eq:F_smalleta}, \eqref{eq:RRE_smalleta} and \eqref{eq:TD_smalleta}. In the first inequality,  $1-F\leq T$ is trivially satisfied as $T$ and $1-F$ have different powers of $\eta$. The other two are satisfied if and only if the following constraint holds on the coefficients,
\begin{equation}
    t_1 \leq \sqrt{2f_2}, ~ t_1 \leq \sqrt{\frac{\log 2}{2} d_1}
\end{equation}
This is numerically found to be the case, where $t_1 \approx 0.28,\sqrt{2f_2} \approx 0.37$ and $\sqrt{\frac{\log 2}{2} d_1} \approx 0.44$. We see that both bounds are satisfied. In particular, $T$ in \eqref{tfbound} is much closer to $\sqrt{1-F^2}$ than to $1-F$. For any two pure states, we have $T= \sqrt{1-F^2}$. This suggests that the relation between $\rho_{ABC}$ and $\tilde \rho_{ABC}$ is somewhat similar to the difference between two pure states. It would be good to understand this more precisely.

\section{Differences in mutual information between $\rho_{ABC}$ and $\tilde \rho_{ABC}$} 
\label{sec:corr}

So far, we have quantitatively studied the differences between 
$\rho_{ABC}$ and $\tilde \rho_{ABC}$ using various distance measures. Let us now try to  understand more qualitatively which properties of the density matrix account for the distance between the two states. As noted in the introduction, the definition of $P^{(\lambda)}$ ensures that the reduced density matrices of $\rho_{ABC}$ and $\tilde \rho_{ABC}$ on each of the subsystems $A$, $B$, $C$, and $BC$ are precisely equal to those of $\rho_{ABC}$. The differences between the two states must  therefore be in the correlations between $A$ and $B$, correlations  between $A$ and $C$, and tripartite correlations among the three systems. 

Let us first understand the difference in the mutual information 
$I(A:B)$ between the two states.
Since $\tilde \rho_{AB}= \Tr_C \,  \sP_{B \rightarrow BC}(\rho_{AB})$, and the channel $\Tr_C \, \sP_{B \rightarrow BC}$ leaves $\rho_{A}\otimes\rho_{B}$ unchanged, we must have 
\be 
 I(A:B)_{\rho}- I(A:B)_{\tilde\rho} = S(\rho_{AB}||\rho_A \otimes \rho_B) - S(\tilde \rho_{AB}||\rho_A \otimes \rho_B) \geq 0 
\ee
by the monotonicity of the relative entropy under quantum channels. 
Since the reduced density matrices in $A$ and $B$ are the same for both states, we have   
\be 
\delta I(A:B) =  I(A:B)_{\rho}- I(A:B)_{\tilde\rho} = S(\tilde \rho_{AB}) - S( \rho_{AB})  \label{ab_diff}
\ee
where $S(\rho_{AB})$ is the entanglement entropy of the vacuum state in a single interval and is given by \eqref{sr_uni}. 
Below we will consider the Renyi version 
\be 
\delta I_q(A:B)= I_q(A:B)_{\rho}- I_q(A:B)_{\tilde\rho} = S_q(\tilde \rho_{AB}) - S_q( \rho_{AB}), \quad S_q(\rho) = -\frac{1}{q-1}\log \Tr[\rho^q] \, ,  \label{r_diff}
\ee
and take its $q \to 1$ limit to get \eqref{ab_diff}. 
The $q$-th Renyi entropy of $\rho_{AB}$ can be expressed as a two-point function of twist operators of dimension $\Delta_{(q)}$ defined in \eqref{dim}, located at $x_1$ and $x_3$, 
\be \label{rq}
\Tr[{(\rho_{AB})}^q] = \frac{r_{q}}{|x_{13}|^{4\Delta_{(q)}}}
\ee
for some constant $r_q$ which is proportional to $\epsilon^{4\Delta_{(q)}}$.  
To find \eqref{ab_diff}, let us then consider the twist operator representation of the $q$-th Renyi entropy of $\tilde \rho_{AB}$, which can again be expressed as a four-point function like the quantities studied in previous sections, 
\begin{align} \label{pren}
 \Tr[{(\tilde \rho_{AB})}^q]  
= \Tr_{AB}[\le(\Tr_C[\rho_{BC}^{m_1}\rho_B^{n_1}\rho_{AB} \rho_B^{n_2}\rho_{BC}^{m_2}]\ri)^q] = \braket{\Sigma_{\tau_A^{-1}}(x_1) \Sigma_{\tau_B^{-1}\tau_A}(x_2) \Sigma_{\tau_C^{-1}\tau_B}(x_3) \Sigma_{\tau_C}(x_4)}
\end{align}
The total number of copies is $q(m_1+n_1+m_2+n_2+1)$, and the permutations appearing in this correlation function can be worked out similarly to previous sections. We find that: 
\begin{itemize}
    \item $\tau_A$ has one cycle of length $q$. 
    \item $\tau_B^{-1}\tau_A$ has $q$ cycles of length $m_1+n_1+m_2+n_2+1$. 
    \item $\tau_C^{-1}\tau_B$ has one cycle of length $q$, and $q$ cycles of length $n_1+n_2+2$. 
    \item $\tau_C$ has $q$ cycles of length $m_1+m_2$. 
\end{itemize}
Putting this cycle structure into the Riemann-Hurwitz formula, we see that this correlation function has genus zero for any $q$. We confirm the resulting universal behaviour of $\delta I(A:B)$ using numerical simulations in the free fermion and Ising CFTs, which have the same central charge $c=1/2$ but different operator content, in Fig. \ref{fig:mutual_diffs}. See appendix A.5. for a discussion of the methods used for the free fermion calculation. On taking the replica values of $m_i, n_i$, we see that the dimensions of the operators at $x_2$ and $x_4$ go to zero, and the dimensions of the operators at $x_1$ and $x_3$ are $\Delta_{(q)}$. The UV divergence therefore cancels out between the two terms in the Renyi version of \eqref{ab_diff}, and hence also in the $q\to 1$ limit. Moreover, this difference depends only on the cross-ratio $\eta$.

\begin{figure}[!h]
    \centering
    \includegraphics[width=0.5\textwidth]{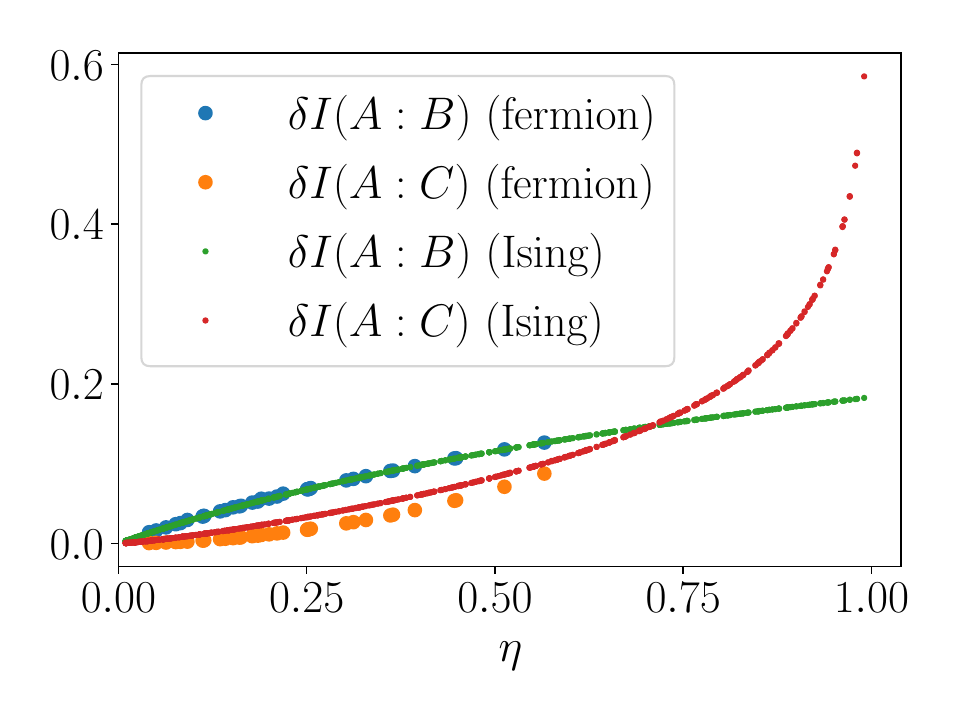}
    \caption{The quantities $\delta I(A:B)$ and $\delta I(A:C)$ evaluated numerically in the Ising and free fermion CFTs.}
\label{fig:mutual_diffs}
\end{figure}

\begin{figure}
    \centering    \includegraphics[width=\textwidth]{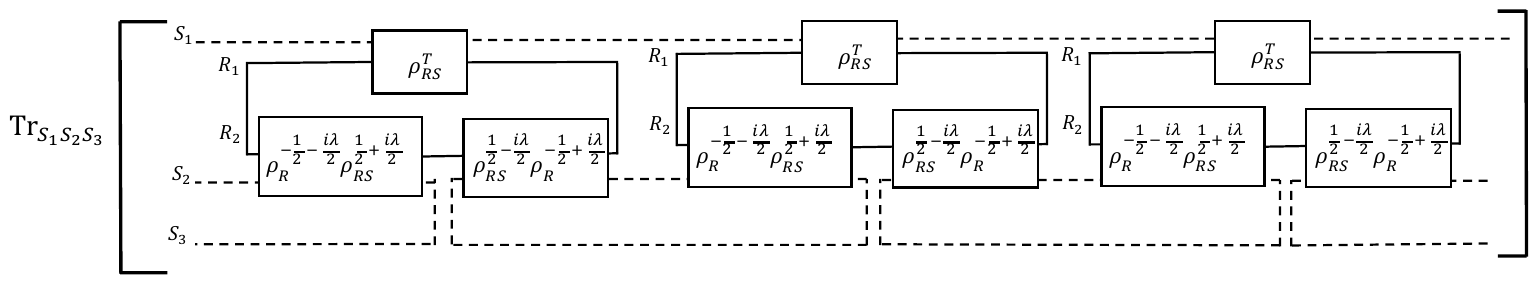}
    \caption{Tensor network representation of the quantity appearing in the $\eta \to 1$ limit of $\delta_qI(A:B)$ for $q=3$. Dashed lines represent index contractions in $S$ and solid lines represent index contractions in $R$.}
    \label{fig:mutual_tn}
\end{figure}

Let us understand the $\eta \to 1$ limit of \eqref{pren}, where we can again use the general form \eqref{series}, and the leading operator is  $O_q =\Sigma_{\tau_C^{-1}\tau_A}$. For the quantity $\Tr[{(\tilde \rho_{AB})}^q]$, $\tau_C^{-1}\tau_A$ has one cycle of length $q$, and $q$ cycles of length $m_1+m_2$. In the replica limit, this operator therefore has dimension $\Delta_{(q)}$, and 
the prefactor $f_1$ defined in \eqref{f1def}
becomes 
\be 
f_1(x_{ij}, \Delta_i) = \frac{1}{|x_{13}|^{4\Delta_{(q)}}} \frac{1}{(1-\eta)^{2\Delta_{(q)}}} \, . 
\ee
We therefore see that in the $\eta \to 1$ limit, the leading behaviour of \eqref{pren} is
\be 
\Tr[(\tilde\rho_{AB})^q] = \frac{1}{|x_{13}|^{4\Delta_{(q)}}} \frac{f_{23 \bar O_q} f_{1 O_q 4}}{r_q} \, . \label{47} 
\ee
We can also observe that the coefficient appearing in the two-point function of $O_q$ in the replica limit is the same as the coefficient  in the Renyi entropy of $\rho_{AB}$ \eqref{rq}. 
Putting \eqref{rq} and \eqref{47} into \eqref{r_diff}, we find that the difference in the two mutual informations approaches a constant in the $\eta \to 1$ limit, 
\be 
\delta I_q(A:B) = - \frac{1}{q-1} \log \le(\frac{f_{23\bar O_q} f_{1O_q4}}{r_q^2} \ri), \quad \eta \to 1\, . 
\ee
We have already identified $r_q$ 
in equation \eqref{rq}, and $f_{23\bar O_q}$
and $f_{1O_q4}$ can also be given expressions in terms of the density matrix using similar logic to the previous sections. In fact, we can see that  $f_{1O_q4}$ is also equal to $r_q$. 
$f_{23\bar O_q}$ is related to
the quantity shown in Fig. \ref{fig:mutual_tn}, which can be written in terms of a matrix $A$ on three copies of $S$ as 
\begin{align} \label{49}
&f_{23\bar O_q} = \Tr[A_{S_1S_2S_3}^q] \,  |y_{23}|^{4\Delta_{(q)}} , \nn &A_{S_1S_2S_3} = \bra{{\rm MAX}_{R_1R_2}} \rho_{R_1S_1}^T \otimes \rho_{R_2}^{-\ha -\frac{i\lambda}{2}} \rho_{R_2S_2}^{\ha+\frac{i\lambda}{2}}\ket{{\rm MAX}_{S_2S_3}} \bra{{\rm MAX}_{S_2S_3}}\rho_{R_2S_2}^{\ha - \frac{i\lambda}{2}}\rho_{R_2}^{-\ha+\frac{i\lambda}{2}} \ket{{\rm MAX}_{R_1R_2}}
\end{align}
Here $R_1, R_2$ are two copies of $R$, and $\ket{{\rm MAX}}$ was defined in \eqref{maxdef}.
For a product state $\rho_{RS}=\rho_R \otimes \rho_S$, we find $\Tr[A_{S_1S_2S_3}^q]= \Tr[\rho_S^q]$. 

 We confirm that the $\eta\to 1$ limit of the $q=1$ case approaches a constant non-zero value approximately equal to 0.18 in Fig. \ref{fig:mutual_diffs}. 
It is interesting that even though we may naively interpret the $\eta \to 1$ limit as a case where $B$ vanishes, the difference in $I(A:B)$ between the two states does not go to zero in this case. Roughly, this seems to come from a competition between the fact that $B$ is becoming smaller, but at the same time the distance between the states $\rho_{ABC}$ and $\tilde \rho_{ABC}$ is diverging according to $-\log F$ or $D$.

Let us now consider the mutual information difference for $A$ and $C$, 
\be \label{deltac}
\delta I(A:C) = I(A:C)_{\rho}-I(A:C)_{\tilde \rho}  = S({\tilde\rho}_{AC})-S(\rho_{AC})  
\ee
Its Renyi version $\delta I_q(A:C)$ can be defined similarly to \eqref{r_diff}. 
Recall that the definition of $\tilde \rho_{ABC}$ was such that any correlations between $A$ and $C$ had to be generated by a map that acts only on the subsystem $B$ in $\rho_{AB}$. It is therefore natural to expect that $\delta I(A:C)$
should be positive, and we find in Fig. \ref{fig:mutual_diffs} that this is indeed the case. Moreover, the dependence on the individual interval lengths in the two terms of \eqref{deltac} cancels out such that \eqref{deltac} depends only on $\eta$. Let us first recall the behaviour of $S(\rho_{AC})$, the entropy of two non-adjacent intervals in the vacuum state. While this quantity is non-universal for general $\eta$, its leading behaviour in the $\eta \to 1$ limit comes from the identity operator in the OPE between the operators at $x_2$ and $x_3$, and is universal: 
\be 
\Tr[\rho_{AC}^q]= \frac{r_q^2}{|x_{23}|^{\Delta_{(q)}} |x_{14}|^{\Delta_{(q)}}}, \quad \eta \to 1 \, . 
\ee
Let us now consider the twist operator representation of $\Tr[\tilde\rho_{AC}^q]$, which again has the general form \eqref{corrf}. In this case, 
\begin{itemize}
    \item $\tau_A$ has one cycle of length $q$. 
    \item $\tau_B^{-1}\tau_A$ has one cycle of length  $(m_1+n_1+m_2+n_2+1)q$. 
    \item $\tau_C^{-1}\tau_B$ has $q$ cycles of length $n_1+n_2+2$, and one cycle of length $q$. 
    \item $\tau_C$ has one cycle of length $q(m_1+m_2)$. 
\end{itemize}
From the cycle structure in this case, we can see that the genus of the covering surface is $q-1$, like in the case of $\Tr[\rho_{AC}^q]$. We can also see that in the replica limit for $m_i$, $n_i$, the dimension of each of the four operators becomes $\Delta_{(q)}$.  In the $\eta \to 1$ limit, we again use the expansion \eqref{series}. The leading operator ${O'}_q = \Sigma_{\tau_C^{-1}\tau_A}$ for this case has one cycle of length $q$, and one cycle of length $(m_1+m_2)q$, so that its dimension in the replica limit is $2\Delta_{(q)}$. Putting this into \eqref{series}, we find that the leading contribution in the $\eta \to 1$ limit is 
\be 
\Tr[{\tilde\rho_{AC}}^q] = \frac{1}{|x_{23}|^{\Delta_{(q)}}|x_{14}|^{\Delta_{(q)}}} \frac{f_{23\bar{O'}_q}f_{1{{O'}}_q4}}{c_{O'_q}} (1-\eta)^{4\Delta_{(q)}} , \quad \eta \to 1
\ee
The leading behaviour of the difference in mutual informations is then 
\be 
\delta I_q(A:C) = -\frac{c}{6}\frac{q+1}{q} \log(1-\eta)  + \sO(1), \quad \eta \to 1 \, . 
\ee
The $q \to 1$ limit is 
\be 
\delta I(A:C) = - \frac{c}{3}\log(1-\eta), \quad \eta \to 1
\ee
This agrees with the fitting in Fig. \ref{fig:mutual_diffs} up to finite size effects.  It also precisely agrees with the behaviour of the conditional mutual information $I(A:C|B)$.  The CMI is often heuristically interpreted as quantifying correlations between $A$ and $C$ that are not mediated by $B$. From way in which  $\tilde \rho_{ABC}$ is constructed, it is natural to view the quantity $\delta I(A:C)$  as another  measure of the unmediated correlations between $A$ and $C$. Such unmediated correlations diverge as we make both $A$ and $C$ much larger than $B$, and both measures of such  correlations turn out to agree in this limit. 

\begin{figure}[!h]
    \centering
\includegraphics[width=0.5\textwidth]{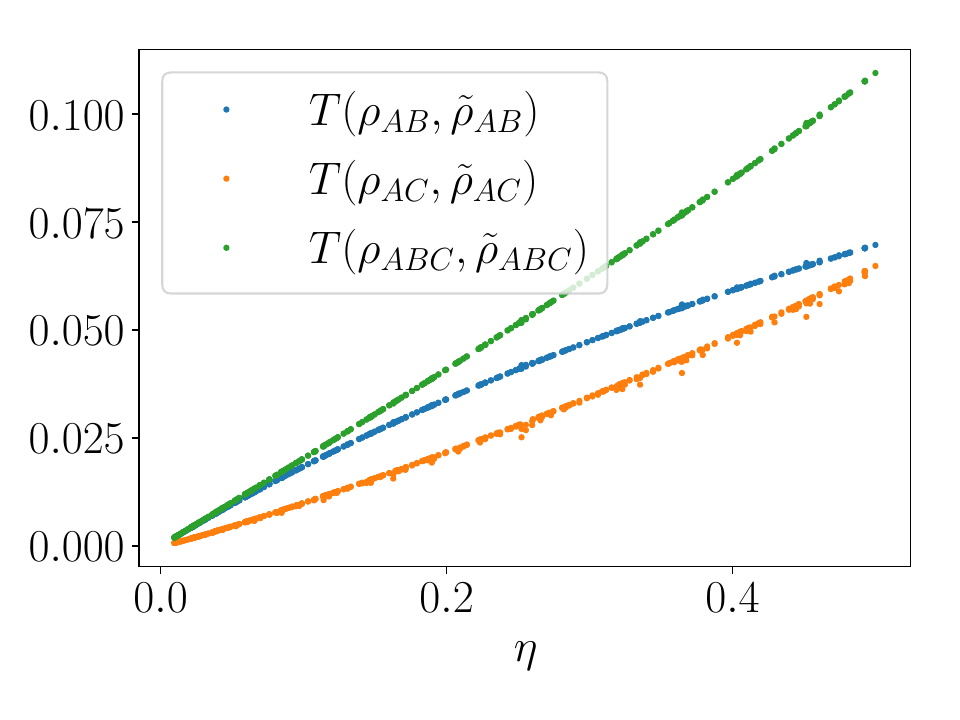}
\caption{Trace distance between the original state $\rho_{ABC}$ and the reconstructed state $\tilde{\rho}_{ABC}$ for various subsystems.}
    \label{fig:trace_sub}
\end{figure}

Let us now try to understand to the $\eta \to 0$ limit of both quantities $\delta I(A:B)$ and $\delta I(A:C)$. Due to the issues with analytic continuation in this OPE limit discussed in previous sections, let us simply interpret the numerical results of Fig. \ref{fig:mutual_diffs} instead of trying to derive them from the twist operator formalism. We see that $\delta I(A:C)$ vanishes faster than $\delta I(A:B)$ in this limit. For small enough $\eta$, this suggests that the difference between $\rho_{ABC}$ and $\tilde \rho_{ABC}$ is mostly accounted for by the difference between the reduced density matrices on $AB$ rather than on $AC$. We confirm this in Fig. \ref{fig:trace_sub}, where we compare the trace distances $T(\rho_{ABC}, \tilde \rho_{ABC})$, $T(\rho_{AB}, \tilde \rho_{AB})$, and $T(\rho_{AC}, \tilde \rho_{AC})$.  Note that we cannot  necessarily interpret this as coming from  $L_C$ becoming much smaller than $L_B$; we can keep $L_B=L_C$ constant and decrease $\eta$ by decreasing $L_A$. 

At small $\eta$, we find from Fig. \ref{fig:mutual_diffs} that 
\be 
\delta I(A:B) \propto \eta \, . 
\ee
The behaviour of $\delta I(A:C)$ is non-universal in this limit, consistent with the fact that it is a difference of two non-universal quantities. We find 
\begin{align} 
\delta I(A:C) \propto \begin{cases}
    \eta^{1.25} ~ \quad \text{Ising model} \\
    \eta^{2} ~ \quad \text{free fermions}
\end{cases}
\end{align} 
which are both consistent with 
\be \label{delac}
\delta I(A:C) \propto \eta^{1 + 2(h + \bar h)} \, . 
\ee
From Fig. \ref{fig:trace_sub}, $T(\rho_{AB}, \tilde \rho_{AB})$ and $T(\rho_{AC}, \tilde \rho_{AC})$ are both linear in $\eta$ for small $\eta$. The quantities $\delta I(A:B)$ and  $\delta I(A:C)$ thus both seem to have a behaviour consistent with some version of the Fannes-Audenaert inequality which bounds differences in von Neumann entropy between two states in terms of their trace distance, 
\be
|S(\rho)-S(\sigma)| \leq \log d \,  T - T \log T - (1-T)\log (1-T) 
\ee
where $d$ is the Hilbert space dimension. Since the Hilbert space dimension is infinite in the present setup, this inequality does not seem to imply any constraints. However, as discussed in \cite{winter_fannes}, there may be versions of the inequality that can be applied to special subsets of states in infinite-dimensional systems.

The behaviour \eqref{delac} of $\delta I(A:C)$ is interesting when compared with the leading behaviour of $I(A:C)_{\rho}$. From the general arguments of \cite{tonni2}, the leading power law of the mutual information between $A$ and $C$ in the vacuum state as $\eta \to 0$ is 
\be 
I(A:C)_{\rho} = \kappa\,  \eta^{2 (h+\bar h)}, \quad \eta \to 0  \label{2hp}
\ee
where $h+\bar h$ is the total dimension of the lowest-dimension primary operator in the theory. For the free fermion CFT, $h+\bar{h}=1/2$, and for the Ising CFT, $h+\bar{h} = 1/8$. 
 The fact that this leading power law does not appear in $\delta I(A:C)$ shows that $I(A:C)_{\tilde\rho}$ has exactly the same leading power law and coefficient.  
Physically, this tells us that we can think of the correlations between $A$ and $C$ captured by the leading power law \eqref{2hp} as  correlations mediated by $B$, since they are present in both $\rho_{ABC}$ and $\tilde \rho_{ABC}$.  This is consistent with the fact that the CMI is insensitive to the operator content of the theory and in particular to the correlations captured by \eqref{2hp}, although we do not find a quantitative matching between these two measures of unmediated correlations in this case. In particular, $I(A:C|B)$ is universal while $\delta I(A:C)$ is not.

\section{Sequential recovery on multiple regions}
\label{sec:multiple}

So far, we have considered a protocol for reconstructing the state by acting with a non-trivial channel from $B$ to $BC$ for arbitrary sizes of $B$ and $C$, as in Fig. \ref{fig:recov}. One natural constraint we can impose on a physical reconstruction process is that in a single step, both the input and the output of the map can have volume no larger than some fixed $V$.  Then consider a state $\rho_{A_1 ... A_n}$, where each $A_i$ has volume $V/2$. We can refer to this as the ``target state'' which we wish to reconstruct. Suppose we start with the state $\rho_{A_1A_2}$, and attempt to reconstruct the target state with a series of $n-2$ steps, as follows. 
At the $j$-th step, we act with $\sP_{A_{j+1} \rightarrow A_{j+1}A_{j+2}}$, where 
$\sP_{R \rightarrow RS}$ for any $R$, $S$ refers to the map 
\be 
\sP_{R \rightarrow RS}(\cdot) = \rho_{RS}^{\ha-\frac{i\lambda}{2}}\rho_R^{-\ha+ \frac{i\lambda}{2}}(\cdot)\rho_R^{-\ha-\frac{i\lambda}{2}}\rho_{RS}^{\ha+\frac{i\lambda}{2}}
\ee
where $\rho_{R}$, $\rho_{RS}$ are the reduced density matrices of the target state. If the state after $n-2$ such steps is $\tilde \rho_{A_1...A_n}$, then the fidelity 
\be 
F(\rho_{A_1,...,A_n},\tilde\rho_{A_1,...,A_n}) \label{multiple_fid}
\ee
provides a new measure of multipartite  correlations among the regions $A_i$ in the target state $\rho$. 

In a state where the correlations decay on some length scale $R$, such as the ground state of a gapped system, it seems reasonable to expect that as long as $V \gg R^d$ in $d$ dimensions, the above sequential recovery procedure works just as well as a single step where we act on $\rho_{A_1A_2}$
with $\sP_{A_2 \rightarrow A_2 A_3 ... A_n}$. This reasoning was used to propose a multiple-step recovery process for systems with finite correlation length in \cite{swingle}. In states with long-range entanglement, such as the vacuum state of a CFT or states obtained from non-equilibrium dynamics, we should expect \eqref{multiple_fid} to be worse than the fidelity of reconstruction by a single-step process, and to contain further information about the entanglement structure of the state.

\begin{figure}[!h]
    \centering
    \includegraphics[width=\textwidth]{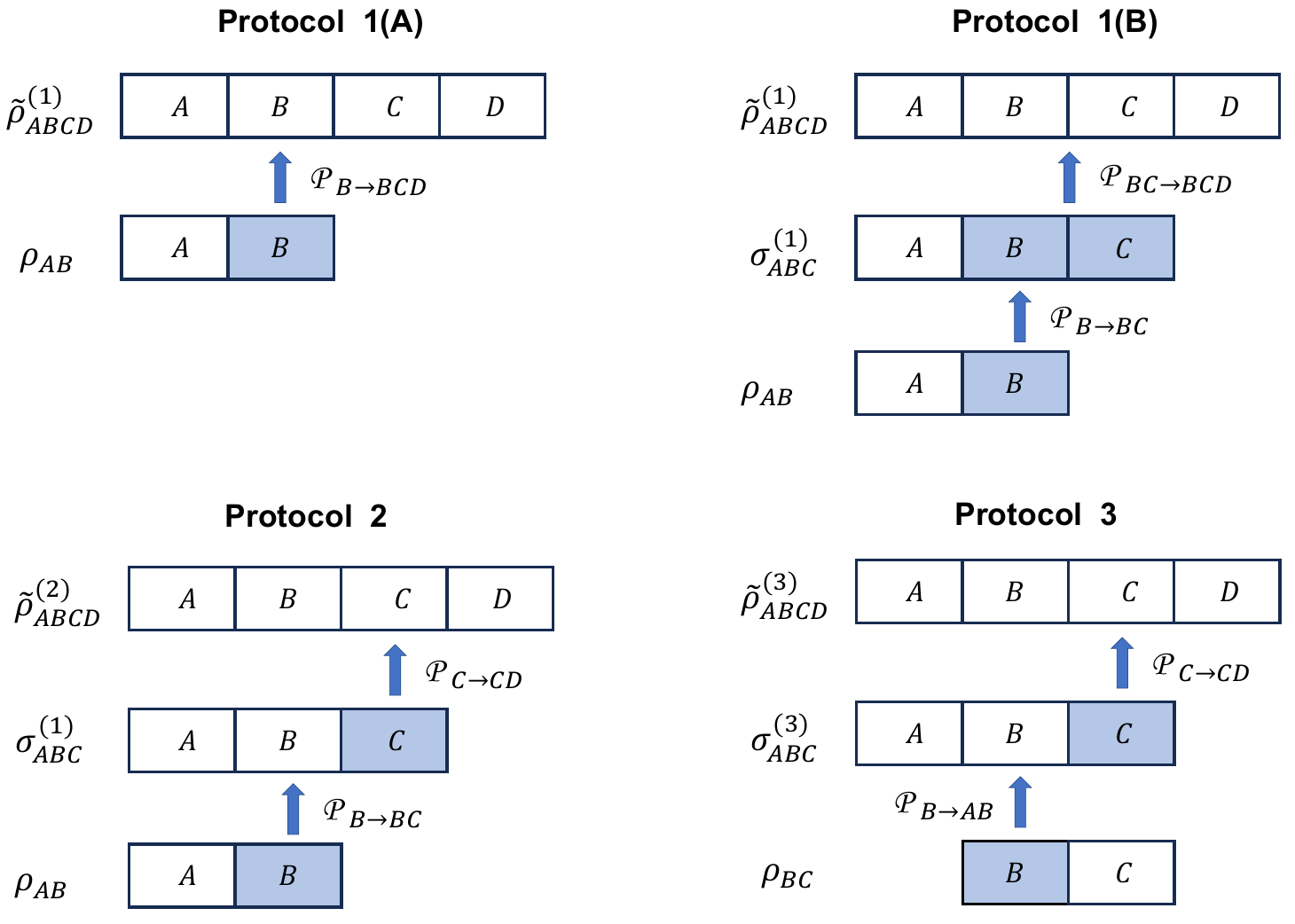}
    \caption{In all cases, we start with the reduced density matrix of the target state on some subsystem. The blue shaded regions indicate where the next stage of the map acts non-trivially.}
    \label{fig:sequential}
\end{figure}

More concretely, let us consider a setup where the full system is divided into four subsystems. A few different recovery protocols we may consider this setup are shown in Fig. \ref{fig:sequential}:
\begin{enumerate}
    \item {\it Protocol 1(A)} is simply the protocol in Fig.~\ref{fig:recov}, which was studied in the previous sections, with $CD$ playing the role of $C$ in that earlier setup. 
   In {\it Protocol 1(B)}, we act with $\sP_{B\rightarrow BC}$ in the first step, forming the output state $\sigma^{\rm (1)}_{ABC}$. In the next step, we act with $\sP_{BC\rightarrow BCD}$ on $\sigma^{\rm (1)}_{ABC}$. While this procedure naively appears different from Protocol 1(A), we can immediately see from the definition of $\sP_{R\rightarrow RS}$ that it gives rise to exactly the same final state on $ABCD$, which we call $\tilde \rho_{ABCD}^{(1)}$.~\footnote{The superscripts here refer to the different protocols and should not be confused with the value of $\lambda$, which we never write explicitly in this section.} 

    Recall from the discussion of Sec. \ref{sec:def} that we have the following lower-bound on the fidelity of this recovery: 
    \be {\rm max}_{\lambda} F^{(1)} = {\rm max}_{\lambda} F(\rho_{ABCD},\tilde \rho_{ABCD}^{(1)})\geq e^{-I(A:CD|B)/2} 
    \ee


    \item In {\it Protocol 2}, we start with the state $\rho_{AB}$, and the first step is the same as in Protocol 1(B). In the next step, we act with the map $\sP_{C \rightarrow CD}$, which is restricted to only act non-trivially on $C$. This gives rise to a new state $\tilde \rho_{ABCD}^{(2)}$. By an extension of the methods used in \cite{wilde}, we show in Appendix \ref{app:proof} that the fidelity of this state with the target state obeys the following lower-bound in any quantum-mechanical system: 
    \be \label{q2}
{\rm max}_{\lambda} F^{(2)} = {\rm max}_{\lambda} F(\rho_{ABCD}, \tilde \rho_{ABCD}^{(2)}) 
 \geq e^{- \ha (I(A:C|B) + I(B:D|C) + I(A:D|BC))} \, . 
\ee
 
\item In {\it Protocol 3}, we start with $\rho_{BC}$ and act with $\sP_{B \rightarrow AB}$, which gives rise to a state $\sigma^{(3)}_{ABC}$ that is distinct from $\sigma^{(1)}_{ABC}$.~\footnote{Note that $\sigma^{(1)}_{ABC}$ has the same reduced density matrix as $\rho_{ABC}$ on $BC$, but different reduced density matrices on $AB$ and $AC$. $\sigma^{(2)}_{ABC}$ has the same reduced density matrix as $\rho_{ABC}$ on $AB$, but different reduced density matrices on $BC$ and $AC$.} In the next step, we act with the same map as in {\it Protocol 2}, ending up with a state $\rho^{(3)}_{ABCD}$ which is general distinct from both  $\rho^{(1)}_{ABCD}$ and $\rho^{(2)}_{ABCD}$. We show in Appendix \ref{app:proof} that the fidelity of this state with $\rho_{ABCD}$ obeys the same lower bound as \eqref{q2}, 
\be \label{q3}
{\rm max}_{\lambda}F^{(3)} = {\rm max}_{\lambda}F(\rho_{ABCD}, \tilde \rho_{ABCD}^{(3)}) 
 \geq e^{- \ha (I(A:C|B) + I(B:D|C) + I(A:D|BC))} \, . 
\ee
\end{enumerate}

To get some intuition for the inequalities \eqref{q2} and \eqref{q3}, note that they imply perfect recovery if 
\begin{align} 
I(A:C|B) &+ I(B:D|C) + I(A:D|BC) =0 
\end{align}
which is equivalent to 
\be 
I(A:C|B)=0 \text{ and } I(B:D|C)=0 \text{ and } I(A:D|BC)=0 \, . \label{equiv}
\ee
The equivalence between  comes from the fact that by the strong subadditivity inequality, each of the terms on the LHS is always non-negative. In Protocol 2, $I(A:C|B)=0$ implies that the recovery in the first step is perfect, so that $\sigma^{(1)}_{ABC}= \rho_{ABC}$. Note further that from the definition of the CMI, $I(B:D|C)+ I(A:D|BC) = I(AB:D|C)$. Such identities are known as the ``chain rule'' of the CMI. The last two conditions on the RHS of \eqref{equiv} are therefore together equivalent to $I(AB:D|C)=0$, which in turn implies that the second step of the recovery process in Protocol 2 yields the state $\rho_{ABCD}$. \eqref{q2} is therefore consistent with our expectations for the conditions for perfect recovery using Protocol 2. The conditions for perfect recovery using Protocol 3 can be understood similarly. 

Let us now turn to understanding the fidelities associated with the sequential recovery procedures for the setup where $A, B, C, D$ are adjacent intervals in the vacuum state of a (1+1)D CFT. By using the replica trick in a similar way to the discussion in previous sections, both $F^{(2)}$ and $F^{(3)}$ can be expressed as  replica limits of five-point functions of twist operators: 
\be 
F^{(i)} = 
\braket{\Sigma_{\tau_A^{-1}}(x_1)\Sigma_{\tau_B^{-1}\tau_A}(x_2)\Sigma_{\tau_C^{-1}\tau_B}(x_3) \Sigma_{\tau_D^{-1}\tau_C}(x_4) \Sigma_{\tau_D}(x_5)}, \quad i = 2, 3  \label{fp}
\ee
In both cases we still have five replica parameters, $k$, $n_1$, $n_2$, $m_1$, $m_2$, and the replica limit is the one defined in \eqref{rep}. The total number of copies is now $s=k(2+2(m_1+n_1+m_2+n_2))$.
The five relevant permutations in the two cases have the following structure: 
\begin{enumerate}
    \item For $F^{(2)}$, 
    \begin{itemize}
    \item $\tau_A^{-1}$ has one cycle with $2k$ elements. 
    \item $\tau_B^{-1}\tau_A$ has $k$ cycles with $m_1+n_1+1$ elements, and $k$ cycles with $m_2+n_2+1$ elements. 
    \item $\tau_C^{-1}\tau_B$ has $k$ cycles with $m_1+n_1+1$ elements, $k$ cycles with $m_2+n_2+1$ elements, and $k$ cycles with $n_1+n_2+2$ elements. 
    \item $\tau_D^{-1}\tau_C$ has $k$ cycles with $m_2+n_2+m_1+n_1+1$ elements. 
    \item $\tau_D$ has 1 cycle with $k(m_1+m_2+1)$ elements. 
    \end{itemize}
    \item For $F^{(3)}$, 
    \begin{itemize}
    \item $\tau_A^{-1}$ has one cycle with $k(m_1+m_2+1)$ elements. 
    \item $\tau_B^{-1}\tau_A$ has $k$ cycles, each  with $n_1+n_2+2$ elements.
    \item $\tau_C^{-1}\tau_B$ has $k$ cycles with $2(m_1+n_1)+1$ elements, and $k$ cycles with $2(m_2+n_2)+1$ elements. 
    \item $\tau_D^{-1}\tau_C$ has $k$ cycles with $n_1+n_2+2$ elements. 
    \item $\tau_D$ has one cycle with $k(m_1+m_2+1)$ elements. 
    \end{itemize}
\end{enumerate}
Putting these cycle structures into the Riemann-Hurwitz formula, we find that both $(\log F^{(2)})/c$ and $(\log F^{(3)})/c$ should be independent of any details of the CFT. The dimensions of all five operators go to zero in the replica limit, and these quantities are independent of the UV cutoff. 

\begin{figure}[!h]
    \centering
    \includegraphics[width=0.8\linewidth]{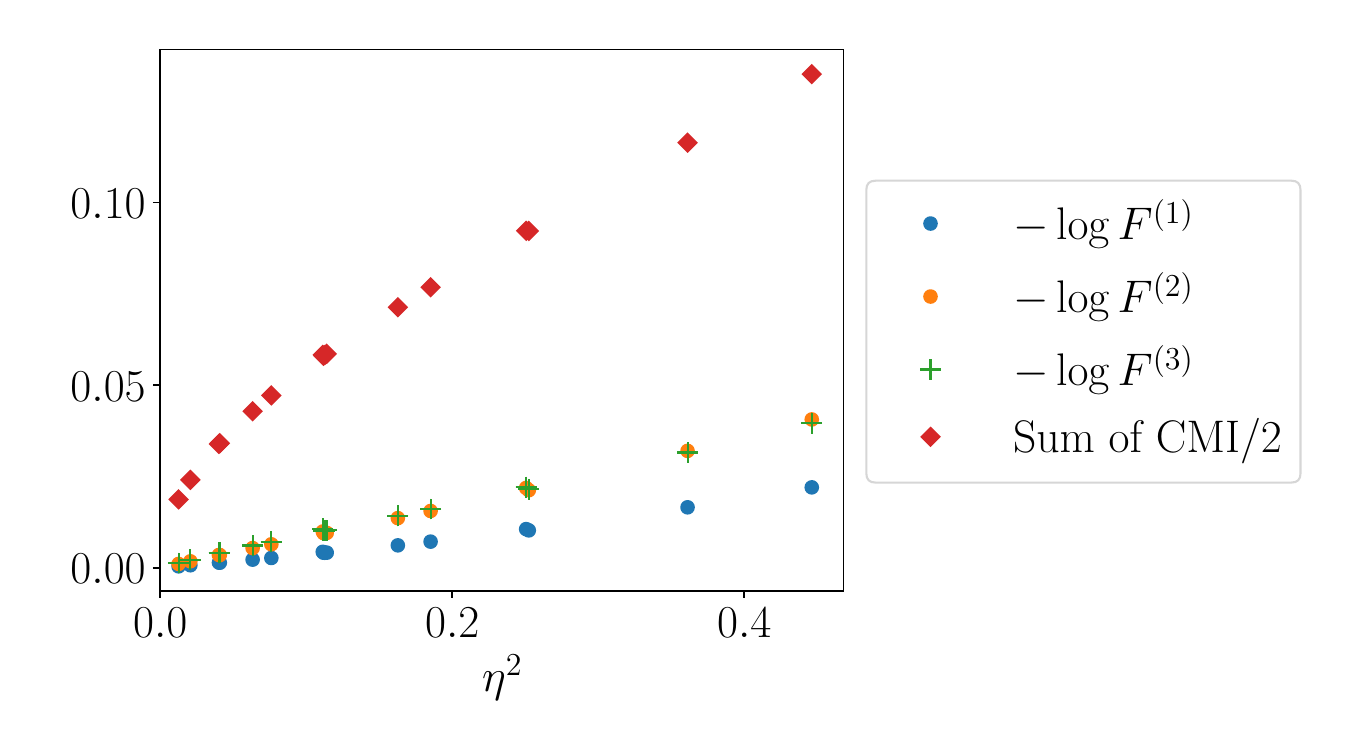}
    \caption{Comparsion of fidelity for the different recovery  schemes, and the lower bounds involving the CMI.}
    \label{fig:multiple_num}
\end{figure}

In Fig. \ref{fig:multiple_num}, we compare $F^{(1)}, F^{(2)}, F^{(3)}$ in the free fermion CFT, taking $\lambda=0$ for each Petz map, for a setup where we fix $L_A = L_D$, $L_B = L_C$, and vary \footnote{Due to numerical instability, we use small sizes with $L_A,L_B\leq 8$.}
\be 
\eta = \frac{L_A (L_C+L_D)}{(L_A + L_B ) (L_B+L_C+L_D)} \,  \, . 
\ee
We consider the regime of small $\eta$, where as discussed in previous sections, $\log F^{(1)} = f_2 c \eta^2$ for $f_2 \approx 0.07$. We find that 
\be 
-\log F^{(2)} \approx -\log F^{(3)} \propto f' \, c\, \eta^2 \, .  \ee
with $f'< f$, confirming that the sequential recovery is less effective than the single-step case.  
For this setup, we have  
\begin{align} 
\ha (I(A:C|B) + I(B:D|C) + I(A:D|BC)) 
& = -\frac{c}{6}\log\le(\frac{1-\eta}{1+\eta} \ri) \approx \frac{c}{3}\eta
\end{align} 
Again, the fidelity is much better than we would expect from the lower bounds \eqref{q2} and \eqref{q3}.

\section{Conclusions and discussion}
\label{sec:disc}


In this paper, we studied the extent to which the reduced density matrix of the CFT vacuum state in 1+1 dimensions on some region can be reconstructed from smaller subregions by an explicit recovery channel called the twirled Petz map. We found that a variety of distance measures between the original and reconstructed states have a universal form that depends only on the central charge and the cross ratio. The recovery as measured by the fidelity turns out to be better than the minimum value expected from general information-theoretic bounds. We found the universal form numerically for all values of $\eta$, and analytically in the limit $\eta \to 1$. We also studied differences in mutual information between the original and recovered states, and multiple-interval generalizations of the universal quantities associated with recovery.

While the universality of the various quantities we considered seems rather mysterious from the arguments based on the covering space and the Riemann-Hurwitz formula, we can provide a simpler  argument as follows: each of the quantities we considered can be written (after using the replica trick for some powers)  as vacuum expectation values of the following form, 
\be 
\braket{0|\rho_1 \rho_2 ...\rho_n |0} \label{vev}
\ee
where each of the $\rho_i$ is a reduced density matrix of the vacuum state on some  single interval. For example, from \eqref{fdef2}, we can write the fidelity between $\rho_{ABC}$ and $\tilde \rho_{ABC}$ as 
\begin{align} 
F(\rho_{ABC}, \tilde \rho_{ABC}) &= \lim_{k \to \ha} \Tr\le[\le(\rho_{ABC} \rho_{BC}^{\ha} \rho_B^{-\ha} \rho_{AB} \rho_B^{-\ha}\rho_{BC}^{\ha}\ri)^k\ri] \nn 
&= \lim_{k \to \ha}  \braket{0|( \rho_{BC}^{\ha} \rho_B^{-\ha} \rho_{AB} \rho_B^{-\ha}\rho_{BC}^{\ha})\, \le(\rho_{ABC} \rho_{BC}^{\ha} \rho_B^{-\ha} \rho_{AB} \rho_B^{-\ha}\rho_{BC}^{\ha}\ri)^{k-1}  |0}  \, .  \label{f_vev}
\end{align}
The relative entropy, trace distance, and the multiple-interval generalizations $F^{(2)}$, $F^{(3)}$, and $F(\rho_{A_1...A_n}, \tilde \rho_{A_1...A_n})$ defined in Section \ref{sec:multiple} can similarly all be expressed in the form \eqref{vev}. Now recall that in a conformal field theory, the density matrix of a ball-shaped region has a universal form in terms of the stress tensor $T^{\mu \nu}$ of the theory~\cite{chm}, which can be written for a single interval $A=[-\frac{a}{2}, \frac{a}{2}]$ in 1+1 dimensions as 
\be 
\rho_{A} = e^{-\sH}, \quad \sH =  2\pi \int_{-\frac{a}{2}}^{\frac{a}{2}} dx^1 \, \frac{(\frac{a}{2})^2 - (x^1)^2}{a} \, T^{00}(0, x^1) +C \label{oneint}
\ee
where $C$ is some constant that ensures normalization. 
Expressions like \eqref{vev} and \eqref{f_vev} thus involve correlation functions of exponentials of the stress tensor in the vacuum state. In general, any operator constructed entirely from the stress tensor is expected to have a universal expectation value in the vacuum state of a (1+1)D CFT that is  determined by the Virasoro algebra.   

One interesting future direction would be to use the expression \eqref{oneint} for the single-interval density matrix, or the even more explicit expressions for the density matrix and correlation matrix in the  free fermion CFT~\cite{casini_free}, in order to analytically evaluate the universal expressions for the fidelity and other quantities 
without having to use the replica trick. The expressions for quantities like $ F(\rho_{ABC}, \tilde \rho_{ABC})$ in terms of the correlation matrix of the free fermion vacuum state are rather complicated
, but it may be possible to simplify these with some further work. Certain expressions derived for the Petz map associated with a general fermion Gaussian channel in \cite{swingle_petz} may also be useful for this purpose.  

An interesting conceptual question would be to understand better  why the state $\tilde \rho_{ABC}$ is able to capture the short-distance entanglement structure of $\rho_{ABC}$, so that the distance measures between $\rho_{ABC}$ and $\tilde \rho_{ABC}$, and the mutual information difference $\delta I(A:B)$, are all well-defined quantities in the continuum limit. From the rigorous perspective of algebraic QFT, it is somewhat challenging even to understand why the mutual information of two non-adjacent intervals turns out to be a well-defined quantity in the continuum limit, as discussed in \cite{longo_cft}. Extending such proofs to the conditional mutual information of three adjacent intervals and the distance measures between $\rho_{ABC}$ and $\tilde \rho_{ABC}$ should lead to a deeper understanding of the structure of the QFT vacuum state. 

Our results also provide a concrete setting for thinking about more general  information-theoretic questions about the twirled Petz map. In all cases with non-zero conditional mutual information that we considered in this paper, the general inequalities \eqref{in}-\eqref{111} were not saturated. We can ask whether or not it is possible to find any quantum states that saturate these inequalities when the CMI is non-zero.  If such states can be found, then do they have some special structure, analogous to the Markov state structure of states that saturate the strong subadditivity inequality~\cite{hayden}? If not, then is there a stronger lower bound on the conditional mutual information in terms of some other recovery operation, or a stronger lower bound on the fidelity in terms of some other information-theoretic quantity in the state $\rho_{ABC}$? Similar questions can be asked about the multiple-interval generalizations of Sec. \ref{sec:multiple}. 

The striking difference in the behaviour of the maximum and the average fidelity for small $\eta$ that we found in Fig. \ref{fig:Flambda_avg} is also worth understanding better. Recall that in this limit, the maximum fidelity showed a different power law from the conditional mutual information lower bound, while the average fidelity only differed from the lower bound by an overall constant.  Due to our inability to understand the small $\eta$ limit using the methods of this paper,  
we were not able to provide an understanding of this observation, but it would be interesting to see if this is an example of more general relations between the CMI and the average, or between the average and the maximum. 

It would also be interesting to make more precise the sense in which tracing out $C$ and then recovering with the Petz map acting on $B$  leads to a loss of correlations between $A$ and $C$ that are not mediated by $B$. Our results in Sec.~\ref{sec:corr} show clearly that the map does not only change the correlations between $A$ and $C$ in the reconstructed state relative to the original state: for small $\eta$, the change in the correlations between $A$ and $B$ is much more significant. It would be instructive to use some simple toy models to better understand the tradeoffs between entanglement of $A$ with $B$ and  with $C$ that are achieved by the map, and how these are related to the entanglement structure of the original state. It should also be interesting to better understand the physical reason for the fact that $\delta I(A:B)$ does not vanish in the $\eta \to 1$ limit. 

The quantities \eqref{qty2}, \eqref{f23_comp}, and \eqref{49} which arise naturally from the $\eta \to 1$ limits of various measures, would also be interesting to understand better, as would the interpretation of the cyclic invariance discussed at the end of Sec. \ref{sec:z1}.

In any relativistic quantum field theory, it is possible to show a non-trivial lower bound on the conditional mutual information $I(A:C|B)$ in setup of Fig.~\ref{fig:regions}~(b), for the case where $L_A = L_C \ll L_B$.  This is simply a restatement of the $c$-theorem of \cite{casini}. The starting point is the strong subadditivity inequality in the setup of Fig.~\ref{fig:regions}~(a), with $R_1 = R_3 = \Delta \ll R_2 = R$. We now consider the case of the vacuum state of a general relativistic QFT, where the strong subaddivitiy inequality is in general not saturated, but we still have $S(AB) = S(BC) = S(\sqrt{R(R+2\Delta)})$, and $S(ABC) = S(R+2\Delta)$. Then expanding LHS of the strong subadditivity inequality in this setup to quadratic order  $\Delta/R$, we find 
\be \label{cthm}
-S''(R) \Delta^2 \geq  \frac{S'(R)}{R} \Delta^2\, . 
\ee
This inequality is interpreted in the literature as the monotonic decay of the $c$-function $c(R)=RS'(R)$ with $R$. Another interpretation is that we can write the LHS of \eqref{cthm} as $S(R+\Delta) + S(R+\Delta) - S(R)- S(R+2\Delta)$, so that it is the conditional mutual information $I(A:C|B)$ for the configuration in Fig. \ref{fig:regions}(b) with three adjacent intervals of lengths $\Delta$, $R$, and $\Delta$ on a spatial slice. So in a relativistic QFT, the usual strong subadditivity inequality CMI $\geq 0$ applied to the setup of Fig.~\ref{fig:regions}(a) gives a stronger lower bound on the CMI for the setup of  Fig.~\ref{fig:regions}(b). We can then ask how this lower bound compares to the general information theoretic bounds of \eqref{in} and \eqref{111}. The results of this paper show that in a conformal field theory, the $c$-theorem is stronger than the general information-theoretic bounds -- the former is saturated, while the latter are not. If there does exist some stronger information-theoretic lower bound on the CMI, it would be interesting to see whether that  coincides with the bound from the $c$-theorem.

Yet another interesting question is whether the distance measures between $\rho_{ABC}$ and $\tilde \rho_{ABC}$ can be identified with some bulk geometric quantities
in holographic CFTs. The reduced density matrix of a subsystem in the boundary CFT is expected to encode the properties of a bulk region known as its entanglement wedge~\cite{dong_harlow, jlms}. Intuitively, the fidelity or relative entropy between $\rho_{ABC}$ and $\tilde \rho_{ABC}$ should be related to the part of the entanglement wedge of $ABC$ that lies outside the entanglement wedges of $AB$, $B$, and $BC$, from which  $\tilde \rho_{ABC}$ 
is constructed. It would be interesting to try to identify a bulk dual by extending the methods of \cite{lewkowycz} used for the entanglement entropy, although this may be quite challenging due to the lack of replica symmetry in the permutations that define these quantities, and the issues with taking the replica limit that we encountered in the CFT calculations in Section \ref{sec:smallz} and Appendix \ref{app:lou}. One starting point may be to look for a bulk quantities which have the limiting behaviours in equations \eqref{18}-\eqref{ds} in empty $AdS_3$.     

Finally, the differences between $\rho_{ABC}$ and $\tilde \rho_{ABC}$ could provide a new way of characterizing the properties of  interesting states with long-range entanglement that appear in other contexts in quantum many-body systems, such as topologically ordered states, or states obtained from nonequilibrium dynamics.

\section{Acknowledgments} 

We would like to thank Gauri Batra, Lorenz Eberhardt, Thomas Faulkner, Abhijit Gadde, Jeongwan Haah, Patrick Hayden, Jonah Kudler-Flam, Raghu Mahajan,  Alex May, Mukund Rangamani, Stephen Shenker, Douglas Stanford, and Jinzhao Wang for helpful discussions. SV is supported by Google and the Stanford Institute for Theoretical Physics. YZ is supported by the Stanford Q-FARM Postdoctoral Fellowship.

\begin{appendix}

\section{Numerical algorithm}
In this section we explain how to compute the  fidelity and other distance measures between $\rho_{ABC}$ and $\tilde \rho_{ABC}$ numerically. For the spin chain  models we use the periodic uniform MPS technique. This section will be mostly concerned about this technique. For the free fermion model, we may alternatively use the correlation matrix techniques, although high numerical precision will be needed due to inversion of an ill-conditioned matrix. 
\label{app:num}
\subsection{Coarse graining of periodic uniform MPS}
Starting with the critial spin chain Hamiltonian with $N$ spins, we can use a periodic uniform matrix product state with finite bond dimension $\chi$ to represent the ground state. Note that $\chi$ has to increase with the system size $N$ \cite{DMRG,MPS}. The MPS can be obtained by energy minimization of the Hamiltonian \cite{puMPS} \footnote{The optimization of a periodic MPS is harder than the usual DMRG. The code can be accessed at https://github.com/FuTen/puMPS.jl.}. Let $A,B,C,D$ be its four parties. We can obtain a coarse-grained state $|\psi\rangle_{ABCD}$ using the standard coarse graining of matrix product state \cite{RGMPS1}, where the dimensions of the coarse grained Hilbert spaces are smaller than $\chi^2$ (the Schimidt rank of the density matrices $\rho_{i},~i=A,B,C,D$). In practice, since most of the Schmidt coefficients are very small, we can further reduce the dimension of the coarse grained Hilbert space $d_i$ by paying a small truncation error. For the detailed algorithm, see the Supplemental Material of \cite{RGMPS2}. We will use this approximation to reduce the numerical cost. With the sizes of the critical system under consideration (Ising model with $L\leq 128$, XXZ model with $L\leq 60$), we find that $d_i = 64$ is enough to reproduce accurate results with truncation error on the order of $10^{-7}$.

\subsection{Fidelity}
As shown in Ref. \cite{Uhlmann}, one may utilize the Uhlmann theorem to efficiently compute the fidelity of tensor network states. Here we will apply similar techniques to the Petz map fidelity.

\begin{figure}
    \centering
    \includegraphics[width = 0.8\linewidth]{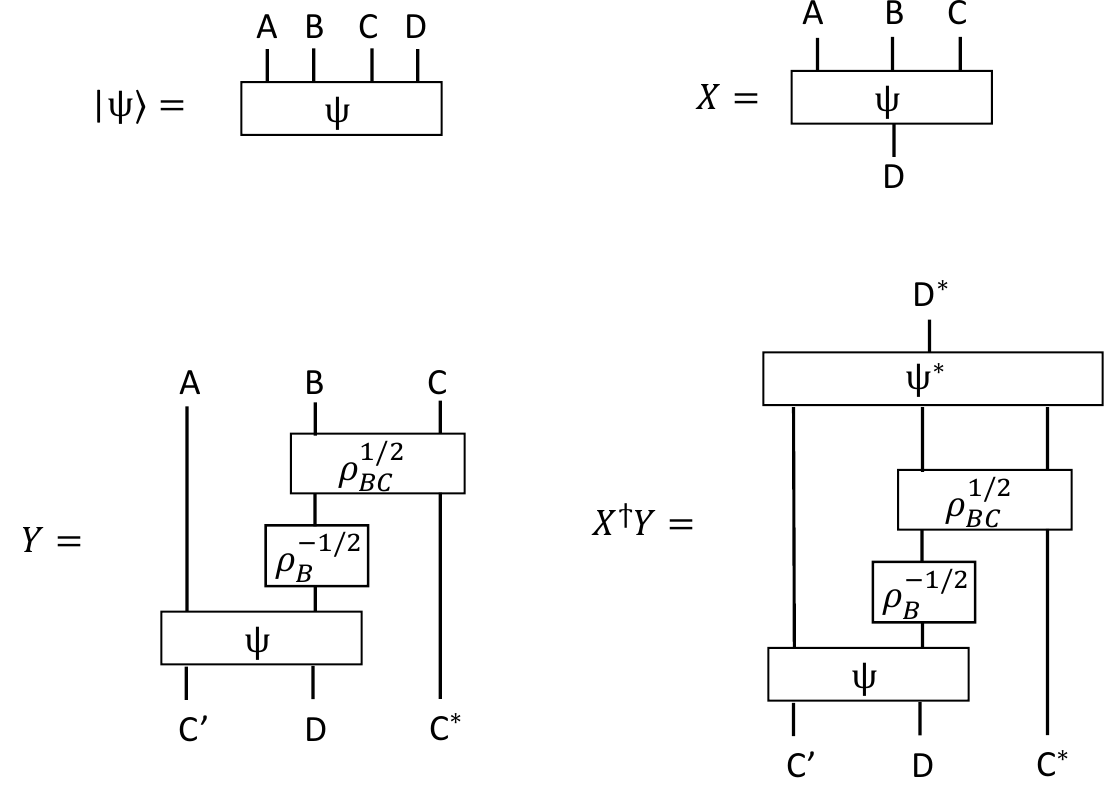}
    \caption{Computation of the Uhlmann fidelity for the Petz recovery map}
    \label{fig:mps}
\end{figure}

The Uhlmann theorem states that the fidelity between two mixed states is the maximal fidelity between their purifications. Let $\rho$ and $\sigma$ be two mixed states, and $|\psi_{\rho}\rangle$ and $|\psi_{\sigma}\rangle$ be two arbitrary purifications, then
\begin{equation}
    F(\rho,\sigma) = \max_{|\psi_{\rho}\rangle,|\psi_{\sigma}\rangle}{|\langle \psi_{\rho}|\psi_{\sigma}\rangle|}.
\end{equation}
In Ref. \cite{Uhlmann}, it has been shown that if we can decompose $\rho = XX^{\dagger}$ and $\sigma = YY^{\dagger}$, then the above expression reduces to 
\begin{equation}
    F(\rho,\sigma) = \mathrm{Tr}|X^{\dagger} Y|,
\end{equation}
where $\mathrm{Tr}|A| = \mathrm{Tr}\sqrt{A^{\dagger} A}$ is the trace norm (sum of singular values of $A$). Now we compute the matrices $X$, $Y$ and $X^{\dagger}Y$. Let $|\psi\rangle_{ABCD}$ be a four-party pure state 
\begin{equation}
    |\psi\rangle_{ABCD} = \sum_{a,b,c,d} \psi_{abcd} |a\rangle_A |b\rangle_B |c\rangle_C |d\rangle_D.
\end{equation}
The state can be obtained using the coarse graining algorithm of the MPS in the last subsection. The reduced density matrix $\rho_{ABC}$ can be expressed as $\rho_{ABC} = XX^{\dagger}$, where
\begin{equation}
    X = \sum_{a,b,c,d} \psi_{abcd} |abc\rangle\langle d|.
\end{equation}
In terms of components, $X$ is exactly the same as the wavefunctions. The Petz recovered state
\begin{equation}
    \tilde{\rho}_{ABC} = \rho^{1/2}_{BC} \rho^{-1/2}_{B} \rho_{AB} \rho^{-1/2}_{B} \rho^{1/2}_{BC}
\end{equation}
can be expressed as $\tilde{\rho}_{ABC} = YY^{\dagger}$, where 
\begin{equation}
    Y = \rho^{1/2}_{BC} \rho^{-1/2}_{B} \sum_{a,b,c,c',d}\psi_{abc'd}|abc\rangle \langle c'dc|,
\end{equation}
where we have introduced an auxiliary Hilbert space $C'$ with basis $|c'\rangle$ and the Hilbert space is isomorphic to $C$. Under the isomorphism between the bra and the ket, the matrix $X$ becomes exactly the original pure state $|X\rangle_{ABCD} = |\psi\rangle_{ABCD}$ and the matrix $Y$ becomes a purification of the Petz recovered state $\tilde{\rho}_{ABC}$, where
\begin{equation}
    |Y\rangle_{ABCC'DC^{*}} = \rho^{1/2}_{BC} \rho^{-1/2}_{B} \sum_{a,b,c,c',d}\psi_{abc'd}|abc\rangle |c'dc\rangle
\end{equation}
In terms of tensor networks, the matrices $X,Y$ and $X^{\dagger} Y$ are shown in Fig.~\ref{fig:mps}.
\subsection{Renyi relative entropy}
The Renyi relative entropy can also be computed efficiently using the MPS at a similar cost. The quantity is defined as
\begin{equation}
    D_{\alpha}(\rho,\tilde{\rho}) = \frac{1}{\alpha-1} \log \tr (\rho^{\alpha} \tilde{\rho}^{1-\alpha})
\end{equation}
for $0<\alpha<1$. The limit of $\alpha\rightarrow 1$ gives the usual relative entropy
\begin{equation}
    D(\rho||\tilde{\rho}) = \tr ({\rho\log \rho -\rho \log \tilde{\rho}}).
\end{equation}
To start with, we use singular value decomposition of the matrix $X$ to get the entanglement spectrum of $\rho_{ABC}$,
\begin{equation}
    |X\rangle = \sum_{i} \sqrt{p_i} |\phi_i\rangle_{ABC} |\phi_i\rangle_D.
\end{equation}
Then,
\begin{equation}
    \rho^{\alpha}_{ABC} = \sum_{i} p^{\alpha}_i |\phi_i\rangle_{ABC}\langle \phi_i|_{ABC}.
\end{equation}
In order to compute the entanglement spectrum of the Petz recovered state $\tilde{\rho}_{ABC}$, we singular value decompose the matrix $Y$, which gives
\begin{equation}
    |Y\rangle = \sum_{j} \sqrt{q_j} |\tilde{\phi}_j\rangle_{ABC} |\tilde{\phi}_j\rangle_{C'DC^{*}}
\end{equation}
Thus,
\begin{equation}
    \tilde{\rho}^{1-\alpha}_{ABC} =\sum_{j} q^{1-\alpha}_j |\tilde{\phi}_j\rangle_{ABC}\langle \tilde{\phi}_j|_{ABC}.
\end{equation}
Finally, we can compute
\begin{equation}
\label{eq:MPSDalpha}
    D_{\alpha}(\rho,\tilde{\rho}) = \frac{1}{\alpha-1}\log \left(\sum_{i,j} p^{\alpha}_{i} q^{1-\alpha}_j |\langle \tilde{\phi}_j|\phi_i\rangle_{ABC}|^2\right)
\end{equation}
The most expensive part of the algorithm is the singular value decomposition of $Y$, which is a 6-leg tensor. One crucial trick to reduce the numerical cost is that we never compute the matrix $Y$ explicitly. Instead, we follow the following steps to compute the singular values and singular vectors, where the intermediate tensors have at most 4 external legs.

\begin{figure}
    \centering
    \includegraphics[width = 0.96\linewidth]{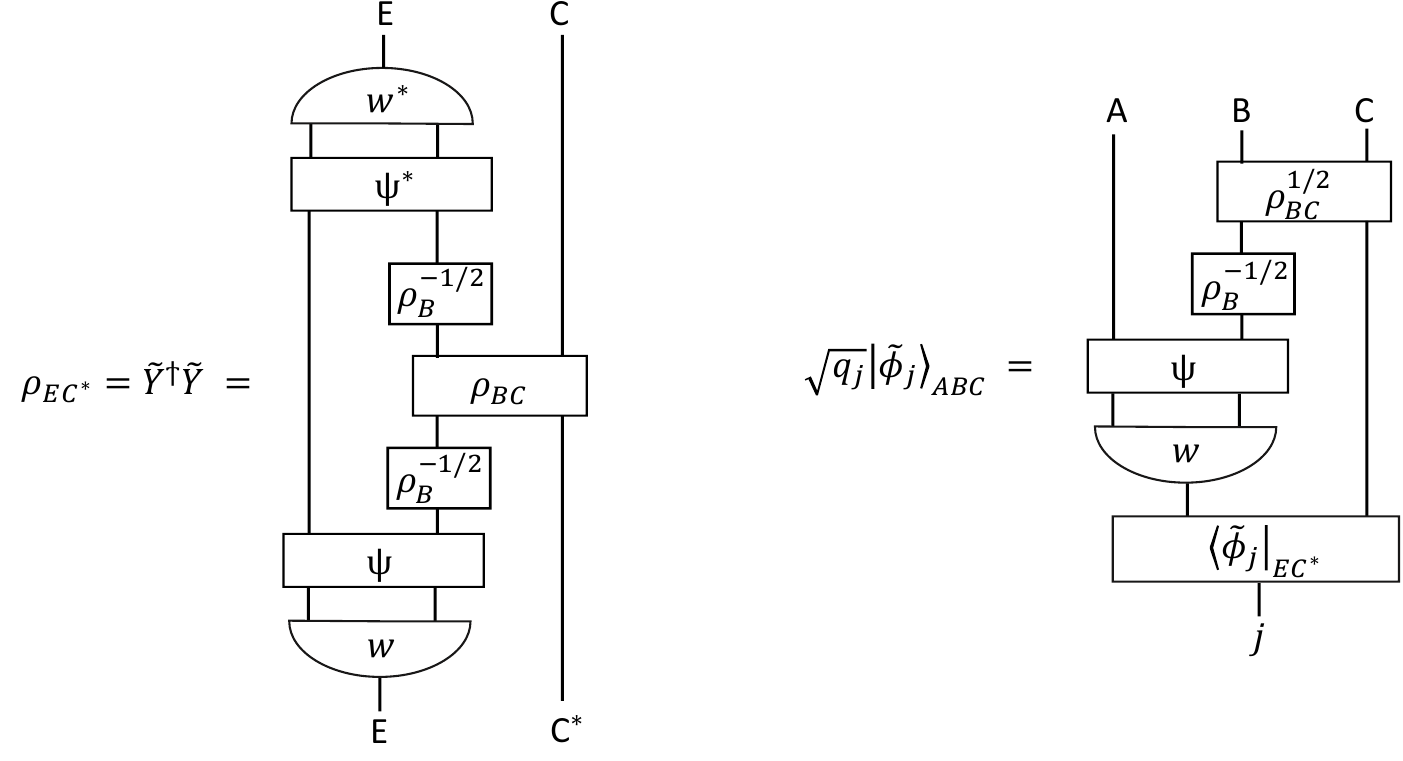}
    \caption{Computation of the Renyi relative entropy for the Petz recovery map}
    \label{fig:mps2}
\end{figure}

First of all, we can use a coarse graining isometry $w:~C'D\rightarrow E$ to reduce the auxiliary dimension in $|Y\rangle$. The isometry projects onto the subspace spanned by the support of $\rho_{C'D}$, neglecting eigenvalues below $\epsilon = 10^{-7}$. The coarse graining can be achieved with a small dimension $d_E$ since $\rho_{C'D}$ has small entanglement, which follows from the fact that $\rho_{C'D}$ isomorphic to the ground state density matrix on $CD$. The resulting tensor $|\tilde{Y}\rangle_{ABCEC^{*}} = w|Y\rangle_{ABCC'DC^{*}}$ is a purification of $\tilde{\rho}_{ABC}$ on $ABC\otimes EC^{*}$. Next, we compute the eigenvalues $q_j$ and eigenvectors $|\tilde{\phi}_j\rangle_{EC^{*}}$ of the density matrix $\rho_{EC^{*}} = \tilde{Y}^{\dagger} \tilde{Y}$. This part is the most expensive part of the algorithm, where the dominant cost is a diagonalization of a matrix of dimension $d_Ed_C$. The order of contraction to obtain $\rho_{EC^{*}}$ is such that we never encounter an intermediate tensor with more than four external legs. Finally, we obtain the singular vectors $|\tilde{\phi}_j\rangle_{ABC}$ by
\begin{equation}
    \sqrt{q_j} |\tilde{\phi}_j\rangle_{ABC} = \langle \tilde{\phi}_j|_{EC^{*}} |\tilde{Y}\rangle_{ABCEC^{*}}.
\end{equation}
Again, the contraction is in an order which we never encounter an intermediate tensor with more than four external legs. Thus we have obtained all $q_j$ and $|\tilde{\phi}_j\rangle_{ABC}$ needed in Eq.~\eqref{eq:MPSDalpha}. The above procedure is summarized in Fig.~\ref{fig:mps2}

\subsection{Trace distance and other norms}
Finally, we can efficiently compute the trace distance and other distance norms with the MPS, where the $L_{\alpha}$ distance is defined as
\begin{equation}
    T_{\alpha} (\rho,\tilde{\rho}) = (\tr |\rho-\tilde{\rho}|^{\alpha})^{1/\alpha}.
\end{equation}
The trace distance is the limit $\alpha\rightarrow 1$, with a conventional $1/2$ factor,
\begin{equation}
    T(\rho,\tilde{\rho}) = \frac{1}{2} \tr |\rho-\tilde{\rho}|.
\end{equation}
In order to compute these norms, we have to obtain the eigenvalues of $\rho-\tilde{\rho}$. Naively, $\rho_{ABC}$ and $\tilde{\rho}_{ABC}$ are both 6-leg tensors, so computing the eigenvalues of $\rho-\tilde{\rho}$ would require a diagonalization of a 6-leg tensor. However, we can use a similar trick as in the previous section to reduce the cost to a diagonalization of 4-leg tensors. The key observation is that the support of $\rho_{ABC}$ is at most $d_D$ dimensional, and the support of $\tilde{\rho}_{ABC}$ is at most $d_Ed_C$ dimensional. Thus, their difference $\rho-\tilde{\rho}$ is supported on a subspace which is at almost $d_Ed_C+d_D$ dimensional (the direct sum of the two supports). In the case of $d_D\ll d_Ed_C$, the dominant cost is similar to diagonalizing $\rho_{EC^{*}}$ in the previous section. However, the support of the two density matrices are not orthogonal subspaces, and in fact they are almost identical at small $\eta$. This introduces an extra complication of diagonalizing an ill-conditioned norm matrix, which increases the error to the order of $10^{-4}$ in the worst case considered (see details below).

We define a norm matrix $N_{ij}$ as 
\begin{equation}
    N_{ij} = \langle \tilde{\phi}_j |\phi_i\rangle_{ABC},
\end{equation}
where $i$ runs in the support of $\rho_{ABC}$, which is at most $d_D$ dimensional, and $j$ runs in the support of $\tilde{\rho}_{ABC}$, which is at most $d_Ed_C$ dimensional. Assuming the supports are linearly independent, we define an extended norm matrix, which encodes the overlap of basis vectors in the direct sum of the two supports.
\begin{equation}
    \tilde{N} =
    \begin{bmatrix}
    I & N \\
    N^{\dagger} & I 
    \end{bmatrix}
\end{equation}
The extended norm matrix is ill-conditioned if the two supports are close. In this basis, the two density matrices can be represented as
\begin{equation}
    \rho = 
    \begin{bmatrix}
    P & PN \\
    N^{\dagger}P & N^{\dagger} P N 
    \end{bmatrix}
\end{equation}
and
\begin{equation}
    \tilde{\rho} = 
    \begin{bmatrix}
    NQN^{\dagger} & NQ \\
    QN^{\dagger} & Q 
    \end{bmatrix},
\end{equation}
where $P$ and $Q$ are diagonal matrices with entries $p_i$ and $q_i$, respectively. In order to obtain the trace distance, we have to find an orthonormal basis of the direct sum of the support. To do this, we diagonalize $\tilde{N} = WDW^{\dagger}$, where $W$ is unitary and $D$ is diagonal. The difference $\rho-\tilde{\rho}$ in the orthonormal basis defined by the columns of $W$ is
\begin{equation}
    \delta \rho = \tilde{N}^{-1/2} (\rho-\tilde{\rho}) \tilde{N}^{-1/2}.
\end{equation}
Thus, the distance norm is
\begin{equation}
    T_{\alpha} (\rho,\tilde{\rho}) = ||\delta {\rho}||_{\alpha} = ||\tilde{N}^{-1/2} (\rho-\tilde{\rho}) \tilde{N}^{-1/2}||_{\alpha} .
\end{equation}
Finding the orthonormal basis is not a numerically stable process if $\tilde{N}$ is ill-conditioned. In the worst case considered, that is, the case with smallest $\eta$, which happens for $L_A=L_C=2, L_B = 62$ with the total length $L=128$ of the Ising model, the total error is roughly escalated to $10^{-4}$, which is roughly $10^{3}$ times the truncation error $10^{-7}$. At larger $\eta$, the escalation is alleviated as expected, where the total error falls onto the same order of the truncation error.  
\end{appendix} 

\subsection{Comments on the correlation matrix method for the free fermion}
For the free fermion CFT, one may alternatively use the correlation matrix method to compute the Petz map fidelity. This allows us to compute the fidelity exactly with the numerical cost scaling as $O(L^3)$, where $L$ is the size of the subsystem $A,B,C$. The reconstruction is given in explicit form in Ref.~\cite{swingle_petz}, see their Eq. (44) and Eq. (51) in particular. Here we provide additional necessary input for the specific CFT example.

The key idea is that the reduced density matrix is completely determined by the two-point correlation functions of the fermionic operator. Let us consider the critical Hamiltonian which is dual to the critical Ising model under the Jordan-Wigner transformation,
\begin{equation}
    H = -i \sum_{i} \gamma_i \gamma_{i+1},
\end{equation}
where $\gamma_i$ is the fermion operator with anticommutation relation $\{\gamma_i,\gamma_j\} = 2\delta_{ij}$. The correlation matrix of any Gaussian state is defined as $G_{jl} = (i/2)\tr ([\gamma_j,\gamma_l]\rho)$. The ground state correlation matrix can be obtained analytically using the Fourier transform,
\begin{equation}
    G_{jl} = \frac{2}{\pi(j-l)}.
\end{equation}
Now we can apply the techniques in Ref.~\cite{swingle_petz} to obtain the Petz map fidelity. Although the formula is straightforward, one has to take extra caution to take the matrix inverse in their Eq. (44). This is because the state is weakly entangled and the majority of the eigenvalues of $G$ of an interval is close to $\pm i$. This then requires very high precision in the numerics. One may use arbitrary precision linear algebra provided in programming language such as Julia or Python. Such a need for high precision is already noted in Ref.~\cite{eisler2017analytical} when they considered the entanglement Hamiltonian. In our numerics, we observe that approximately $5L$ digits of floating point precision is needed for diagonalization of a correlation matrix on $2L$ contiguous fermions.

\section{Covering maps and Liouville actions}
\label{app:lou}
We review the general setup and formulas for the Liouville action in Sec.~\ref{app:lou_rev}, and then derive various specific correlation functions of twist operators relevant for the main text in the subsequent sections. 

\subsection{Review of Liouville action}
\label{app:lou_rev}

Let us review the methods of \cite{lunin} for calculating a correlation function of twist operators by mapping the original space to a covering space. We will closely follow the notation of Appendix D of \cite{avery}, and we refer the reader to that reference for a clear and detailed derivation of the formulas briefly summarized below. 

Suppose we have $M$ copies of the CFT, and a $K$-point function of twist operators in $\sS_M$,  
\be
\braket{\Sigma_{\tau_1}(z_1) \Sigma_{\tau_2}(z_2) ...  \Sigma_{\tau_K}(z_K)} = Z_{\epsilon, \delta}^{(M)}[\tau_1(z_1), ...,\tau_K(z_K) ] \label{b1def}
\ee
The right-hand side should be seen as a more precise version of the definition of the twist operators from \eqref{sigmadef}. It refers to a partition function on $M$ copies of the CFT, with a circle of radius $\epsilon$ cut out around each of the $z_j$, at which we put the boundary conditions that 
a field $\Psi_I$ on the $I$'th copy is sent to $\Psi_{\tau_j(I)}$ on going around the circle. The definition also includes an IR cutoff $\delta$, so the metric on the original space is given by 
\be
(ds^2)_{g} = \begin{cases} 
d z  d \bar z & |z|< \frac{1}{\delta} \\
d\tilde z d\bar{\tilde z} & |\tilde z|< \frac{1}{\delta} 
\end{cases}, \quad \tilde z = \frac{1}{\delta^2} \frac{1}{z} \label{dsg} 
\ee
which describes two large discs glued together at radius $1/\delta$. 

The main idea of \cite{lunin} is that if we can find a map $z(t)$ such that the fields as a function of the coordinate $t$ become single-valued, then on the space labelled by $t$, known as the covering space, \eqref{b1def} can be equated with a partition function of the theory without any twisted boundary conditions. We will refer to the original space labelled by $z$ as the base space below. The covering space can in general be some higher genus surface, but we can restrict to the case where it is genus zero for the correlation functions of interest in this paper.  Even in this case, the covering space path integral is non-trivial due to the fact that  the metric as a function of the coordinate $t$ is no longer the flat metric. Instead, if $(ds^{2})_{\tilde g}$ is the flat metric $dt d\bar t$ on the covering space, then the curved metric induced from the base space is  
\be 
(ds^2)_g = dz d \bar z = \frac{dz}{dt} \frac{d\bar z}{d\bar t}\, dt d \bar t = e^{\phi} (ds^2)_{\tilde g}
\ee
where the last equality can be seen as the definition of $\phi(t)$, and we have not explicitly written down a similar equation for the second half of the sphere in \eqref{dsg}. Then by the Weyl anomaly, 
\be
\braket{\Sigma_{\tau_1}(z_1) \Sigma_{\tau_2}(z_2) ...  \Sigma_{\tau_K}(z_K)} =  \int_g D\psi e^{-I_{E,g}[\psi]}  = e^{S_L} \int_{\tilde g} D\psi e^{- I_{E, \tilde g}[\psi]} \, . \label{sl} 
\ee
where $S_L$ is the Liouville action associated with $\phi$, 
\be \label{sldef}
S_L = \frac{c}{96\pi} \int d^2 t \, \sqrt{- \tilde g} \,[\partial_{\mu} \phi \, \partial_{\nu} \phi \, {\tilde g}^{\mu \nu} + 2 \tilde R \phi] 
\ee
where $\tilde R$ refers to the curvature on the covering space. We see that the main ingredient for computing $S_L$ is to find the covering map $z(t)$, and use it to find $\phi(t)$. 

Suppose the twist operator $\tau_1$ consists of a single cycle $(1, 2, ..., p)$, so that on going around $z_1$, we have $\psi_1 \to \psi_2 \to ... \to \psi_p \to \psi_1$. This means that to return to $\psi_1$, we need to go around $z_1$ $p$ times  by an angle of $2\pi$ on the base space. For the fields to be single-valued on the covering space, going an angle of $2\pi p$ around  $z_1$ should correspond to going around an angle of $2\pi$ around the image $t_1$ of $z_1$. So close to $t_1$, the map should have the form 
\be 
z(t) \approx z_1 + a_1(t-t_1)^p , \quad  t \to t_1
\ee
for some constant $a_1$. Now if $\tau_1$ consists of multiple cycles which have lengths $p_1, ..., p_{N_1}$, we should have $N_1$ distinct images $t_1, ..., t_{N_1}$ of $z_1$, such that 
\be \label{ztb}
z(t) \approx  z_1 + a_i(t-t_i)^{p_i}, \quad t \to t_i 
\ee
for some constants $a_i$.  All such images of $z_j$ with $p_i>1$, which correspond to the presence of non-trivial cycles on the base space, are referred to as ``ramified points'' on the covering space, and $p_i$ is referred to as the ramification index. 
Now if $\tau_2$ has $N_2$ cycles of length $p_{N_1+1}, ..., p_{N_1+2}$, then $z_2$ should have $N_2$ ramified images, close to which we should have a similar behaviour to \eqref{ztb} with $z_1$ replaced with $z_2$, and so on for other $z_j$ up to $j=K-1$. It is convenient to set the last point $z_K=\infty$ using global conformal invariance. (More precisely, this means it is placed at the center of the second disc in \eqref{dsg}.) Let $\tau_K$ have $N_K$ cycles, which we label $q_0, ..., q_{K-1}$. The first cycle can be mapped to infinity on the covering space without loss of generality, 
\be 
z \approx b_0 t^{q_0}, \quad t \to \infty
\ee
and the remaining cycles correspond to multiple poles of the covering map at $t= t_i^{\infty}$, for $i = 1, ..., N_{K}-1$
\be 
z \approx \frac{b_i}{(t-t_i^{\infty})^{q_i}}, \quad t \to t_i^{\infty} \, . \label{pole}
\ee
In addition to this, the map may also have simple poles at points $\ell_i$, which do not correspond to any cycles of the permutation but do contribute to the Liouville action. We can add these poles to the set of  $\{t_i^{\infty}\}$, with $q_i=1$, and also label their residues with $b_i$. 

The above discussion gives us the parameters $\{a_i, p_i\}$, $i = 1, ..., N_J$, where $N_J$ is the total number of cycles in all permutations at finite points $z_j$, and $\{b_i, q_i\}$, $i = 0, ..., N_{K}+N_P-1$, where $N_K$ is the total number of cycles of the twist operator placed at infinity and $N_P$ is the total number of simple poles. In order to find the $a_i, b_i$, we have to find the explicit form of $z(t)$ which is consistent with the above expansions close to each of the $t_i$ and $t_{i}^{\infty}$. As shown in \cite{avery}, once we do find these parameters, $S_L$ can be  expressed entirely in terms of them and in terms of the IR and UV cutoffs. After dividing the twist operators by an appropriate normalization factor so that their two-point function at unit separation becomes 1, we get an expression that depends only on $\{a_i, p_i\}$, $\{b_i, q_i\}$, and the IR cutoff $\delta$: 
\begin{align}
&\braket{\Sigma_{\tau_1}^{\rm norm}(z_1) \Sigma_{\tau_2}^{\rm norm}(z_2) ...  \Sigma^{\rm norm}_{\tau_K}(z_K)} \nn
&= \delta^{4\Delta_{\tau_K}}\le(\prod_{i=1}^{N_J} p_i^{-\frac{c}{12}(p_i+1)} \ri) \le(\prod_{i=0}^{N_K-1} q_i^{\frac{c}{12}(q_i-1)} \ri) \le( \prod_{i=1}^M |a_i|^{-\frac{c}{12}\frac{p_i-1}{p_i}} \ri)
\le(\prod_{i=0}^{N_K+N_P-1} |b_i|^{-\frac{c}{12}\frac{q_i+1}{q_i}} \ri)|b_0|^{\frac{c}{6}} q_0^{\frac{c}{6}} \label{master}
\end{align}
Some general methods for finding a genus zero covering map were developed in \cite{lunin} and \cite{roump}. We will make use of these methods in our discussion of various examples below.

\subsection{Liouville action for fidelity of $\rho_{RS}$ and $\rho_R \otimes \rho_S$} 
\label{app:lou_fid}

Let us evaluate  the quantity $\frac{f_{1O_a4}}{\sqrt{c_{O_a}}}$, which appeared in the coefficient of the $\eta \to 1$ limit of the $F(\rho_{ABC}, \tilde \rho_{ABC})$ in Sec. \ref{sec:z1}. The main purpose of this calculation is to show a simple example of a  quantity appearing in this paper where the Liouville action can be computed explicitly and analytically continued, before discussing more complicated cases in later sections.

Recall that in the replica limit, $\frac{f_{1O_a4}}{\sqrt{c_{O_a}}}$ is UV finite. It can therefore be written as a limit of the appropriately normalized twist operators written in the previous section. In the discussion of Sec. \ref{sec:z1},  $f_{1O_a4}$ was related to $\Tr[(\rho_R \rho_S^{m_1+m_2} \rho_{RS})^k]$ for two adjacent intervals $R$ and $S$, which became $\Tr[(\rho_R  \rho_S \rho_{RS})^{\ha}]$ on taking the replica limits of $m_1, m_2, k$. Since we are eventually interested in this replica limit, let us simplify the calculation by writing 
$\Tr[(\rho_R  \rho_S \rho_{RS})^{\ha}] = \lim_{k \to \ha}\Tr[(\rho_R  \rho_S \rho_{RS})^{k}]$. The latter quantity for integer $k$ can be related to a three-point function of the following permutations on $3k$ copies:
\begin{align} 
&\tau_1 = (1, 3, 4, 6, ..., 3(k-1)+1, 3k) \\
&\tau_2 = (1, 2, 3k) \, (3, 4, 5)\, ... \,  (3(k-1),3(k-1)+1 , 3(k-1)+2)  \\
& \tau_3 = (2, 3, 5, 6, ..., 3k-1, 3k) 
\end{align}
$\tau_1$ and $\tau_3$ each have one cycle of length $2k$, and $\tau_2$ has $k$ cycles of length 3. In calculating $\Tr[(\rho_R  \rho_S \rho_{RS})^{k}]$,  $\tau_1$, $\tau_2$, $\tau_3$ appear in order from left to right at the three endpoints of $R$ and $S$.   To find the covering map, it is convenient to use  the cyclic invariance of the three-point function mentioned in Section \ref{sec:z1} to put ${\tau_3}$, ${\tau_1}$, and ${\tau_2}$ at 0, 1, and $\infty$ respectively. 
We then have 
\be 
\frac{f_{10_a4}}{\sqrt{c_{O_a}}}= \lim_{k\to \ha} \frac{\braket{\Sigma^{\rm norm}_{\tau_3}(0) \Sigma^{\rm norm}_{\tau_1}(1) \Sigma^{\rm norm}_{\tau_2}(\infty)}}{\delta^{4\Delta_{\tau_2}}} \label{3p_r}
\ee

To use the formula \eqref{master} for the correlation function of normalized twist operators, we need to find the covering map $z(t)$ and in particular the parameters $a_i$, $b_i$. From the cycle structures of $\tau_3$, $\tau_1$, and  $\tau_2$, $z=0$ and $z=1$ should have one ramified image each with ramification order $2k$, and these can be mapped to $t=0$ and $t=1$ using global conformal transformations. $z=\infty$ should have $k$ ramified images with ramification order 3. We can use global conformal invariance to map one of these images to infinity, but the remaining images $t_1^{\infty}$, ..., $t_{k-1}^{\infty}$ cannot be arbitrary. While it is in general complicated to evaluate three-point functions involving multiple-cycle twist operators for this reason, we can use a trick inspired by Appendix C of \cite{faulkner_ref} to find the covering map for this particular case.

We first look for the covering map $z(w)$ for a three-point function of single-cycle twist  operators on three copies of the theory of length 2, 2, and 3 respectively placed at $z=0$, $z=1$, and $z=\infty$. Say we fix the images of $z=0, 1, \infty$ under this map to be $t=0, 1, \infty$ respectively. Then we can  compose this with a map $w(t)$ for a two-point function of single-cycle twist operators of length $k$ at the points $w=0$ and $w=1$ respectively. By expanding both $w(t)$ and $z(w)$ close to their ramified points, we can convince ourselves that the map $z(t)$ resulting from this composition has precisely the structure we need for the three-point function \eqref{3p_r}. 

Now the two steps of finding $z(w)$ and $w(t)$ are both simple. We want $z(w)$ to have the following behaviours: 
\begin{align} 
&z = a_0' w^2 , & w \to 0 \label{sim1} \\
&z = 1 + a_1' (w-1)^2,  & w \to 1 \label{sim2} \\
& z = b_0' w^3, & w \to \infty \label{sim3} 
\end{align}
where the constants $a_0', a_1', b_0'$ still need to be determined by finding the full expression for $z(w)$. To derive this, note that from \eqref{sim1} and \eqref{sim2}, $z'(w)$ must be proportional to both $w$ and $w-1$. The simplest guess is 
\be 
z'(w) = C w (w-1) \label{zp_sim}
\ee
for some constant $C$. This also has the behaviour $z'(w)\to w^2$ as $w \to \infty$ required from \eqref{sim3}. Now by integrating \eqref{zp_sim} and requiring $z(w=0)=0$ and $z(w=1)=1$, we can fix the map to be~\footnote{We could in principle have derived this from more complicated formulas of the covering map for a general three-point function of single-cycle twist operators in \cite{lunin}, but we found it simpler to use the method of \cite{roump} involving the derivative of $z(t)$.} 
\be 
z(w) = 3 w^2 - 2w^3
\ee
This has the form \eqref{sim1}-\eqref{sim2}, with $a_0'=3$, $a_1'=-3$, and $b_0'=-2$. 

Next, note that the covering map $w(t)$ for a two-point function of twist operators of length $k$ is well known: 
\be 
w(t) = \frac{t^k}{t^k- (t-1)^k}
\ee

Now let us check that $z(t)=z(w(t))$ has the required behaviours for $z=0, 1, \infty$, and identify the parameters $a_i$, $b_i$ needed for the formula \eqref{master}.  $z=0$ has one image at $t=0$. Close to $t=0$, we can expand
\be 
z(t) = 3(-1)^{2k} t^{2k} , \quad t\to 0 \,   
\ee
so we have 
\be
p_1=2k, \quad  a_1 = 3(-1)^{2k}. \label{p1}
\ee
$z=1$ has one image at $t=1$, and 
\be 
z(t) = 1 - 3(t-1)^{2k}, \quad t\to 1  
\ee
so 
\be 
p_2 = 2k, \quad  a_2=-3 \, . \label{p2}
\ee
$z=\infty$ has $k$ images, one of which is at $t=\infty$, and the others are at 
\be 
t_m = \frac{1}{1-e^{2\pi i m/k}}, \quad m = 1, ..., k-1 \, . 
\ee
Near these points, the map has the following behaviours:
\begin{align}
z(t) = -\frac{2}{k^3} t^3, \quad t \to \infty
\end{align}
and 
\be 
z(t) = -\frac{2}{k^3}\le( \frac{t_m^k}{t_m^{k-1}-(t_m-1)^{k-1}}\ri)^3 \frac{1}{(t-t_m)^3}, \quad t \rightarrow t_m \, . 
\ee
We therefore have 
\begin{align}
& q_0 = 3, \quad b_0 = -\frac{2}{k^3} \label{q0}\\
& q_m = 3, \quad b_m =  -\frac{2}{k^3}\le( \frac{t_m^k}{t_m^{k-1} - (t_{m}-1)^{k-1}}\ri)^3, \quad m =1, ..., k-1 \, . \label{qm}
\end{align}

Now let us put \eqref{p1}, \eqref{p2}, \eqref{q0}, \eqref{qm} into the formula \eqref{master}. We get~\footnote{Note that $\prod_{m=1}^{k-1}\frac{t_m^k}{t_m^{k-1}-(t_m-1)^{k-1}} = \le(\frac{1}{\prod_{m=1}^{k-1}(1-e^{2\pi i m/k})}\ri)^2 = 1/k^2$\, . } 
\be 
S_L - 4 \Delta_{\tau_2}\log \delta = c\le[-\frac{1}{6}(2k+1)\log 2k + \frac{k+1}{6}\log 3 +\le(\frac{1}{6}-\frac{k}{9}\ri)\log \frac{2}{k^3} + \frac{1}{3}\log k^2\ri]
\ee
Putting $k=1/2$ in this expression and substituting into \eqref{3p_r}, we find 
\be 
\frac{f_{1O_a4}}{\sqrt{c_{O_a}}} = e^{0.12062 \, c} \, . 
\ee
which agrees with numerical results for this quantity, as discussed in the main text. 

\subsection{Covering map for $\frac{f_{23\bar O_a}}{\sqrt{c_{O_a}}}$}
\label{app:lou_23}

Let us now consider the harder case of the quantity $\frac{f_{23\bar O_a}}{\sqrt{c_{O_a}}}$, which from \eqref{m1n1} we can relate to the replica limit of the quantity 
\be 
Q= \Tr_S[\Tr_R[\rho_{RS}^{m_1+1} \rho_R^{n_1}] \Tr_R[\rho_{RS}^{m_2+1} \rho_R^{n_2}]]
\ee
The $k$-dependence appears as an overall power of this expression. The appropriately normalized version of $Q$ can be expressed in terms of a three-point function 
\be 
Q \propto \braket{\Sigma^{\rm norm}_{\tau_1}(0) \Sigma^{\rm norm}_{\tau_2}(1) \Sigma^{\rm norm}_{\tau_3}(\infty)} \label{3p}
\ee
where $\tau_1$ has one cycle of length $n_1+n_2+2$, $\tau_2$ has one cycle of length $m_1+m_2+2$, and $\tau_3$ has one cycle of length $m_1+n_1+1$, and one cycle of length $m_2+n_2+1$. (We have again cyclically reordered the twist operators to put the multiple-cycle permutation at infinity.) From this cycle structure, the covering map for \eqref{3p} should have the following behaviour: 
\begin{align}
z(t) = \begin{cases}
a_0 \,  t^{n_1+n_2+2}, & t \to 0  \\
1+ a_1 \, (t-1)^{m_1+m_2+2}, & t \to 1 \\
b_0 \, t^{m_1+n_1+1}, & t \to \infty \\
b_1/(t-\ell)^{m_2+n_2+1}, & t \to \ell 
\end{cases} \label{ab4}
\end{align}
The point $\ell$ needs to be determined. Now from the behaviour of the derivative $z'(t)$ close to 0, 1, and $\ell$, it must have the form 
\be \label{zp_comp}
z'(t) = C \frac{t^{n_1+n_2+1}(t-1)^{m_1+m_2+1}}{(t-\ell)^{m_2+n_2+2}} \, . 
\ee
This also has the right behaviour $z'(t) \propto t^{m_1+n_1}$ as $t\to\infty$, so we can take $C$ to be a $t$-independent constant. Now to determine the parameters $\ell$ and $C$ and the overall constant on integrating \eqref{zp_comp}, we need to impose the following conditions: 
\begin{enumerate} 
\item Since $z'(t)$ is a total derivative, its residue at the multiple pole $\ell$
should vanish~\cite{roump, dei_eb}. This determines $\ell$ in terms of $m_i, n_i$. 
\item $z(t=0) = 0$, and $z(t=1) = 1$. These conditions determine $C$ and the constant of integration in terms of $m_i, n_i$. 
\end{enumerate}

After solving for these parameters, we can find the $a_i$, $b_i$ in \eqref{ab4} in terms of $m_i, n_i$ and substitute into \eqref{master} to find $Q$, and find $\frac{f_{23\bar O_a}}{\sqrt{c_{O_a}}}$ by taking the replica limit. 

It turns out, however, that it is not possible to find an analytic expression for $\ell$ in terms of $m_i, n_i$ from point 1. The condition is 
\be 
{\rm Res}_{t=\ell} z'(t) =0  \label{reszero}
\ee
where 
\begin{align}
&{\rm Res}_{t=\ell} z'(t) = \frac{d^{m_2+n_2+1}}{dt^{m_2+n_2+1}} \le(t^{n_1+n_2+1}(t-1)^{m_1+m_2+1}\ri)|_{t = \ell}\\
&= \sum_{p=0}^{m_1+m_2+1} {m_1+m_2+1 \choose p}(-1)^p (m_1+n_1+2-p)_{m_2+n_2+1}\ell^{m_1+n_1+1-p} \label{poly}\\
= 
&\frac{\ell^{1 + m_1 + n_1}
    \Gamma(3 + m_1 + m_2 + n_1 + n_2) {}_2F_1(-1 - m_1 - m_2, -1 - 
     m_1 - n_1, -2 - m_1 - m_2 - n_1 - n_2, 1/\ell)}{\Gamma(2 + m_1 + n_1)} \label{hyp}
\end{align}
where $(x)_n$ refers to the Pochhammer symbol $x(x+1)...(x+n-1)$, and $\Gamma$ and ${}_2F_1$ to the Gamma function and the hypergeometric function. The final expression \eqref{hyp} is not particularly illuminating, but we can already see from \eqref{poly} that the condition for the vanishing of the residue is a polynomial equation in $\ell$ whose degree increases with $m_1, n_1$. This means that we cannot obtain an analytic expression for $\ell$ in terms of arbitrary integers $m_i$, $n_i$.  As a result, we cannot obtain $S_L$ and $\frac{f_{23\bar O_a}}{\sqrt{c_{O_a}}}$ in the replica limit using analytic continuation in this case. 

One reason we should not be too surprised that there did not turn out to be an analytic solution in this case is that all expressions above depend on $m_i, n_i$ only through the combinations $m_2+n_2$, $m_1+n_1$, $n_1+n_2$, $m_1+m_2$. The dependence on the parameter $\lambda$ of the twirled Petz map cancels out in each of these combinations. So any analytic expression we had found from the above procedure would have resulted in a value of $\frac{f_{23\bar O_a}}{\sqrt{c_{O_a}}}$ that did not depend on $\lambda$. This would be inconsistent with the nontrivial $\lambda$-dependence that we observe in Fig. \ref{fig:ope_comparison}.

\subsection{Covering map for the fidelity of $\rho_{ABC}$ and $\tilde \rho_{ABC}$}
\label{app:lou_4pt}

Recall that we equated the replica version of the fidelity to a four-point function of twist operators: 
\be 
F_{k, n_1, n_2, m_1, m_2} = \braket{\Sigma_{\tau_A}^{-1}(x_1) \Sigma_{\tau_B^{-1}\tau_A}(x_2) \Sigma_{\tau_C^{-1}\tau_B}(x_3) \Sigma_{\tau_C}(x_4)} \, , 
\ee
where $\tau_A^{-1}$ has one cycle with $2k$ elements, $\tau_B^{-1}\tau_A$ has $k$ cycles with $m_1+n_1+1$ elements and $k$ cycles with $m_2+n_2+1$ elemets, $\tau_C^{-1}\tau_B$ has $k$ cycles with $n_1+n_2+2$ elements, and $\tau_C$ has one cycle with $(m_1+m_2+1)k$ elements. It is convenient to first use cyclic invariance and then put three of the points at 0, 1, and $\infty$, so that we instead need to evaluate 
\be 
\braket{\Sigma_{\tau_C^{-1}\tau_B}(0) \Sigma_{\tau_C}(\eta) \Sigma_{\tau_A^{-1}}(1) \Sigma_{\tau_B^{-1}\tau_A}(\infty)} \, . \label{cor2}
\ee
Next, using a composition trick similar to \cite{faulkner_ref}
and Sec. \ref{app:lou_fid}, above, we can reduce the problem to two steps: first, we need to find the covering map $z(w)$ for a four-point function 
\be 
\braket{\Sigma_{\tau_1}(0) \Sigma_{\tau_2}(\eta) \Sigma_{\tau_3}(1) \Sigma_{\tau_4}(\infty)} \label{b42}
\ee
for certain permutations $\tau_i$ on $N= n_1+m_1+n_2+m_2+2$ copies, and then we compose this with a covering map $w(t)$ for single-cycle twist operators of length $k$ at the images of $\eta$ and $1$ in $w$-space under the first map $z(w)$. In trying to do the first step of finding $z(w)$, we will encounter a more complicated version of the same difficulty as in the previous subsection, which will prevent us from obtaining an analytic expression. 

In \eqref{b42}, $\tau_1$ has one cycle with $n_1+n_2+2$ elements, $\tau_2$ has one cycle with $m_1+m_2+1$ elements, $\tau_3$ has one cycle with 2 elements, and $\tau_4$ has two cycles: one with $m_1+n_1+1$ elements, and one with $m_2+n_2+1$. So the map $z(w)$ should have the following behaviours: 
\begin{align} 
&z \approx a_1 w^{n_1+n_2+2}, \quad w \rightarrow 0 \\
&z \approx  \eta + a_2(w-w_1)^{m_1+m_2+1}, \quad w \rightarrow w_1 \\
&z \approx 1 + a_3(w-1)^{2}, \quad w \rightarrow 1 \\
&z \approx  \frac{b_1}{(w-w_2)^{m_2+n_2+1}}, \quad w \rightarrow w_2 \\ 
& z \approx b_0 w^{m_1+n_1+1}, \quad w \rightarrow \infty \label{winf}
\end{align} 
Here $w_1, w_2$ are points in the $w$ plane which need to be determined. 
As in the previous sections, based on the above behaviours we can write  
\be  \label{dz}
z'(w) = C \frac{w^{n_1+n_2+1} (w-w_1)^{m_1+m_2} (w-1)}{(w-w_2)^{m_2+n_2+2}}
\ee
As $w \rightarrow \infty$, \eqref{dz} implies that $z'(w) \propto w^{m_1+m_2}$, which is the expected behaviour from \eqref{winf}, so there is no need to include additional poles, and $C$ is a constant with respect to $w$. To determine $w_1$, $w_2$, and $C$, we must use the following conditions: 
\begin{enumerate}
\item From the fact that 
$z'$ is a total derivative, 
\be 
\text{Res}_{w = w_2} z'(w) = 0 \, . \label{res_4p}
\ee
\item 
$z(0)=0, \quad  z(1) = 1, \quad z(w_1)= \eta, \quad z(w_2) = \infty \, .$
\end{enumerate}
Again, the difficulties will arise from the residue condition \eqref{res_4p}. Note that 
\be 
\text{Res}_{w = w_2} z'(w) 
 = \frac{d^{m_2+n_2+1}}{dw^{m_2+n_2+1}}\le(w^{n_1+n_2+1} (w-w_1)^{m_1+m_2} (w-1) \ri)|_{w=w_2}
\ee
Using the binomial expansion for $(w-w_1)^{m_1+m_2}$, we find 
\begin{align} 
&
\text{Res}_{w = w_2} z'(w) \nn 
& = \sum_{p=0}^{m_1+m_2} {m_1+m_2 \choose p} (-w_1)^{p} \big[ (n_1+m_1-p+2)_{m_2+n_2+1} w_2^{n_1+m_1+1-p} \nn
& \quad \quad  \quad \quad  \quad \quad  - (n_1+m_1-p+1)_{m_2+n_2+1} w_2^{n_1+m_1-p} \big] \label{poly4}\\
& = \frac{w_2^{1+m_1+n_1}\Gamma(3+m_1+m_2+n_1+n_2) {}_2F_{1}(-m_1-m_2, -1-m_1-n_1, -2-m_1-m_2-n_1-n_2, \frac{w_1}{w_2})}{\Gamma(2+m_1+n_1)} \nn
& - \frac{w_2^{m_1+n_1}\Gamma(2+m_1+m_2+n_1+n_2) {}_2F_1(-m_1-m_2, -m_1-n_1, -1-m_1-m_2-n_1-n_2, \frac{w_1}{w_2})}{\Gamma(m_1+n_1+1)} \label{res2}
\end{align}
Now we need to solve for the above expression being equal to zero to express $w_2$ in terms of $w_1$ or vice versa, and then impose the requirements in point 2. But again from \eqref{poly4}, we see that we have a polynomial equation of arbitrarily high degree in either $w_1$ or $w_2$ for general $m_i, n_i$, which cannot be solved analytically. So again, we cannot obtain an analytic expression for the Liouville action using this method. 

Once again, note that all expressions in this section involved $\lambda$-independent combinations of the $m_i, n_i$. So not being able to solve the above equations and analytically continue is important for consistency with the non-trivial $\lambda$-dependence we saw in this quantity in Sec. \ref{sec:num}.

\section{Proof of lower bounds for multiple-step recovery}
\label{app:proof}

Let us show the general bounds of \eqref{q2} and \eqref{q3} for the multiple-step recovery process. We discuss the case \eqref{q2} in detail, and briefly mention the similar proof of \eqref{q3}. 

Let us start with  the twirled Petz map for a general channel $\sN$, following \cite{wilde}. If $\sN$ is a map $S \to B$, then the twirled Petz map $\sP_{\sN, \sigma}$ (for some reference state $\sigma \in S$) is defined as 
\be
\sP_{\sigma, {\cal N}}^{(\lambda)}(\rho)=\sigma^{\frac{1}{2}-\frac{i\lambda}{2}}{\cal N}^{\dagger}\left({\cal N}(\sigma)^{-\frac{1}{2}+\frac{i\lambda}{2}
}\, \rho \, {\cal N}(\sigma)^{-\frac{1}{2}-\frac{i\lambda}{2}
}\right)\sigma^{\frac{1}{2}+\frac{i\lambda}{2}}\, . 
\ee
We get the map \eqref{twirled} considered for most of this paper by taking $\sN = \Tr_C$ and $\sigma = \rho_A \otimes \rho_{BC}$. $\sN^{\dagger}:B \to S$ refers to the adjoint channel of $\sN$. Any $\sN: S \to B$ can be given an isometric extension, 
\be 
\sN (\cdot) = \Tr_E[U_{S \to B E} (\cdot) U_{S \to B E}^{\dagger} ], \quad  U_{S \to B E}^{\dagger}U_{S \to B E} = \mathbf{1}_S \, \label{iso}
\ee
in terms of which the adjoint can be expressed simply as 
\be 
\sN^{\dagger}(\cdot) = U_{S \to B E}^{\dagger} \,  [(\cdot)_B \otimes \mathbf{1}_E] \, U_{S \to B E} \, . 
\ee

\begin{figure}[!h]
    \centering
    \includegraphics[width=0.6\textwidth]{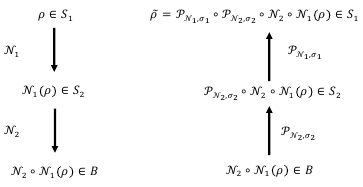}
    \caption{Two-step recovery for general channels.}
    \label{fig:twostep}
\end{figure}

Now we consider the composition of two maps, $\sN_1: S_1 \to S_2$, and $\sN_2: S_2 \to B$, as shown in Fig. \ref{fig:twostep}. For each of these we can introduce isometric extensions, using the isometries $U_{S_1 \to S_2 E_1}$ and $V_{S_2 \to B E_2}$ respectively. Then we can express the adjoints as follows: 
\begin{align} 
\sN_2^{\dagger}(\cdot) = V^{\dagger}_{S_2\to BE_2}\, (\cdot)_B \otimes \mathbf{1}_{E_2} \, V_{S_2\to BE_2}, \quad B \to S_2 \\
\sN_1^{\dagger}(\cdot) =U^{\dagger}_{S_1\to S_2E_1}\, (\cdot)_{S_2} \otimes \mathbf{1}_{E_1}\, U_{S_1\to S_2E_1}, \quad S_2 \to S_1 
\end{align} 

\begin{figure}[!h]
    \centering
    \includegraphics[width=0.8\textwidth]{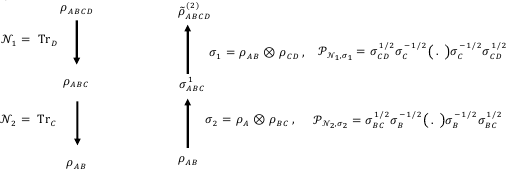}
    \caption{Two-step recovery for the setup of ``Protocol 2.''}
    \label{fig:twostep_2}
\end{figure}

Now we are interested in the fidelity $F(\rho, \tilde \rho)$, where 
\be 
\tilde \rho = \sP_{N_1, \sigma_1} \circ \sP_{\sN_2, \sigma_2} \circ \sN_2 \circ \sN_1(\rho)\, . 
\ee
See Fig. \ref{fig:twostep}. 
For the choices of $\sN_{1,2}$ and $\sigma_{1,2}$ shown in Fig. \ref{fig:twostep_2}, we get the fidelity $F^{(2)}= F(\rho_{ABC}, \tilde \rho_{ABCD}^{(2)})$ of protocol 2.

Using the isometric extensions, the fidelity can be expressed in the form 
\be 
F(\rho, \tilde \rho) = || [\sN_2 \circ \sN_1(\rho)^{\frac{z}{2}} (\sN_2(\sigma_2)^{-\frac{z}{2}} \otimes \mathbf{1}_{E_2}) V_{S_2 \to B E_2} \sigma_2^{\frac{z}{2}} \sN_1(\sigma_1)^{-\frac{z}{2}}]\otimes \mathbf{1}_{E_1} U_{S_1 \to S_2 E_1}\sigma_1^{\frac{z}{2}} \sqrt{\rho}||_1 \label{fiddef}
\ee
with $z = 1- i\lambda$. Now we define the operator appearing inside the one-norm in \eqref{fiddef} as an operator-valued function on the complex plane: 
\be 
G(z) = [\sN_2 \circ \sN_1(\rho)^{\frac{z}{2}} (\sN_2(\sigma_2)^{-\frac{z}{2}} \otimes \mathbf{1}_{E_2}) V_{S_2 \to B E_2} \sigma_2^{\frac{z}{2}} \sN_1(\sigma_1)^{-\frac{z}{2}}]\otimes \mathbf{1}_{E_1} U_{S_1 \to S_2 E_1}\sigma_1^{\frac{z}{2}} \sqrt{\rho} 
\ee
We then use the following theorem about functions of this kind, which appears as Theorem 2.1 of \cite{Wilde_2015}: 
\begin{theorem}\label{hadamard}
    Let
\[
S\equiv\{z\in\mathbb{C}:0\leq\text{Re}(z)\leq 1\}
\]
and let $G: S\rightarrow L({\cal H})$ a bounded map that is holomorphic on the interior of $S$ and continuous on the boundary. Let $\theta\in (0,1)$ and define
\[
\frac{1}{p_\theta}\equiv\frac{1-\theta}{p_0}+\frac{\theta}{p_1}
\]
where $p_0, p_1\in [1, \infty]$. For $k=0,1$, define $M_k\equiv \sup_{t\in\mathbb{R}}\|G(k+it)\|_{p_k}$. Then
\[
\|G(\theta)\|_{p_\theta}\leq M_0^{1-\theta}M_1^{\theta},
\]
where we use the norm $\|A\|_p^p=\text{Tr}(A^{\dagger}A)^{p/2}$.
\end{theorem}

Now let us take $p_0=2$, $p_1=1$, and $\theta\in(0, 1)$, so that $p_{\theta}=2/(1+\theta)$. Then note that 
\begin{align}
    M_0&=\sup_{\lambda\in\mathbb{R}}\|G(i\lambda)\|_2\nonumber\\
    &=\sup_{\lambda\in\mathbb{R}}\left\|\left(\left([{\cal N}_2\circ{\cal N}_1(\rho)]^{i\lambda/2}[{\cal N}_2(\sigma_2)]^{-i\lambda/2}\otimes \mathbf{1}_{E_2}\right)V\sigma_2^{i\lambda/2}[{\cal N}_1(\sigma_1)]^{-i\lambda/2}\otimes \mathbf{1}_{E_1}\right)U\sigma_1^{i\lambda/2}\rho^{1/2}\right\|_{2}\nonumber\\
&\leq \|\rho^{1/2}\|_2=1
\end{align}
and
\begin{align}
M_1=\sup_{\lambda\in\mathbb{R}}\|G(1-i\lambda)\|_1 =\sup_{\lambda\in\mathbb{R}}F(\rho, \tilde \rho^{(\lambda)}) \, 
\end{align}
from \eqref{fiddef}. 
Applying Theorem \ref{hadamard}, we obtain
\be
\|G(\theta)\|_{2/(1+\theta)}^{1/\theta}\leq\sup_{\lambda\in\mathbb{R}}F(\rho, \tilde \rho^{(\lambda)}).
\ee
Setting $\theta=(1-\alpha)/\alpha$, we have
\be\tilde{\Delta}_{\alpha}\geq -2\log \sup_{\lambda\in\mathbb{R}}F(\rho,\tilde \rho^{(\lambda)})). \label{c12}
\ee
for all $\alpha \in (1/2, 1)$,  where we have defined 
\be  \label{delgen}
\tilde \Delta_{\alpha}= - \frac{2\alpha}{1-\alpha} \log ||\sN_2 \circ \sN_1(\rho)^{\frac{1-\alpha}{2\alpha}} (\sN_2(\sigma_2)^{\frac{\alpha-1}{2\alpha}}\otimes \mathbf{1}_{E_2}) V \sigma_2^{\frac{1-\alpha}{2\alpha}} \sN_1(\sigma_1)^{\frac{\alpha-1}{2\alpha}}\otimes \mathbf{1}_{E_2} U \sigma_1^{\frac{1-\alpha}{2\alpha}} \sqrt{\rho}||_{2\alpha} 
\ee
We expect that for general channels $\sN_1$ and $\sN_2$, 
\be 
\lim_{\alpha \to 1}\tilde \Delta_{\alpha} = D(\rho||\sigma_1) - D(\sN_1(\rho)||\sN_1(\sigma_1)) + D(\sN_1(\rho)||\sigma_2) - D(\sN_2\circ\sN_1(\rho)||\sN_2(\sigma_2) )
\ee
For the particular case of Fig. \ref{fig:twostep_2}, this corresponds to 
\begin{align} 
\lim_{\alpha \to 1}\tilde \Delta_{\alpha} &=  I(AB:CD)-I(AB:C) + I(A:BC)-I(A:B) \label{lima} \\
&= I(A:C|B) + I(B:D|C) + I(A:D|BC) \\
& 
=  -S(ABCD)+S(CD)-S(C)+S(BC)-S(B)+S(AB) \, . \label{c17}
\end{align}
and therefore \eqref{c12} implies \eqref{q2}. Let us explicitly show \eqref{lima}. For the choices of Fig. \ref{fig:twostep_2}, we have 
\be \label{del}
\tilde\Delta_{\alpha} =  \frac{2\alpha}{\alpha-1}\log ||\rho_{AB}^{\frac{1-\alpha}{2\alpha}} \rho_B^{\frac{\alpha-1}{2\alpha}} \rho_{BC}^{\frac{1-\alpha}{2\alpha}}\rho_C^{\frac{\alpha-1}{2\alpha}} \rho_{CD}^{\frac{1-\alpha}{2\alpha}} \sqrt{\rho_{ABCD}} ||_{2\alpha}
\ee
On the other hand, we can write the combination of entropies in \eqref{c17} as 
\begin{align}
    &-S(ABCD)+S(CD)-S(C)+S(BC)-S(B)+S(AB)\nonumber\\
    &\qquad=D\left(\rho_{ABCD}||\exp\left(\log\rho_{CD}-\log\rho_C+\log\rho_{BC}-\log\rho_B+\log\rho_{AB}\right)\right)\nonumber\\
&\qquad=\lim_{\alpha\rightarrow 1}\tilde{D}_{\alpha}\left(\rho_{ABCD}||\exp\left(\log\rho_{CD}-\log\rho_C+\log\rho_{BC}-\log\rho_B+\log\rho_{AB}\right)\right) \label{c19}
\end{align}
where $\tilde D_{\alpha}$ is the Sandwiched Renyi relative entropy \cite{sand}, 
\be
\tilde{D}_{\alpha}(\rho||\sigma)\equiv\frac{1}{\alpha-1}\log\text{Tr}\left[\left(\sigma^{(1-\alpha)/2\alpha}\rho\sigma^{(1-\alpha)/2\alpha}\right)^{\alpha}\right] = \frac{2\alpha}{\alpha-1} ||\sqrt{\rho} \sigma^{\frac{1-\alpha}{2\alpha}} ||_{2\alpha}
\ee
Next, note that for any matrices $A_i$, 
\be 
\lim_{n \to \infty} \le(\prod_i A_i^{1/n}\ri)^{n} \approx \exp\le(\sum_i \log A_i\ri) \label{id21}
\ee
where the product on the left-hand side can be taken in any order. Using this for the exponential in \eqref{c19}, with $n = (2\alpha)/(1-\alpha)$, we find that the final expression in \eqref{c19} is precisely the $\alpha \to 1$ limit of \eqref{del} (using the fact that $X^{\dagger}X$ has the same eigenvalues as $X X^{\dagger}$ for any $X$).

The proof of \eqref{q3} can be seen very similarly by taking $\sN_1 = \Tr_D$, $\sigma_1= \rho_{AB} \otimes \rho_{CD}$, and $\sN_2 = \Tr_A$, $\sigma_2= \rho_{AB} \otimes \rho_{C}$ in the above discussion. In particular, for \eqref{delgen} in this case, we get a product of the same set of matrices $\rho_{AB}^{\frac{1-\alpha}{2\alpha}}$,  $\rho_B^{\frac{\alpha-1}{2\alpha}}$,  $\rho_{BC}^{\frac{1-\alpha}{2\alpha}}$, $\rho_C^{\frac{\alpha-1}{2\alpha}}$,  $\rho_{CD}^{\frac{1-\alpha}{2\alpha}}$ that appear in \eqref{del} in a different order. Recall that in the $\alpha \to 1$ limit, from \eqref{id21}, we can choose to put the matrices in the exponent in \eqref{c19} in any order. 

 \bibliographystyle{jhep.bst}
\bibliography{markov.bib}
\end{document}